\documentclass[paper]{JHEP3}
\usepackage{epsfig}
\usepackage{amsmath}
\usepackage{amssymb}
\usepackage{amsbsy}
\usepackage{enumerate}
\usepackage{bbm}
\usepackage{url}

\usepackage[hang,sf]{subfigure}

\newtheorem{property}{Property}

\newcommand{\tmmathbf}[1]{\ensuremath{\boldsymbol{#1}}}
\newcommand{\tmop}[1]{\ensuremath{\operatorname{#1}}}

\newcommand{\tmtextit}[1]{{\itshape{#1}}}
\newcommand{\tmtexttt}[1]{{\ttfamily{#1}}}
\newcommand{\tmtextup}[1]{{\upshape{#1}}}

\newenvironment{enumeratenumeric}{\begin{enumerate}[1.] }{\end{enumerate}}
\newenvironment{enumerateroman}{\begin{enumerate}[i.] }{\end{enumerate}}
\newenvironment{itemizedot}{\begin{itemize} }{\end{itemize}}
\newenvironment{itemizeminus}{\begin{itemize} }{\end{itemize}}

\def\beq{\begin{equation}}
\def\beqn{\begin{eqnarray}}
\def\eeq{\end{equation}}
\def\eeqn{\end{eqnarray}}
\def\abs#1{\left|#1\right|}

\newcommand\HERWIG{{\tt HERWIG}}

\newcommand\PYTHIA{{\tt PYTHIA}}

\newcommand\MADGRAPH{{\tt MADGRAPH}}
\newcommand\TOPDRAWER{{\tt TOPDRAWER}}

\def\lq{\left[} 
\def\rq{\right]}

\def\({\left(} 
\def\){\right)}

\newcommand\sss{\mathchoice%
{\displaystyle}%
{\scriptstyle}%
{\scriptscriptstyle}%
{\scriptscriptstyle}%
}

\newcommand\nplus{\oplus}
\newcommand\nminus{\ominus}

\newcommand\splus{{\sss \nplus}}
\newcommand\sminus{{\sss \nminus}}

\newcommand\alr{{\alpha_{\sss\rm r}}}
\newcommand\Nrr{{N_{f_b}^{\tmop{rr}}}}
\newcommand\Phin{{\bf \Phi}_n}
\newcommand\Phinpo{{\bf \Phi}_{n+1}}

\newdimen\hbigcirc
\newdimen\wbigcirc

\newdimen\figwidth

\figwidth=\textwidth

\newcommand\ep{\epsilon}

\newcommand\as{\alpha_{\sss\rm S}}

\newcommand\pt{p_{\sss\rm T}}

\newcommand\kt{k_{\sss\rm T}}
\newcommand\ktvec{\vec{k}_{\sss\rm T}}

\newcommand\mur{\mu_{\sss\rm R}}
\newcommand\muf{\mu_{\sss\rm F}}

\newcommand\matB{{\cal B}}

\newcommand\MCatNLO{{\tt MC@NLO}}

\newcommand     \MSB            {\ifmmode {\overline{\rm MS}} \else
                                 $\overline{\rm MS}$  \fi}

\newcommand\CA{C_{\sss\rm A}}

\newcommand\CF{C_{\sss\rm F}}
\newcommand\TF{T_{\sss\rm F}}

\newcommand\NF{n_{\rm f}}

\newcommand\POWHEG{{\tt POWHEG}}
\newcommand\POWHEGBOX{{\tt POWHEG BOX}}

\newcount\minutes 
\newcount\scratch 
 
\def\timestamp{%
\scratch=\time 
\divide\scratch by 60 
\edef\hours{\the\scratch} 
\multiply\scratch by 60 
\minutes=\time 
\advance\minutes by -\scratch 
---$\,$\hours:\null 
\ifnum\minutes< 10 0\fi 
\the\minutes} 
 
\preprint{
DESY 10-018\\
SFB/CPP-10-22\\
IPPP/10/11\\
DCPT/10/22}

\title{A general framework for implementing NLO calculations 
in shower Monte Carlo programs: \\the POWHEG BOX}
\author{Simone Alioli\\
Deutsches Elektronen-Synchrotron DESY\\ 
Platanenallee 6, D-15738 Zeuthen, Germany\\
  E-mail: \email{simone.alioli@desy.de}}
\author{Paolo Nason\\
  INFN, Sezione di Milano-Bicocca,
  Piazza della Scienza 3, 20126 Milan, Italy\\
  E-mail: \email{Paolo.Nason@mib.infn.it}}
\author{Carlo Oleari\\
  Universit\`a di Milano-Bicocca and INFN, Sezione di Milano-Bicocca\\
  Piazza della Scienza 3, 20126 Milan, Italy\\
  E-mail: \email{Carlo.Oleari@mib.infn.it}}
\author{Emanuele Re\\
  Institute for Particle Physics Phenomenology, Department of Physics\\
  University of Durham, Durham, DH1 3LE, UK\\
  E-mail: \email{emanuele.re@durham.ac.uk}}
\abstract{
In this work we illustrate the \POWHEGBOX{}, a general computer code
framework for implementing NLO calculations in shower Monte Carlo
programs according to the \POWHEG{} method. Aim of this work is to
provide an illustration of the needed theoretical ingredients, a view
of how the code is organized and a description of what a user should
provide in order to use it.
}
\keywords{QCD, Monte Carlo, NLO Computations, Resummation, Collider Physics
\vfill 
\vfill 
}

\begin{document}

\section{Introduction}
The \POWHEG{} method is a prescription for interfacing NLO calculations with
parton shower generators. It was first suggested in ref.~\cite{Nason:2004rx},
and was described in great detail in ref.~\cite{Frixione:2007vw}.  Until now,
the \POWHEG{} method has been applied to the following processes: $Z$ pair
hadroproduction~\cite{Nason:2006hf}, heavy-flavour
production~\cite{Frixione:2007nw}, $e^+ e^-$ annihilation into
hadrons~\cite{LatundeDada:2006gx} and into top
pairs~\cite{LatundeDada:2008bv}, Drell-Yan vector boson
production~\cite{Alioli:2008gx,Hamilton:2008pd}, $W'$
production~\cite{Papaefstathiou:2009sr}, Higgs boson production via gluon
fusion~\cite{Alioli:2008tz,Hamilton:2009za}, Higgs boson production
associated with a vector boson (Higgs-strahlung)~\cite{Hamilton:2009za},
single-top production~\cite{Alioli:2009je}, $Z+1$~jet
production~\cite{POWHEG_Zjet}, and, very recently, Higgs production in vector
boson fusion~\cite{Nason:2009ai}.  Unlike \MCatNLO{}~\cite{Frixione:2002ik},
\POWHEG{} produces events with positive (constant) weight, and, furthermore,
does not depend on the Monte Carlo program used for subsequent showering.  It
can be easily interfaced to any modern shower generator and, in fact, it has
been interfaced to \HERWIG{}~\cite{Corcella:2000bw, Corcella:2002jc} and
\PYTHIA{}~\cite{Sjostrand:2006za} in refs.~\cite{Nason:2006hf,
  Frixione:2007nw, Alioli:2008gx, Alioli:2008tz, Alioli:2009je,
  Nason:2009ai}.

The present work introduces a computer framework that implements in practice
the theoretical construction of ref.~\cite{Frixione:2007vw}. We call this
framework the \POWHEGBOX{}. The aim of the \POWHEGBOX{} is to construct a
\POWHEG{} implementation of an NLO process, given the following ingredients:
\begin{enumerateroman}
\item The list of all flavour structures of the Born processes.

\item The list of all flavour structures of the real processes.
 
\item  The Born phase space.

\item \label{item:born} The Born squared amplitudes ${\cal B}$, the color
  correlated ones ${\cal B}_{ij}$ and spin correlated ones ${\cal B}_{\mu\nu}$.
 These are common ingredients of NLO calculations performed
  with a subtraction method.
  
\item \label{item:real} The real matrix elements squared for all relevant
  partonic processes.

\item \label{item:virtual} The finite part of the virtual corrections
  computed in dimensional regularization or in dimensional
  reduction.
    
\item The Born colour structures in the limit of a large  number of colours.
\end{enumerateroman}

With the exception of the virtual corrections, all these ingredients are
nowadays easily obtained. A matrix element program (like \MADGRAPH) can be
used to obtain~(\ref{item:born}) and~(\ref{item:real}).  The
colour-correlated and spin-correlated Born amplitudes are also generated
automatically by programs like MadDipole~\cite{Frederix:2008hu} and
AutoDipole~\cite{Hasegawa:2009tx}.  Recent progress in the automatization of
the virtual cross section calculation may lead to developments where even the
virtual contribution~(\ref{item:virtual}) may be obtained in a painless
way~\cite{Binoth:2010xt, Hahn:1998yk, Kurihara:2002ne, Belanger:2003sd,
Ellis:2007br, Ossola:2007ax, Binoth:2008uq, Berger:2008sj, Lazopoulos:2008ex,
Winter:2009kd}.  Given the ingredients listed above, the \POWHEGBOX{} does
all the rest.  It automatically finds all the singular regions, builds the
soft and collinear counterterms and the soft and collinear remnants, and then
generates the radiation using the \POWHEG{} Sudakov form factor.

The purpose of this work is twofold. Our first aim is to complete here the
theoretical work of ref.~\cite{Frixione:2007vw}, by explaining several
variants of the procedure illustrated there, that have turned out to be
necessary to fulfill our goal. In doing so, we will refer often to formulae
given in ref.~\cite{Frixione:2007vw}, that is thus a prerequisite for reading
the present work. Our second aim is to illustrate specifically how the code
really works. It is true to some extent that well-written codes are self
explanatory, and, in fact, we tried to write the \POWHEGBOX{} as
transparently as possible. However, what may not be so easy to understand
from the code is its global organization. We believe that this document,
together with the source code, could be used by others to understand the code
up to the point of being able to modify it.

Strictly speaking, the present work is neither the documentation of the
\POWHEGBOX{} code, nor a description of its theoretical basis. So, for
example, we do not include a rigorous description of all the subroutines in
the code, and, as we already said, we refer to ref.~\cite{Frixione:2007vw}
for the theoretical bases of our method. In general, one expects that a
theoretical paper should include the description of how a given calculation
has been performed. This normally includes the illustration of some algebraic
steps, and, maybe, an indication of what part of the calculation was
performed numerically, so that the reader should be able, on the basis of the
given indications, to verify its content. In the present case, the
calculation is performed relying heavily on computer algorithms, and the only
realistic way to verify its content is to understand the code. Thus, here we
explain the algorithms and give sufficient indications on the code structure
for the reader to understand it and verify its correctness. We believe that a
detailed documentation of the \POWHEGBOX{} is not necessary for this purpose,
since the details are better understood by directly studying the
code. Furthermore, the code itself will unavoidably evolve with time, so that
detailed documentation may not be so helpful after all. In summary, this
paper should be simply seen as a description of our calculations, that, being
performed essentially by a computer program, must, to some extent, coincide
with a description of the program itself.

Researchers wishing to use the \POWHEGBOX{} should not need, in principle, to
study or understand this whole paper. They only need to know in which format
they have to supply the ingredients that are listed above. These are
summarized in section~\ref{sec:interface}.  Looking at the various
implementation examples included in the code should also help with this
task. In the near future, we will provide a manual that documents in detail
the user interface. The remaining part of this work should be useful in order
to understand better certain features that the \POWHEGBOX{} provides, and
that the user may need, or to implement new features that are not yet
available.

The paper is organized as follows: in section~\ref{sec:interface} we report
all the information needed to interface an NLO program to the \POWHEGBOX{}.
Thus, for example, we specify how the flavour structure of scattering
processes is represented, how to specify their kinematics, and the format
that the Born, virtual and real cross section subroutines must have. In
section~\ref{sec:find_regions} we describe the algorithm used to generate all
the singular regions of the process, and how these are represented in the
computer code. A simple example is also discussed in detail. In
section~\ref{sec:btilde} we describe how the $\tilde{B}$ function is
constructed. This function, when integrated over the radiation variables,
yields the inclusive cross section at fixed underlying Born configuration,
and thus is crucial for the first stage of the event generation, i.e.~the
generation of the underlying Born structure. The computation of $\tilde{B}$
is quite complex, so this section is divided into several subsections,
dealing with the Born contribution, the soft-virtual contribution, the real
contribution and its soft and collinear limits.  In section~\ref{sec:damprem}
we describe the mechanism provided in \POWHEGBOX{} for tuning the part of the
real cross section that is dealt with using the shower Monte Carlo method,
and how the remaining finite part is treated.  This separation into a
singular and a finite part, besides being useful for tuning the \POWHEG{}
output, also provides a method to improve generation efficiency in certain
cases.  In section~\ref{sec:bbinit} we give an overview of the initialization
stage of \POWHEG{}. At this stage the inclusive cross section is computed,
and the appropriate grids are set up for the generation of the underlying
Born configurations. In section~\ref{sec:rad} we describe the second stage of
the event generation, i.e.~the generation of radiation, and in
section~\ref{sec:leshouches} we describe how the \POWHEG{}-generated event is
prepared for further showering by a standard shower Monte Carlo
program. Finally, in section~\ref{sec:conc}, we give our conclusions.

Several appendixes collect further analytical and technical details.  In
appendix~\ref{app:soft} we report the formulae for the soft integrals that
are used in the \POWHEGBOX{}. In appendix~\ref{app:colllim} we report the
collinear limits of the real cross section, that are used in the \POWHEGBOX{}
to build the collinear counterterms. In appendixes~\ref{app:ubfsr}
and~\ref{app:ubisr} we describe the upper bounding functions used in the
generation of radiation, both for final-state and for initial-state
radiation. In appendix~\ref{app:scales} we describe how the renormalization
and factorization scales are set in the \POWHEGBOX{}, and how the strong
coupling constant is computed. Finally, in appendix~\ref{app:misc}, we give a
discussion of a few miscellaneous topics that are useful for using and
understanding the program.

\section{The format of the user subroutines}\label{sec:interface}
By flavour structure of a process we mean the type of all incoming and
outgoing particles. In the program, the flavour type is denoted by an integer
number, so that the flavour structure is a list of integers. The ordering and
numbering of particles follow the rules:
\begin{enumerate}
  \item first particle: incoming particle with positive rapidity
  
  \item second particle: incoming particle with negative rapidity
  
  \item from the third particle onward: final-state particles ordered as
  follows
  \begin{itemizeminus}
    \item colourless particles first,
    
    \item massive coloured particles,
    
    \item massless coloured particles.
  \end{itemizeminus}
\end{enumerate}
The flavour is taken incoming for the two incoming particles and outgoing for
the final-state particles.

The flavour index is assigned according to the Particle Data Group
conventions~\cite{Amsler:2008zzb}, except for the gluons, where 0 (rather
than 21) is used. \\

\noindent
{\bf Example}: if we are interested in the
associated production of a $t$ quark and a vector boson $Z$ plus two jets
\begin{equation}
pp \to Z \, t + 2 {\rm \ jets}\,,
\eeq
then one of its contributing subprocesses is 
\beq
\label{eq:exa1}
b\, u \to Z\, t\, s \,g \,, 
\eeq
whose flavour structure, according to the previous rules, is given by
\beq
\label{eq:exa1_flst}
\lq  5,\, 2, \,23, \,6,\, 3,\, 0 \rq.
\end{equation}
In QCD calculations, the colourless particles and the massive coloured
particles will remain the same at the Born and NLO level. In the NLO
calculation, the flavour structure of real graphs will have one more light
parton in the final state. The virtual term, being the interference of the
Born and of the one-loop amplitude, has the same flavour structure of the
Born term.

In the \POWHEGBOX, the header file \tmtexttt{pwhg\_flst.h}, in the
\tmtexttt{include} subdirectory, contains all arrays and parameters having to
do with flavour structures. They depend upon the parameters
\tmtexttt{nlegborn} and \tmtexttt{nlegreal} that are set to the number of
legs (incoming plus outgoing) of the Born (or virtual) and real graphs. The
user of the \POWHEGBOX{} will have to set explicitly \tmtexttt{nlegborn} in
the \tmtexttt{nlegborn.h} file, that is thus included before the
\tmtexttt{pwhg\_flst.h} file in all program units that need to access the
flavour structures.  In the example~(\ref{eq:exa1}), the user should set
\tmtexttt{nlegborn=6}.  The variable \tmtexttt{nlegreal} is always set to
\tmtexttt{nlegborn+1}.  The user should also set the variables
\tmtexttt{flst\_nborn} and \tmtexttt{flst\_nreal} to the number of
inequivalent flavour structures for the Born and real graphs, and should also
fill the arrays

\tmtexttt{flst\_born(k=1:nlegborn,\,j=1:flst\_nborn)}

\tmtexttt{flst\_real(k=1:nlegreal,\,j=1:flst\_nreal)}\\
in an appropriate initialization subroutine, named
\tmtexttt{init\_processes}.  Notice that flavour structures that are
equivalent under a permutation of final-state particles should never appear
in the list. Thus, in the example of eqs.~(\ref{eq:exa1})
and~(\ref{eq:exa1_flst}), either $\lq 5,\, 2, \,23, \,6,\, 3,\, 0 \rq$ or
$\lq 5,\, 2, \,23, \,6,\, 0,\, 3 \rq$ may appear as flavour structures, but
not both.

The user should also set the value of \tmtexttt{flst\_lightpart}, the
position of the first light coloured parton. Then, in the example previously
described, \tmtexttt{flst\_lightpart=5}.

\subsection{Tagging parton lines}
At times it is convenient to treat lines with the same flavour as if they
were different. 

\begin{figure}[tbh]
\begin{center}
  \epsfig{file=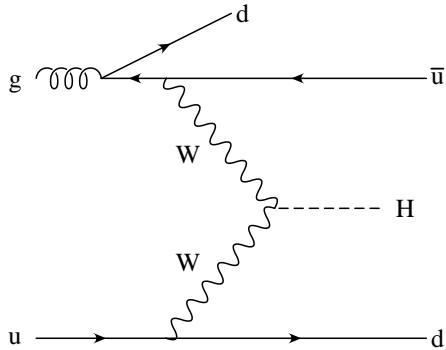,width=0.4\textwidth}
\end{center}
\caption{\label{fig:wwfus} Example of NLO gluon-initiated correction to Higgs
  boson production in VBF: $ g u\rightarrow H \bar{u} d d$.}
\end{figure}

\begin{figure}[tbh]
\begin{center}
  \epsfig{file=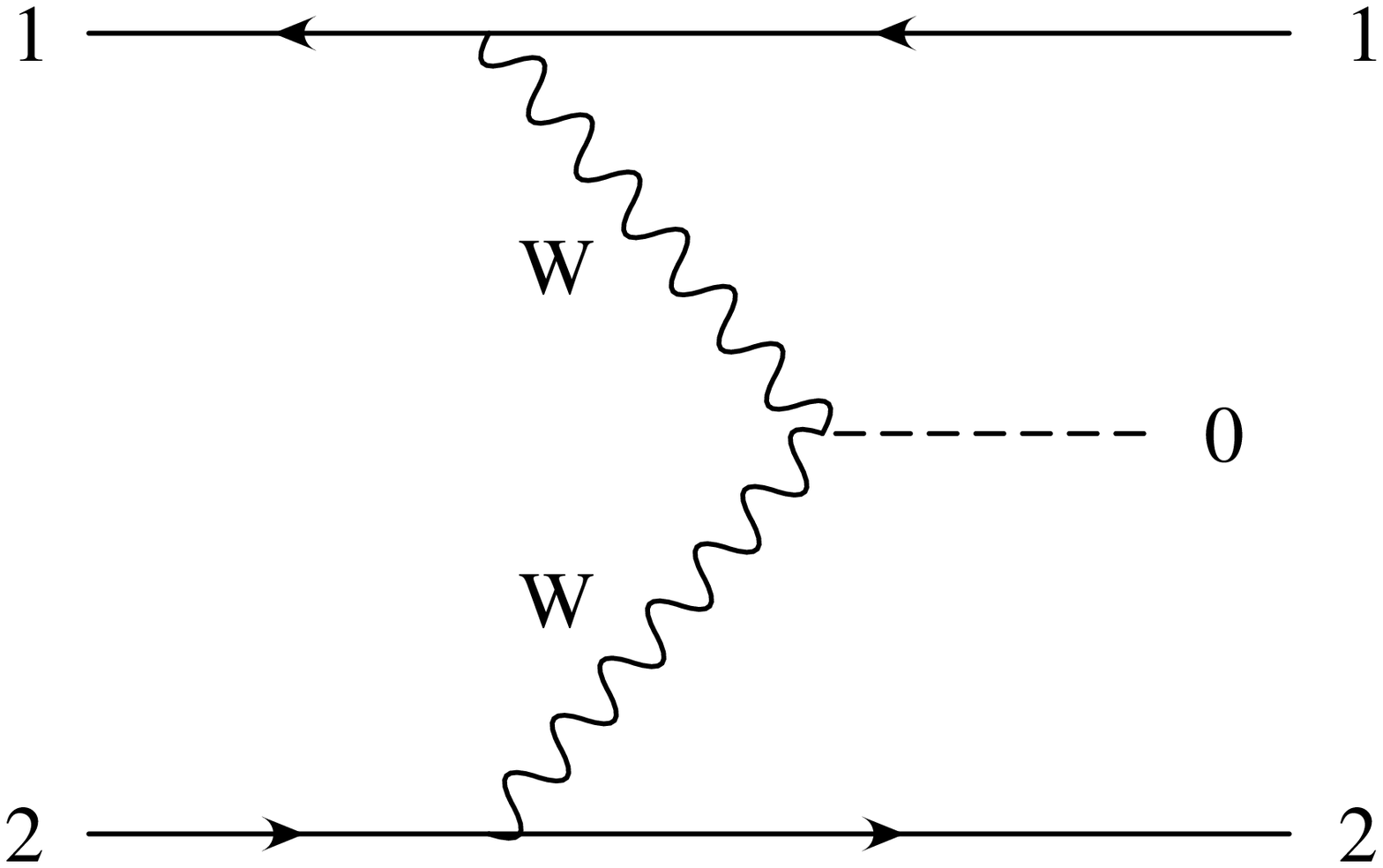,width=0.35\textwidth}\qquad\quad
  \epsfig{file=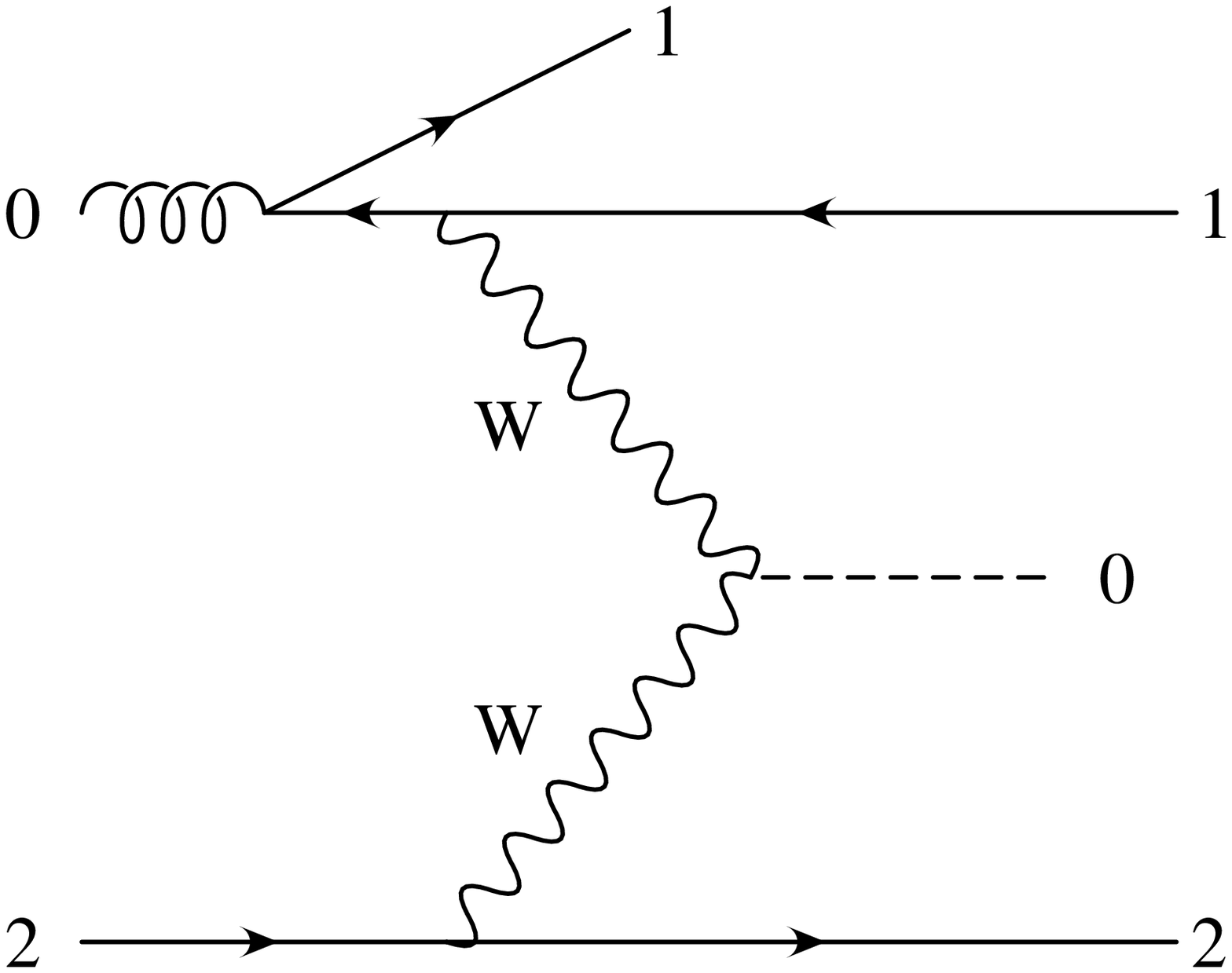,width=0.35\textwidth}
\end{center}
  \caption{\label{fig:wwfustags} Tag assignment for the underlying Born graph
  $\bar d u \to H \bar u d$ and its gluon-initiated real diagram $ g u\to H
  \bar{u} d d$. }
\end{figure}

One such example is Higgs boson production via vector boson fusion~(VBF).
The fermion lines attached to the vector bosons may be treated as being
distinct. Of course, in $W^+ W^-$ fusion, they may also be effectively
distinct, with the $W^+$ coming from a $u$ quark turning into a $d$, and the
$W^-$ coming from a $c$ quark turning into an $s$. Consider however the real
graph depicted in fig.~\ref{fig:wwfus}.
It corresponds to a gluon-initiated next-to-leading correction to VBF Higgs
boson production: $g u \rightarrow H \bar{u} d d$. It is clear that the two
$d$ quarks in the final state have a very different role, and should be kept
distinct. However, as far as the flavour combinatorics is concerned, they are
considered identical in the \POWHEGBOX{}, that assumes that the graphs are
already symmetrized with respect to identical final-state particles. Thus,
the combinatoric algorithm will generate two regions for this graph,
corresponding to either $d$ being collinear to the incoming gluon. In order
to overcome this problem, the \POWHEGBOX{} allows the possibility to
attribute a tag to each line, so that lines with the same flavour but
different tags will be treated differently from the combinatoric point of
view. In the example at hand, one assigns the tags according to the scheme in
fig.~\ref{fig:wwfustags}. We arbitrarily assign a tag equal to zero to
particles that we do not need to tag (the initial-state gluon and the
produced Higgs boson).
Within this scheme, the two final-state $d$ quarks will be treated as
different from the combinatoric point of view. Only the quark tagged as $1$
in the real graph will generate a singular region, since if quark 2 were
collinear to the incoming gluon, the associated underlying Born would have an
incoming antiquark tagged as 2, and thus would not be present.

In the code, tagged lines are treated by generating an internal flavour index
that replaces the real flavour index, in such a way that the internal flavour
is different for lines with different flavour or different tag. The
combinatorics is carried out with these internal flavour numbers. At the end,
internal flavour numbers are replaced with the original real flavour numbers.
From this point on, tags are ignored.  The tags are set to zero during
\POWHEG{} initialization. If needed, the user's initialization routine should
appropriately set the arrays \tmtexttt{flst\_borntags} and \tmtexttt{\tt
  flst\_realtags}.

The task of changing the flavour indexes to account for different tags, and
of changing them back to the original state, are carried out by the routines
\tmtexttt{mapflavours} and \tmtexttt{unmapflavours}, that are called near the
beginning and near the end of the \tmtexttt{genflavreglist} routine, the
subroutine that finds the different singular regions, as described in
section~\ref{sec:find_regions}.

\subsection{The Born phase space}
\label{sec:born}
The Born phase space (provided by the user of the \POWHEGBOX) is a subroutine
named \tmtexttt{born\_phsp(xborn)} that fills the four-momenta of
Born-process particles (both in the laboratory and in the center-of-mass
frame), the Bjorken $x$ of the two incoming partons, the minimal mass of the
Born system and the phase space volume.

It receives as input \tmtexttt{xborn}, an array of real numbers, in the range
$[0,1]$. If no resonances are present in the final state, the dimension of
this array is $3\,n - 2$, $n$ being the number of final-state particles,
i.e.~$n=$\tmtexttt{nlegborn}$-2$ . In case some resonances are present, their
virtuality will require one more variable, and the user will have to take
account of them too, increasing accordingly the numbers of elements of this
array.

In the paper, we will refer to this set of numbers as $X_{\tmop{born}}$. We
recall that the Born phase space $\tmmathbf{\Phi}_n$, defined
in~\cite{Frixione:2007vw}, is given by
\begin{equation}
  d \tmmathbf{\Phi}_n = d x_{\oplus} \,d x_{\ominus} (2 \pi)^4 \delta^4 \left(
  k_{\oplus} + k_{\ominus} - \sum_{i = 1}^n k_i \right) \prod_{i = 1}^n
  \frac{d^3 k_i}{(2 \pi)^3 2 k^0_i}\, .
\end{equation}
The \tmtexttt{born\_phsp} routine should perform the following tasks:
\begin{enumeratenumeric}
  \item Set \tmtexttt{kn\_pborn(mu=0:3,\,k=1:nlegborn)} and
    \tmtexttt{kn\_cmpborn(mu=0:3,\,k=1:nlegborn)}{\footnote{All variables
	with the \tmtexttt{kn\_} prefix are defined in the header file
	\tmtexttt{pwhg\_kn.h}.}} to the Born momenta in the laboratory frame
    and in the center-of-mass~(CM) frame. The Lorentz index $\mu=0$ denotes
    the time component, $1, 2$ the transverse components $x, y$, and 3 the
    longitudinal component $z$.  Set the variables \tmtexttt{kn\_xb1} and
    \tmtexttt{kn\_xb2} to the value of the parton momentum fraction
    $x_\oplus$ and $x_\ominus$.  Set the variable \tmtexttt{kn\_sborn} to the
    squared CM energy of the Born process.
  \item The array \tmtexttt{kn\_masses} should be filled with the masses of
    the legs of the process. Furthermore, the variable \tmtexttt{kn\_minmass}
    should be set to a fixed (i.e.~independent upon the kinematics) lower
    bound on the mass of the final state. Thus, if no resonances are present,
    it is typically set to the sum of the masses of the final-state
    particles. If there are resonances, it will be set to the sum of the
    lower limits of the windows imposed around the resonances.
  
  \item Set the variable \tmtexttt{kn\_jacborn} to the Jacobian
  \begin{equation}
    J_{\tmop{born}} = \left| \frac{\partial \tmmathbf{\Phi}_n}{\partial
      X_{\tmop{born}}} \right| .
  \end{equation}
\end{enumeratenumeric}

\subsection{The Born and Born-correlated squared amplitudes}
The user of the \POWHEGBOX{} should provide the routine

\tmtexttt{setborn(p(0:3,1:nlegborn),\,bflav(1:nlegborn),\,born,\,}

\tmtexttt{\qquad\qquad
  bornjk(1:nlegborn,1:nlegborn),\,bmunu(0:3,0:3,1:nlegborn))}.\\ 
Given the four-momenta \tmtexttt{p} and the flavour structure
\tmtexttt{bflav} of a Born subprocess, the routine should return the Born
squared matrix element $2s_b\mathcal{B}$ in \tmtexttt{born}, the colour
correlated one in \tmtexttt{bornjk} and the spin correlated one in
\tmtexttt{bmunu}. The flux factor $1/(2\,s_b)=$\tmtexttt{1/(2*kn\_sborn)}
(where $s_b$ is the center-of-mass energy squared of the Born process) should
not be included,\footnote{In the notation of ref~\cite{Frixione:2007vw},
  $\mathcal{B}$ includes the flux factor} since it is supplied by the
\POWHEGBOX.

The colour correlated Born amplitude is defined in eq.~(2.97) of
ref.~\cite{Frixione:2007vw}. We report it here for completeness
\begin{equation}
\label{eq:colourcorr}
2s_b\matB_{ij}=-N
\sum_{\stackrel{{\mbox{\rm
      \scriptsize 
      spins}}}{{\mbox{\rm \scriptsize colours}}}}
{\cal M}_{\{c_k\}}\, \left( {\cal M}^\dagger_{\{c_k\}} \right)_{
\stackrel{\scriptstyle c_i \to c'_i}{c_j \to c'_j}}\;
 T^a_{c_i,c'_i} \, T^a_{c_j,c'_j}\,.
\end{equation}
Here ${\cal M}_{\{c_k\}}$ is the Born amplitude, and $\{c_k\}$ stands for the
colour indexes of all external coloured particles in the amplitude.  The
suffix on the parentheses that enclose ${\cal M}^\dagger_{\{c_k\}}$ indicates
that the colour indexes of partons $i,j$ are substituted with primed indexes
in ${\cal M}^\dagger_{\{c_k\}}$.  The factor $N$ is the appropriate
normalization factor including averages over initial spin and colour and
symmetry factors.  We assume summation over repeated colour indexes ($c_k$,
$c'_i$, $c'_j$ and $a$) and spin indexes.  For gluons $T^a_{cb}=if_{cab}$,
where $f_{abc}$ are the structure constants of the $SU(3)$ algebra. For
incoming quarks $T^a_{\alpha\beta}=t^a_{\alpha\beta}$, where $t$ are the
colour matrices in the fundamental representation (normalized as ${\rm
  Tr}[t\,t]=1/2$). For antiquarks $T^a_{\alpha\beta} =-t^a_{\beta\alpha}$. It
follows from colour conservation that $\matB_{ij}$ satisfy
\begin{equation}\label{eq:colcons}
\sum_{i,i\ne j}\matB_{ij}=C_{f_j}\matB\,,
\end{equation}
where $i$ runs over all coloured particles entering or exiting the process,
and $C_{f_j}$ is the Casimir constant for the colour representation of
particle $j$.  The spin correlated Born squared amplitude $\mathcal{B}^{\mu
  \nu}_j$ is defined to be non-zero if the $j^{\tmop{th}}$ Born leg is a
gluon, and is basically the Born cross section obtained by leaving the gluon
indexes of the $j^{\tmop{th}}$ leg uncontracted. More precisely, we can write
\begin{equation}
  \mathcal{B}_j^{\mu \nu} = N \sum_{\{i\}, s_j, s'_j} \mathcal{M} \(\{i\},
  s_j\)\, \mathcal{M}^{\dagger} \(\{i\}, s'_j\)\, (\epsilon^{\mu}_{s_j})^*\,
  \epsilon^{\nu}_{s'_j}\,,
\end{equation}
where $\mathcal{M} \(\{i\}, s_j\)$ is the Born amplitude, $\{i\}$ represent
collectively all remaining spins and colours of the incoming and outgoing
particles, and $s_j$ represents the spin of the $j^{\tmop{th}}$ particle. The
$\epsilon^{\mu}_{s_j}$ are polarization vectors, normalized as
\begin{equation}
  \sum_{\mu, \nu} g_{\mu \nu}\, (\epsilon^{\mu}_{s_j})^*\, \epsilon^{\nu}_{s'_j}
  = - \delta_{s_j s'_j}\, .
\end{equation}
Thus
\begin{equation}
 \sum_{\mu, \nu} g_{\mu \nu} \, \mathcal{B}_j^{\mu \nu} = -
 \mathcal{B}\, . 
\end{equation}
Notice that the Born squared amplitude is requested for each individual
flavour structure of the contributing subprocesses. Many different flavour
structures will return identical or proportional values of the Born cross
section. For example $d \bar{d} \rightarrow Z$ is identical to $s \bar{s}
\rightarrow Z$, and $u \bar{u} \rightarrow \gamma^*$ is proportional to $d
\bar{d} \rightarrow \gamma^*$. The \POWHEGBOX{} identifies these identical
contributions initially, and stores the proportionality constants. When
computing the Born cross section for all needed flavour structures, it
computes only the minimum number of squared amplitudes it needs, and obtains
the others using the proportionality relations found initially.

\subsection{The virtual amplitudes}
\label{sec:virtual}
The user should provide a subroutine 

\tmtexttt{setvirtual(p(0:3,1:nlegborn),vflav(1:nlegborn),virtual)},\\
that returns in \tmtexttt{virtual} the finite part $\mathcal{V}_{\sss\rm
  fin}$ of the virtual cross section for the process with flavour structure
\tmtexttt{vflav} and external momenta \tmtexttt{p}.  The
$\mathcal{V}_{\sss\rm fin}$ contribution is defined, in conventional
dimensional regularization (after renormalization), by
\begin{equation}\label{eq:virtform}
\mathcal{V}_b=\mathcal{N} \, \frac{\as}{2 \pi} \Bigg[ \frac{1}{\ep^2}\,a\,
\mathcal{B} +\frac{1}{\ep}\sum_{i,j} c_{ij}\,
\mathcal{B}_{ij}+\mathcal{V}_{\sss\rm fin}\Bigg],
\end{equation}
where the $a$ and $c_{ij}$ coefficients do not depend upon $\ep$.  Their
explicit form can be found, for example, in ref.~\cite{Frederix:2009yq}. The
normalization factor is
\begin{equation}\label{eq:Ndef}
  \mathcal{N} = \frac{(4 \pi)^{\epsilon}}{\Gamma (1 - \epsilon)} \left(
  \frac{\mur^2}{Q^2} \right)^{\epsilon}.
\end{equation}
Here it is important to remark that, in the conventional dimensional
regularization, $\mathcal{B}$ and $\mathcal{B}_{ij}$ in
formula~(\ref{eq:virtform}) are evaluated in $(4-2\ep)$ dimensions, and thus
do depend upon $\ep$.  In fact, all formulas for the soft contributions and
the collinear remnants used in the \POWHEGBOX{} are computed in the \MSB
scheme, dropping the $1/\ep^2$ and $1/\ep$ contributions written in terms of
the $(4-2\ep)$-dimensional expression of the Born squared amplitudes and with
the normalization factor of eq.~(\ref{eq:Ndef}) (see for example
appendix~\ref{app:soft}). Thus, if the virtual contribution is written as in
formula~(\ref{eq:virtform}), the divergent terms dropped in the \POWHEGBOX{}
cancel exactly the $1/\ep$ and $1/\ep^2$ terms in~(\ref{eq:virtform}),
leaving the finite contribution $\mathcal{V}_{\sss\rm fin}$.

If the NLO calculation has been performed in the Dimensional Reduction~(DR)
scheme, in order to use it within the \POWHEGBOX{}, we do the following.
First of all, we assume that the NLO result is expressed in terms of the
standard \MSB coupling constant. If this is not the case, the appropriate
straightforward corrections need to be applied. Then one defines
$\mathcal{V}_{\rm fin}^{\rm DR}$ using the same formula~(\ref{eq:virtform})
where now $\mathcal{B}$ and $\mathcal{B}_{ij}$ are evaluated in four
space-time dimensions. In other words, $\mathcal{V}_{\rm fin}^{\rm DR}$ is
defined as the zeroth order coefficient of the Taylor expansion of
$\mathcal{V}_b^{\rm DR}2\pi/(\as \mathcal{N})$ in $\ep$.  The expression to
be used in the \POWHEGBOX{} is~\cite{Kunszt:1993sd}
\begin{equation}
  \mathcal{V}_{\rm fin} = \mathcal{V}_{\rm fin}^{\rm DR} - \mathcal{B}
  \,\sum_i \tilde{\gamma} (f_i)\,,
\end{equation}
where $\tilde{\gamma} (g) = N_c / 6$, and $\tilde{\gamma} (q) = (N_c^2 - 1) /
(4 N_c)$, with the index $i$ running over all coloured massless partons of
the amplitude and where $N_c$ is the number of colours.

The scale $Q$ in eq.~(\ref{eq:Ndef}) is completely arbitrary in this context.
It may be chosen for convenience equal to $s$ or equal to the
renormalization/factorization scale. Within the \POWHEGBOX{} it has been
fixed to be equal to the renormalization scale $\mur$, and so also the user
should set it to this value.

\subsection{The real squared amplitudes}
Together with the Born and Born-correlated squared amplitudes, the user of
the \POWHEGBOX{} should provide the routine

\tmtexttt{setreal(p(0:3,nlegreal),\,rflav(1:nlegreal),\,amp2)}\\
that computes, given the momenta \tmtexttt{p} of the external particles, the
squared amplitude\footnote{Averaged over spin and colours; as for the Born
  and the virtual contributions, also in this case the flux factor should not
  be included by the user.} for the real process specified by the flavours
\tmtexttt{rflav}, stripped off by a factor $\as/(2\pi)$.

As for the Born contribution, also the real squared amplitude is requested
for each individual flavour structure contributing to the real cross section.
Many different flavour structures will return identical or proportional
values for the real squared amplitude. The \POWHEGBOX{} identifies these
identical contributions initially, and stores the proportionality
constants. When computing the real cross section for all needed flavour
structures, it calculates only the minimum number of squared amplitudes it
needs, and obtains the others using the proportionality relations found
initially.

\subsection{Analysis routines}
\label{sec:analysis}
In the analysis file, \tmtexttt{pwhg\_analysis.f}, the user provides her/his
own analysis routine to plot kinematic distributions.

In order to have a unique analysis file, able to deal with events at the
partonic or at the hadronic level (i.e.~after the shower), we pass all the
kinematics and properties of the particles via the \tmtexttt{HEPEVT} common
block.  In this way, the \tmtexttt{analysis} subroutine has only to have
access to this common block and to the value of the differential cross
section.  This last subroutine is then completely independent from the
program that has filled the common block, and can be called by \PYTHIA{} or
\HERWIG{} too.

The \POWHEGBOX{}, during the integration of the $\tilde{B}$ function, can
produce plots with fixed NLO accuracy. This is done via a call to the routine
\tmtexttt{analysis\_driver}, that fills the common block with the kinematic
momenta. This routine receives as input parameters the value of the cross
section at that specific kinematic phase-space point and the integer variable
\tmtexttt{ikin}.  If \tmtexttt{ikin} is set to 0, the event is treated as a
Born-like one, and the \tmtexttt{nlegborn} momenta in the
\tmtexttt{kn\_pborn} kinematic common block are copied on the
\tmtexttt{HEPEVT} common block. Otherwise, the subroutine copies on the
common block the \tmtexttt{nlegreal} momenta stored in \tmtexttt{kn\_preal}.
As final step, it invokes the routine \tmtexttt{analysis}, that receives as
input parameter only the value of the cross section.\footnote{In the
  \POWHEGBOX{} package, we have included our own \tmtexttt{analysis} routine,
  that uses \TOPDRAWER{} as histogramming tool for our plots.}

\section{The singular regions}
\label{sec:find_regions}
For each real flavour structure, the \POWHEG{} method requires that one
decomposes the real cross section into the sum of contributions that are
divergent in one singular region only. In the notation of
ref.~\cite{Frixione:2007vw} we write
\begin{equation}
  R = \sum_{\alr} R^{\alr} . \label{eq:alphareg}
\end{equation}
Each $\alr$ is thus associated with a single flavour structure, and a single
singular region. Sometimes we will refer to it as $\alr$ region (or
\tmtexttt{alr} region, which is the name of the variable that we often use to
represent it in the code). The separation in eq.~(\ref{eq:alphareg}) is
illustrated in detail in section 2.4 of ref.~{\cite{Frixione:2007vw}}.

We have investigated two methods for obtaining such separation: the first one
is based on the subtraction method proposed by Catani and
Seymour~(CS)~\cite{Catani:1996vz}, and the second one on the method proposed
by Frixione, Kunszt and Signer~(FKS)~\cite{Frixione:1995ms, Frixione:1997np}.
It was found that the FKS subtraction method is better suited for our
purpose, and so we focused upon it. The difficulties with the CS dipole
method are due to the large number of dipoles and dipole types, and to the
fact that it is not easy to separate the real cross section into a sum of
terms having the same singularity structure of the CS dipoles. It is in fact
not possible to simply weight the real cross sections with factors that
vanish in all but one singular region, since for each singular region there
are several CS dipoles, depending upon the choice of the spectator.

The FKS method is slightly more cumbersome than the CS method when
counterterms are computed using the collinear and soft plus-distributions. It
turns out, however, that this difficulty remains the same for all processes.
The procedure to disentangle the plus-distributions can be coded simply in a
general way once and for all.

In the FKS framework, singular regions are characterized as follows:
\begin{itemizedot}
  \item A final-state parton $i$ is becoming collinear to either
    initial-state partons $j=1,2$, or soft
  
  \item A final-state parton $i$ is becoming collinear to a final-state
    parton $j$, or soft.
\end{itemizedot}
We will call $i$ the \tmtextit{emitted or radiated parton}, and $j$ the
\tmtextit{emitter}. If we replace the pair emitted-emitter with a single
parton of the appropriate flavour, we obtain the flavour configuration of the
underlying Born.

Given the list of flavours of the real graphs, one is faced with the
combinatoric problem of finding all singular regions. In the \POWHEG{}
framework, this task is carried out keeping in mind that we should be able to
group easily all singular regions that share the same underlying Born.
In order to ease this task, it is convenient to choose a standard ordering
for the flavour structure of each \tmtexttt{alr}.  One can easily demonstrate
that the flavour of all the \tmtexttt{alr} can be ordered in such a way that
the two following properties are satisfied:
\begin{property} \label{prop1}
The emitted parton is always the last parton.
\end{property}
\begin{property} \label{prop2}
The underlying Born configuration, obtained by removing the emitted parton,
and replacing the emitter with a parton of the appropriate flavour, is
exactly equal to one of the Born flavour structures present in the list of
Born processes \tmtexttt{\tt flst\_born}.
\end{property}
Property~\ref{prop2} is non-trivial, since in general the underlying Born
structure obtained from a generic \tmtexttt{alr} will be equivalent up to a
permutation to a flavour structure present in the \tmtexttt{\tt flst\_born}
list.

It is clear that putting all the \tmtexttt{alr} in a standard form with the
properties \ref{prop1} and \ref{prop2} simplifies the \POWHEG{}
implementation. In fact, since real contributions sharing the same underlying
Born are often grouped together in \POWHEG{}, it is better if the underlying
Born flavour structures are unique.  This will be illustrated more clearly in
the example described in section~\ref{sec:example}.

In case of initial-state collinear singularities, the emitter will be
assigned the value 1~(2) to distinguish collinear emissions from incoming
line 1~(2) (the $\nplus$ and $\nminus$ directions of
ref.~\cite{Frixione:2007vw}).  If the emitted parton is a gluon, the emitter
will be assigned the conventional value 0, that means that both 1 and 2 may
have emitted it. These distinctions are such to minimize the number of
regions, maintaining however the fact that, to each region, we associate a
unique underlying Born configuration.

The FKS framework, for final-state radiation~(FSR), distinguishes between the
emitter and the emitted parton (often called the FKS parton), in the fact
that only the emitted parton leads to soft singularities. Thus, the emitter
can be a quark, with the emitted parton being a gluon, but not viceversa. On
the other hand, the emitter-emitted can be a quark-antiquark pair. In this
case, it does not matter what we choose to be the emitted parton. By
convention, we will always choose it to be the antiquark. If the emitter and
radiated partons are both gluons, when computing the corresponding $\alr$, we
supply a damping factor that removes the soft singularity of the emitter,
with an appropriate compensating coefficient. For example, if we call $E_{\rm
  em}$ and $E_{\rm r}$ the CM energies of the emitter and emitted parton, the
damping factor may be chosen equal to $E_{\rm em} / (E_{\rm em} + E_{\rm
  r})$, and an extra factor of 2 is supplied to account for the region where
the role of emitter and emitted partons are exchanged.

In the \POWHEGBOX, the task of finding all regions associated with a given
flavour structure is performed by the routine

\tmtexttt{find\_regions(nleg,rflav,nregions,iregions)}\\
where \tmtexttt{nleg=nlegreal} and \tmtexttt{rflav(1:nlegreal)} is the input
flavour structure.  It returns in \tmtexttt{nregions} the number of regions
found, and, for every found region \tmtexttt{j}, it returns the positions (in
the string of flavours) of the emitter and of the emitted parton (arbitrarily
ordered) in \tmtexttt{iregions(1:2,j)}.

The algorithm for finding the final-state regions is the following:
\begin{enumeratenumeric}
  \item Loop over all massless parton pairs
    \tmtexttt{i=flst\_lightpart:nlegreal}, \tmtexttt{j=i+1:nlegreal}.

  
  \item Check if \tmtexttt{i} and \tmtexttt{j} can come from the splitting of
    the same parton (i.e.~if they have opposite flavours, or if they are both
    gluons, or one of them is a gluon).
  
  \item If they cannot come from the same parton flavour, skip them.
  
  \item If they can come from the same parton flavour, build up a flavour
    list with \tmtexttt{nlegborn} elements, obtained from \tmtexttt{rflav} by
    deleting partons \tmtexttt{i,j} and adding a parton with the appropriate
    flavour (i.e.~if \tmtexttt{i,j} have opposite flavour, or are both
    gluons, add a gluon; if one is a gluon add the flavour of the other
    parton). Check if the newly built flavour list is an admissible flavour
    structure for the Born cross section. This is done by checking if the
    Born flavour structure at hand is equivalent, up to a permutation of the
    final-state partons, to any element of the list \tmtexttt{flst\_born}.
 
  \item If the underlying Born flavour structure is valid, increase
    \tmtexttt{nregions} and set

    \tmtexttt{iregions(1,nregions) = i} and
    \tmtexttt{iregions(2,nregions)=j}.
\end{enumeratenumeric}
The initial-state regions are treated similarly. We check, for each
final-state light coloured parton \tmtexttt{j}, if it may come from the
splitting of an initial-state parton.  If it comes from the first (second)
incoming line, then the number of regions \tmtexttt{nregions} is increased
and we set \tmtexttt{iregions(1,nregions) = 1 (2)} and
\tmtexttt{iregions(1,nregions) = j}. If the emitted parton \tmtexttt{j} is a
gluon, then only one region is generated, and we set
\tmtexttt{iregions(1,nregions) = 0}.

The first task of the \POWHEGBOX{} is to build the list of all $\alr$ in the
standard form specified in Properties~\ref{prop1} and~\ref{prop2}.  This task
is performed by the subroutine \tmtexttt{genflavreglist}, that, more
specifically, does the following:
\begin{itemizedot}
  \item It sets the variable \tmtexttt{flst\_nalr} to the total number of
    inequivalent singular regions $\alr$ found.
  
  \item It sets the variable \tmtexttt{flst\_nregular} to the number of real
    flavour structures that do not have any singular region,{\footnote{An
	example of such configurations is given by the $q \bar{q} \rightarrow
	H g$ flavour structure in Higgs boson production via gluon fusion.}}
    and it fills the array

  \tmtexttt{flst\_regular(k=1:nlegreal,\,alr=1:flst\_nregular)}

   with the corresponding flavour structures. If \tmtexttt{flst\_nregular} is
   greater than 0, also the flag \tmtexttt{flg\_withreg} (defined in the
   header file \tmtexttt{pwhg\_flg.h}) is set to true.
  
  \item It fills the array
    \tmtexttt{flst\_alr(k=1:nlegreal,\,alr=1:flst\_nalr)} with the flavour
    structure corresponding to the given \tmtexttt{alr} region. The ordering
    is guaranteed to respect Properties~\ref{prop1} and~\ref{prop2}.
  
  \item It sets the array \tmtexttt{flst\_emitter(alr)} to the emitter of the
    \tmtexttt{alr} structure.
  
  \item It sets the \tmtexttt{flst\_mult(alr)} array to the multiplicity of
    the \tmtexttt{alr}$^{\tmop{th}}$ structure. This number arises because
    regions may be found that are equivalent by permutations of the external
    legs. Only one is retained in this case, with the correct multiplicity.
  
  \item It sets the array \tmtexttt{flst\_uborn(k=1:nlegborn,\,alr)} to the
    flavour structure of the underlying Born of the
    \tmtexttt{alr}$^{\tmop{th}}$ structure.
  
  \item It sets the array \tmtexttt{flst\_alr2born(alr)} to an index in the
    \tmtexttt{flst\_born} array, that points to the underlying Born flavour
    structure of the \tmtexttt{alr}$^{\tmop{th}}$ region.
  
  \item It sets up a pointer structure from the underlying Born index to the
    set of \tmtexttt{alr}'s that share it: it sets the array
    \tmtexttt{flst\_born2alr(0,jborn)} to the number of \tmtexttt{alr}
    regions that have \tmtexttt{jborn} as underlying Born,
    i.e.~\tmtexttt{flst\_alr2born(alr)=jborn}, and sets
    \tmtexttt{flst\_born2alr(i,jborn)}, with
    \tmtexttt{i=1:flst\_born2alr(0,jborn)}, to the corresponding
    \tmtexttt{alr} index.
  
  \item For each \tmtexttt{alr}, a list of all singular regions for the
    corresponding flavour structure is also build. These are needed in order
    to compute $\mathcal{R}_{\alr}$, as we will see further on.  The array
    element \tmtexttt{flst\_allreg(1,0,alr)} is set to the number of singular
    regions of the flavour structure of the \tmtexttt{alr}$^{\rm th}$ region,
    i.e.~\tmtexttt{flst\_allreg(1,0,alr)=nregions} of that particular
    \tmtexttt{alr}. Then \tmtexttt{flst\_allreg(i=1:2,\,k=1:nregions,alr)} is
    set to the list of pairs of indexes characterizing the emitter and the
    emitted parton for each singular region.
\end{itemizedot}
In order to perform this task, the subroutine \tmtexttt{genflavreglist} loops
through the \tmtexttt{flst\_real} list of real flavour structures, calling
for each of them the routine \tmtexttt{find\_regions}. Each region found is
first transformed with a permutation, in such a way that the emitted parton
is always the last in the list. In the case of final-state radiation, the
emitter is also moved near the emitted parton (i.e.~at the
\tmtexttt{nlegreal-1} position) with a permutation. At this stage, one also
makes sure that, if the emitter-emitted pair is made by a quark (antiquark)
and a gluon, the gluon is always the emitted parton, and if the pair is a
quark and antiquark, the antiquark is always the emitted parton (if this is
not the case, emitter and emitted partons are exchanged). The lists
\tmtexttt{flst\_alr} and \tmtexttt{flst\_emitter} are updated
accordingly. Once this procedure is completed, the \tmtexttt{alr} list is
complete, but each element may appear more than once. The list is thus
searched for equivalent elements, it is collapsed in such a way that each
element appears only once, and a multiplicity factor \tmtexttt{flst\_mult} is
setup to keep track of how many occurrences of a given contribution are
present. At the end of this procedure, only inequivalent \tmtexttt{alr}'s
remain, with an associated multiplicity factor. But it may still happen that
the same underlying Born configuration may appear in different \tmtexttt{alr}
with different ordering. At this stage the \tmtexttt{alr} list is scanned
again. If at any point one finds and underlying Born that differs from one
appearing in the \tmtexttt{flst\_born} list, the flavour structure of the
current \tmtexttt{arl} (and its underlying Born flavour structure) are
permuted, in such a way that one recovers the same ordering of one of the
elements appearing in the \tmtexttt{flst\_born} list.  In case of final-state
radiation, the emitter may end up to be no longer the \tmtexttt{nlegreal-1}
leg of the process, and so, the array \tmtexttt{flst\_emitter} is updated
accordingly.

\subsection{Example}
\label{sec:example}
Due to the intrinsic complexity of the procedure for finding the singular
regions, it might be useful to examine it on a simple example.
\begin{table}[tbh]
\begin{center}
  \begin{tabular}{|c|l|l|}
\hline
    \tmtexttt{jborn}&processes &\tmtexttt{flst\_born}\\
\hline
1 & $sc \to gud$        & $[3,4,0,2,1]$ \\
2 & $gu \to \bar{s} s c$  & $[0, 2, -3,3,4]$\\
\hline
  \end{tabular}
  \caption{\label{tab:born} Flavour structure of two Born subprocesses and
    the corresponding \POWHEGBOX{} notation, in the third column.}
\end{center}
\end{table}  
Let us consider a process that, at the Born level, has only two flavour
structures (\tmtexttt{flst\_nborn=2}) and five coloured massless partons
(\tmtexttt{nlegborn=5} and \tmtexttt{flst\_lightpart=3}).  In the second
column of table~\ref{tab:born}, we list two partonic Born subprocesses, with
the corresponding \POWHEGBOX{} flavour structure \tmtexttt{flst\_born}.

\begin{table}[tbh]
\begin{center}
  \begin{tabular}{|c|l|l|}
\hline
    \tmtexttt{jreal}&processes &\tmtexttt{flst\_real}\\
\hline
1 & $sc \to gudg$            & $[3,4,0,2,1,0]$\\
2 & $sc \to s\bar{s}ud$      & $[3,4,3,-3,2,1]$\\
3 & $gc \to gud\bar{s}$      & $[0,4,0,2,1,-3]$\\
4 & $d u \to d \bar{s} s c $ & $[1, 2,1, -3,3,4]$\\
\hline
  \end{tabular}
  \caption{\label{tab:real} Flavour structure of four real subprocesses and
    the corresponding \POWHEGBOX{} notation, in the third column.}
\end{center}
\end{table}  
Suppose now that the real-process contributions are only the four ones in
table~\ref{tab:real}, so that \tmtexttt{flst\_nreal=4}. Since the real
contributions have one more parton with respect to the Born diagrams, we have
also \tmtexttt{nlegreal=6}. Note that we deliberately have not included
subprocesses such as $sg \to gud \bar{c}$, $gg \to \bar{s} s c \bar{u}$ and
$gu \to \bar{s} s c g$, in order to keep the example short.

%
%

The subroutine \tmtexttt{find\_regions} is called on each flavour structure
of the list \tmtexttt{flst\_real}. This subroutine returns the list of
emitter-radiated pairs, and their total number.  For example, when the first
flavour structure $[3,4,0,2,1,0]$ is passed to the subroutine, this returns
the list of 7 pairs: $\{(3,4),(3,5),(3,6),(4,6),(5,6),(0,3),(0,6)\}$.  The
$(3,4)$ pair means that the gluon in the third position can be emitted by the
up quark in the fourth position, and that, by removing this gluon, the so
obtained underlying Born is a valid Born, since its flavour structure is
equivalent (up to a permutation of the final-state lines) to the first
flavour structure in \tmtexttt{flst\_born}.  The $(0,3)$ pair represents the
singular region associated with gluon 3 being emitted by both initial-state
partons, and, once removed, we get a valid flavour Born.  When the subroutine
is called on the second flavour structure in \tmtexttt{flst\_real},
$[3,4,3,-3,2,1]$, it returns a single pair $(3,4)$, meaning that the
$s\bar{s}$ pair can come from the splitting of a gluon, and the diagram
obtained by replacing the $s\bar{s}$ pair by a gluon is a valid Born.
Similarly the call of \tmtexttt{find\_regions} on the third element of
\tmtexttt{flst\_real} returns the pair $(1,6)$ that signals the fact that
this diagram is compatible with the splitting of an initial-state gluon into
an $s \bar{s}$ pair.

\begin{table}[tbh]
\begin{center}
  \begin{tabular}{|c|c|c|l|}
\hline
    \tmtexttt{alr}&\tmtexttt{iregions}&\tmtexttt{flst\_emitter}&\tmtexttt{flst\_alr}\\
\hline
 1 & (3,4) & 5 & $[3,4,1,0,2,0]$ \\
 2 & (3,5) & 5 & $[3,4,2,0,1,0]$ \\
 3 & (3,6) & 5 & $[3,4,2,1,0,0]$ \\
 4 & (4,6) & 5 & $[3,4,0,1,2,0]$ \\
 5 & (5,6) & 5 & $[3,4,0,2,1,0]$ \\
 6 & (0,3) & 0 & $[3,4,2,1,0,0]$ \\
 7 & (0,6) & 0 & $[3,4,0,2,1,0]$ \\
 8 & (3,4) & 5 & $[3,4,2,1,3,-3]$\\
 9 & (1,6) & 1 & $[0,4,0,2,1,-3]$\\
10 & (1,3) & 1 & $[1,2,-3,3,4,1]$\\
\hline
  \end{tabular}
  \caption{\label{tab:example1a} List of all the 10 singular regions found up
    to this stage of the program, of the emitter-radiated pairs, of the
    position of the emitter, \tmtexttt{flst\_emitter}, and of the
    corresponding flavour structure, \tmtexttt{flst\_alr}.}
\end{center}
\end{table}  
In total we have then 10 singular regions, so that, at this stage of the
program, \tmtexttt{flst\_nalr=10}.  We have listed them in the second column
of table~\ref{tab:example1a}.  The flavour structure of these singular
regions is then saved in the array \tmtexttt{flst\_alr} after a suitable
permutation of the final-state partons, such that the emitted parton is in
the last position (\tmtexttt{nlegreal}), and, in case of final-state
radiation, the emitter is in the \tmtexttt{nlegreal-1} position. At the same
time, the position of the emitter is recorded in the array
\tmtexttt{flst\_emitter} (see the third and fourth column of
table~\ref{tab:example1a}).

\begin{table}[tbh]
\begin{center}
  \begin{tabular}{|c|l|c|l|}
\hline
    \tmtexttt{alr}&\tmtexttt{flst\_alr}&\tmtexttt{flst\_mult} 
    & \tmtexttt{flst\_uborn}\\
\hline
 1  & $[3,4,1,0,2,0]$  & 2 & $[3,4,1,0,2]$  \\
 2  & $[3,4,2,0,1,0]$  & 2 & $[3,4,2,0,1]$  \\
 3  & $[3,4,2,1,0,0]$  & 1 & $[3,4,2,1,0]$  \\
 4  & $[3,4,2,1,0,0]$  & 2 & $[3,4,2,1,0]$  \\
 5  & $[3,4,2,1,3,-3]$ & 1 & $[3,4,2,1,0]$  \\
 6  & $[0,4,0,2,1,-3]$ & 1 & $[3,4,0,2,1]$  \\
 7  & $[1,2,-3,3,4,1]$ & 1 & $[0,2,-3,3,4]$ \\
\hline
  \end{tabular}
  \caption{\label{tab:example1b} List of all the 7 singular regions found, of
    their flavour structure, \tmtexttt{flst\_alr}, of the multiplicity of
    each singular region, \tmtexttt{flst\_mult}, and of the underlying Born
    flavour, \tmtexttt{flst\_uborn}.}
\end{center}
\end{table}  
At this stage, the list of singular regions is scanned, and, if two elements
are equivalent (up to a permutation of the final-state partons) and have
equal emitted and radiated parton, one of them is removed from the list and
the multiplicity factor \tmtexttt{flst\_mult} of that singular region is
increased.  By referring to table~\ref{tab:example1a}, the fourth
\tmtexttt{alr} flavour list, $[3,4,0,1,2,0]$, is equivalent to the first one,
$[3,4,1,0,2,0]$, so that it is removed and the multiplicity factor of the
first singular region is set to~2. This is illustrated in
table~\ref{tab:example1b}, where the fifth and the seventh singular regions
of table~\ref{tab:example1a} have been removed and the multiplicity factors
of the second and fourth singular region in table~\ref{tab:example1b} are set
to~2.

In the last column of table~\ref{tab:example1b}, we collect the underlying
Born flavour structures, \tmtexttt{flst\_uborn}, corresponding to each
\tmtexttt{alr} singular region. Each of these underlying Born must be
equivalent (up to a permutation of the final-state partons) to one of the
Born in the list of valid Born flavour structures, \tmtexttt{flst\_born}.

At this point, we scan the list of the underlying Born flavour structures,
\tmtexttt{flst\_uborn}, and permute the flavours in such a way to obtain
exactly one element of the \tmtexttt{flst\_born} list.  Correspondingly, we
reorder the flavours of the corresponding \tmtexttt{alr} singular region.
For example, consider the first element of \tmtexttt{flst\_uborn},
$[3,4,1,0,2]$. This element becomes equal to $[3,4,0,2,1]$ (the first element
of the \tmtexttt{flst\_born} list), only after the exchange of the element in
the third position with the one in the forth position ($1\leftrightarrow 0$)
followed by the exchange of the element in the fourth position with the one
in the fifth position ($2\leftrightarrow 1$). We perform the same exchanges
on the first element of \tmtexttt{flst\_alr}, i.e.~$[3,4,1,0,2,0]$, and we
obtain $[3,4,0,2,1,0]$, and, after these permutations, the emitter is the
parton in the fourth position, so that \tmtexttt{flst\_emitter=4}.
\begin{table}[tbh]
\begin{center}
  \begin{tabular}{|c|l|c|c|c|c|}
\hline
    \!\tmtexttt{alr}\!&\tmtexttt{flst\_alr}&\!\tmtexttt{emi}\!
    &\!\tmtexttt{a2b}\!& \! \tmtexttt{nreg}\! & \tmtexttt{flst\_allreg}\\
\hline
 1 & $[3,4,0,2,1,0]$  & 4 & 1 & 7 & $\{(3,4),(3,5),(3,6),(4,6),(5,6),(0,3),(0,6)\}$ \\
 2 & $[3,4,0,2,1,0]$  & 5 & 1 & 7 & $\{(3,4),(3,5),(3,6),(4,6),(5,6),(0,3),(0,6)\}$ \\
 3 & $[3,4,0,2,1,0]$  & 3 & 1 & 7 & $\{(3,4),(3,5),(3,6),(4,6),(5,6),(0,3),(0,6)\}$ \\
 4 & $[3,4,0,2,1,0]$  & 0 & 1 & 7 & $\{(3,4),(3,5),(3,6),(4,6),(5,6),(0,3),(0,6)\}$ \\
 5 & $[3,4,3,2,1,-3]$ & 3 & 1 & 1 & $\{(3,6)\}$ \\
 6 & $[0,4,0,2,1,-3]$ & 1 & 1 & 1 & $\{(1,6)\}$ \\
 7 & $[1,2,-3,3,4,1]$ & 1 & 2 & 1 & $\{(1,6)\}$ \\
\hline
  \end{tabular}
  \caption{\label{tab:example2} List of the flavours of the singular regions
    after they have been reordered so that the corresponding underlying Born
    is equal to one of the flavour subprocess in the \tmtexttt{flst\_born}
    list. From the third column on: the position \tmtexttt{emi} of the
    emitter (\tmtexttt{emi=flst\_emitter}), the underlying Born flavour
    structure pointer \tmtexttt{a2b} (\tmtexttt{a2b=flst\_alr2born}), the
    number of singular regions \tmtexttt{nreg} associated with that
    particular \tmtexttt{alr} region (\tmtexttt{nreg=flst\_allreg(1,0,alr)})
    and the list of the corresponding emitter-radiated pairs,
    \tmtexttt{flst\_allreg(1:2,1:nreg,alr)}.}
\end{center}
\end{table}  
By performing this task on every underlying Born flavour structure, we obtain
the second and third column of table~\ref{tab:example2}.

As final tasks, we fill the arrays with the pointers to go from an
\tmtexttt{alr} region to its underlying Born flavour structure and viceversa.
The first 6 elements in the \tmtexttt{flst\_alr} list have the first Born
flavour structure as underlying Born, and only the last region has the second
Born flavour structure, so that the elements of the array
\tmtexttt{flst\_alr2born} are as in the fourth column of
table~\ref{tab:example2}.
\begin{table}[tbh]
\begin{center}
  \begin{tabular}{|c|c|c|}
\hline
    \tmtexttt{jborn}&\tmtexttt{flst\_born2alr(0,jborn)}& \tmtexttt{flst\_born2alr}\\
\hline
 1 & 6 & $\{1,2,3,4,5,6\}$ \\
 2 & 1 & $\{7\}$\\
\hline
  \end{tabular}
  \caption{\label{tab:example3} In the second column, the number of singular
    \tmtexttt{alr} regions that have the \tmtexttt{jborn}$^{\rm th}$ Born
    subprocess as underlying Born and the corresponding list of these
    \tmtexttt{alr} regions.}
\end{center}
\end{table}  
Viceversa, 6 \tmtexttt{alr} singular regions have the first Born flavour
structure as underlying Born, while the seventh has the second one, so that
\tmtexttt{flst\_born2alr(0,1)=6} and \tmtexttt{flst\_born2alr(0,2)=1} (see
table~\ref{tab:example3}). The list of these singular regions is shown in the
third column: \tmtexttt{flst\_born2alr(1:6,1)=}$\{1,2,3,4,5,6\}$ and
\tmtexttt{flst\_born2alr(1:1,2)=}$\{7\}$.

Finally, for every flavour structure in \tmtexttt{flst\_alr}, we build the
list of all the associated singular regions. This is done by simply calling
again \tmtexttt{find\_regions} on every element of the second column of
table~\ref{tab:example2}. For every \tmtexttt{alr}, the total number of
singular regions, \tmtexttt{nreg}, is saved in
\tmtexttt{flst\_allreg(1,0,alr)} (see the fifth column of
table~\ref{tab:example2}), and all the pairs of emitted-radiated partons is
saved in the array \tmtexttt{flst\_allreg(1:2,1:nreg,alr)} (see sixth
column).

\section{The $\boldsymbol{\tilde{B}}$ function}\label{sec:btilde}
In order to generate an event, \POWHEG{} first generates a Born kinematic and
flavour configuration, with a probability proportional to (see eq.~(4.13) in
ref.~{\cite{Frixione:2007vw}})
\begin{equation}
  \bar{B} ( \tmmathbf{\Phi}_n) \, d \tmmathbf{\Phi}_n = \left[ \sum_{f_b}
  \bar{B}^{f_b} ( \tmmathbf{\Phi}_n) \right] d \tmmathbf{\Phi}_n,
\end{equation}
where
\begin{eqnarray}
\label{eq:Bbar}
  \bar{B}^{f_b} ( \tmmathbf{\Phi}_n) & = & \left[ B ( \tmmathbf{\Phi}_n) + V (
  \tmmathbf{\Phi}_n) \right]_{f_b} + \sum_{\alr \in \{\alr |f_b \}}
  \int \left[ d \Phi_{\tmop{rad}} \, \hat{R} (\Phinpo)
  \right]_{\alr}^{\bar{\Phi}^{\alr}_n = \Phin} \nonumber\\
  & + & \sum_{\alpha_{\oplus} \in \{\alpha_{\oplus} |f_b \}} \int \frac{d
  z}{z}\, G_{\oplus}^{\alpha_{\oplus}} ( \tmmathbf{\Phi}_{n, \oplus}) +
  \sum_{\alpha_{\ominus} \in \{\alpha_{\ominus} |f_b \}} \int \frac{d z}{z}\,
  G_{\ominus}^{\alpha_{\ominus}} ( \tmmathbf{\Phi}_{n, \ominus}) \,. 
\end{eqnarray}
The square brackets with a suffix represent a context: everything inside is
relative to that suffix. The symbol $f_b$ labels a Born flavour structure.
Thus, due to this context notation, $B$ and $V$ in the square bracket refer
to the contribution of the Born and soft-virtual cross section having the
flavour structure $f_b$. The suffix in the first summation represents a sum
over all $\alr$ that have $f_b$ as underlying Born. The corresponding square
bracket under the integral sign means that we integrate in the radiation
phase space of the current $\alr$, keeping the underlying Born variables
$\bar{\Phi}^{\alr}_n$ fixed and equal to $\Phin$. The real contribution $R$
has been properly regularized using the plus distributions (see eq.~(2.88)
and~(2.89) in~\cite{Frixione:2007vw}), so that it is written with a hat, and
has to be handled properly. The way we deal with the generation of the
underlying Born kinematics in \POWHEG{} is the following. For each singular
region, we parametrize the radiation variables $\Phi_{\tmop{rad}}$ as
function of three variables (\tmtexttt{xrad} in the code) in the unit cube,
\begin{equation}
  X_{\tmop{rad}} = \left\{ X_{\tmop{rad}}^{(1)}, \,X_{\tmop{rad}}^{(2)},\,
  X_{\tmop{rad}}^{(3)} \right\},
\end{equation}
and we also parametrize the $z$ variable in eq.~(\ref{eq:Bbar}) as a function
of $X_{\tmop{rad}}^{(1)}$ in the $[0,1]$ interval. We then introduce the
$\tilde{B}$ function, defined as
\begin{equation}
\label{eq:btilde}
  \tilde{B}(\Phin, X_{\tmop{rad}}) = \sum_{f_b} \tilde{B}^{f_b}
  (\Phin, X_{\tmop{rad}}), 
\end{equation}
with
\begin{eqnarray}
  \tilde{B}^{f_b} ( \tmmathbf{\Phi}_n, X_{\tmop{rad}}) & = & \left[ B (
  \tmmathbf{\Phi}_n) + V ( \tmmathbf{\Phi}_n) \right]_{f_b} + \sum_{\alr
  \in \{\alr |f_b \}} \left[ \left| \frac{\partial
      \Phi_{\tmop{rad}}}{\partial 
  X_{\tmop{rad}}} \right|  \hat{R} ( \tmmathbf{\Phi}_{n + 1})
  \right]_{\alr}^{\bar{\Phi}^{\alr}_n = \Phin} \nonumber\\
  & + & \sum_{\alpha_{\oplus} \in \{\alpha_{\oplus} |f_b \}} \frac{1}{z} \left|
  \frac{\partial z}{\partial X_{\tmop{rad}}^{(1)}} \right|
  G_{\oplus}^{\alpha_{\oplus}} ( \tmmathbf{\Phi}_{n, \oplus}) +
  \sum_{\alpha_{\ominus} \in \{\alpha_{\ominus} |f_b \}} \frac{1}{z} \left|
  \frac{\partial z}{\partial X_{\tmop{rad}}^{(1)}} \right|
  G_{\ominus}^{\alpha_{\ominus}} ( \tmmathbf{\Phi}_{n, \ominus}).\phantom{aaaa} 
\end{eqnarray}
One now integrates $\tilde{B} ( \tmmathbf{\Phi}_n, X_{\tmop{rad}})$ in the
full $(\tmmathbf{\Phi}_n, X_{\tmop{rad}})$ phase space, using an integration
routine that implements the possibility of generating the integrand with a
uniform weight after a single integration. The routine that we use is
\tmtexttt{mint}~\cite{Nason:2007vt}, that was explicitly built for
application in \POWHEG. Since \tmtexttt{mint} is designed to integrate in the
unit hypercube, also the Born phase space has to be mapped in a hypercube,
spanned by the variables $X_{\tmop{born}}$, as described in
section~\ref{sec:born}.

\subsection{The \tmtexttt{btilde} function}
The \tmtexttt{btilde(xx,www0,ifirst)} function implements the function
$\tilde{B}$ in eq.~(\ref{eq:btilde}). The first elements of the array
\tmtexttt{xx} correspond to \tmtexttt{xborn} and the last 3 correspond to
\tmtexttt{xrad}.  The weight \tmtexttt{www0} is passed by the integration
routine, and equals the weight factor (arising from integration volume and
importance sampling) supplied by the integration routine.  The flag
\tmtexttt{ifirst} has a rather involved use (described in detail in
ref.~\cite{Nason:2007vt}) that is needed in order to implement the folding of
some integration variable.  The subroutine \tmtexttt{mint} may be in fact
requested to use more points for some selected integration variable, keeping
fixed all the others, for each single random contribution to the integral. In
practice, one may require that more points for the variables $X_{\tmop{rad}}$
are used for a single value of $X_{\tmop{born}}$. Upon the first call with a
new value of $X_{\tmop{born}}$, the function \tmtexttt{btilde} is called with
\tmtexttt{ifirst=0}. In all subsequent calls with the same $X_{\tmop{born}}$,
but different $X_{\tmop{rad}}$, \tmtexttt{btilde} is called with
\tmtexttt{ifirst=1}. After all calls with \tmtexttt{ifirst=1}, a last call
with \tmtexttt{ifirst=2} is performed, where \tmtexttt{btilde} accumulates
all the values computed since the last \tmtexttt{ifirst=0} call. The
quantities that depend only upon $X_{\tmop{born}}$ are computed only once
when the call with \tmtexttt{ifirst=0} is performed. Thus, the Born phase
space is generated at this stage. The Born cross sections for all Born
flavour structures are computed and made available in appropriate common
blocks at this stage, by calling the subroutine \tmtexttt{allborn}. The
virtual contributions to \tmtexttt{btilde} are also computed here.  The
subroutines \tmtexttt{btildeborn(resborn,www)} and
\tmtexttt{btildevirt(resvirt,www)} fill the arrays \tmtexttt{resborn} and
\tmtexttt{resvirt} with the contributions for each Born flavour
structure. The contributions from the collinear remnants and from the real
cross section (\tmtexttt{btildecoll} and \tmtexttt{btildereal}) are computed
both for \tmtexttt{ifirst=0} and \tmtexttt{ifirst=1}. Notice that, in these
cases, the index in the arrays \tmtexttt{btildecoll} and
\tmtexttt{btildereal} refers to a given underlying Born. Thus, for each array
entry in \tmtexttt{btildereal}, for example, several contributions from
different $\alr$ regions (sharing the same underlying Born) are summed
up. All the contributions to \tmtexttt{btilde} are accumulated in the array
\tmtexttt{results}. When called with \tmtexttt{ifirst=2}, the behaviour of
the program depends upon the setting of the flag \tmtexttt{negflag}. If true,
\tmtexttt{btilde} should only compute the contribution of negative
weights. Thus, the positive entries in \tmtexttt{results} are zeroed, and the
negative entries are replaced with their absolute value. If
\tmtexttt{negflag} is false (normal behaviour), the negative entries are
zeroed. The array \tmtexttt{results} is also stored in the
\tmtexttt{rad\_btilde\_arr} array, defined in the header file
\tmtexttt{pwhg\_rad.h}. It is needed at the stage of event generation, where
the underlying Born flavour structure will be chosen with a probability
proportional to its entries.

\subsection{The Born cross section}
The Born contribution to the \tmtexttt{btilde} function is evaluated as
follows. When \tmtexttt{btilde} is called with \tmtexttt{ifirst=0}, the Born
phase space is generated with a call to the \tmtexttt{gen\_born\_phsp}
subroutine, that, in turn, calls the user provided \tmtexttt{born\_phsp}
subroutine. For cases when the Born cross section is itself infrared
divergent (for example, in the $Z+{\rm jet}$ production case), we need either
a generation cut for the underlying Born configuration, or a Born suppression
factor. The \POWHEGBOX{} provides a standard form for the latter.  We do not
further discuss this issue here; in the forthcoming
reference~\cite{POWHEG_Zjet} we will illustrate this problem in great detail.
The factorization and renormalization scales are set with a call to
\tmtexttt{setscalesbtilde}, and then the routine \tmtexttt{allborn} is
called.  The routine computes the Born contributions for all Born flavour
structures, and stores them in the arrays \tmtexttt{br\_born} for the cross
section, \tmtexttt{br\_bornjk} for the colour correlated Born cross section,
and \tmtexttt{br\_bmunu} for the spin correlated one. Many subsequent calls
make use of the Born cross sections, so it is mandatory that this call is
performed first. In particular, soft and collinear remnants need the Born
terms, and so do the singular limits of the real cross section.

In order to understand the code of the \tmtexttt{allborn} routine, it is
better to assume first that the flag \tmtexttt{flg\_smartsig} is set to
false. The role of this flag is explained in detail in section
\ref{sec:flg_smartsig}.

The call to \tmtexttt{btildeborn(resborn)}, in the \tmtexttt{btilde}
function, fills the array \tmtexttt{resborn} with the Born cross section for
each Born flavour configuration, including the corresponding parton
distribution function~(pdf) factors. In case the \tmtexttt{flg\_nlotest} is
set, while computing the integral of the \tmtexttt{btilde} function, an
analysis routine is also called to perform a bare NLO calculation for several
user-provided distributions. In the case of the Born result, this analysis
routine is called within \tmtexttt{btilde} at the end of a given folding
sequence.

\subsection{The soft-virtual cross section}
The soft-virtual amplitude is described in detail in section~2.4.2 of
ref.~\cite{Frixione:2007vw} in the case of massless coloured partons.  We
complete here the formulae given in~\cite{Frixione:2007vw} in order to
include the case of massive partons. All these formulae are implemented in
the \POWHEGBOX{}, that can also deal with massive coloured partons.  When
massive coloured partons are present, formula~(2.99)
in~\cite{Frixione:2007vw} becomes
\begin{equation}
\label{eq:soft_virt}
  \mathcal{V} = \frac{\as}{2 \pi} \left( \mathcal{Q}\,  \mathcal{B} +
  \sum_{i \neq j} \mathcal{I}_{i j} \, \mathcal{B}_{i j} - \mathcal{B} \sum_i
  \mathcal{I}_i + \mathcal{V}_{\tmop{fin}} \right) .
\end{equation}
The quantity $\mathcal{I}_{i j}$ is non vanishing for $i$ and $j$ denoting
any coloured initial and final-state partons. $\mathcal{I}_i$ is non zero for
$i$ denoting any massive coloured parton. They both arise from soft
radiation, and are reported in appendix~\ref{app:soft} in the same form that
appears in the code. The $\mathcal{Q}$ term has the same expression given in
ref.~\cite{Frixione:2007vw}
\begin{eqnarray}
\label{eq:Q_def}
  \mathcal{Q} & = & \sum_i \left[ \gamma'_{f_i} - \log \frac{s \delta_0}{2
  Q^2} \left( \gamma_{f_i} - 2 C_{f_i} \log \frac{2 E_i}{\xi_c \sqrt{s}}
  \right) \right. \nonumber\\
  &  & \left. + 2 C_{f_i} \left( \log^2 \frac{2 E_i}{\sqrt{s}} - \log^2 \xi_c
  \right) - 2 \gamma_{f_i} \log \frac{2 E_i}{\sqrt{s}} \right] \nonumber\\
  &  & - \log \frac{\muf^2}{Q^2} \left[ \gamma_{f_{\oplus}} + 2
  C_{f_{\oplus}} \log \xi_c + \gamma_{f_{\ominus}} + 2 C_{f_{\ominus}} \log
  \xi_c \right].
\end{eqnarray}
$E_i$ denotes the energy of parton $i$ in the partonic CM frame,
and $f_i$ denotes the flavour, i.e.\ $g$ for a gluon,
$q$ for a quark and $\bar{q}$ for an antiquark. In addition we have
\begin{eqnarray}
&&C_g=\CA\,,\phantom{aaaaaaaaaaaaaaaaaaaaaaa\!}
C_q=C_{\bar{q}} = \CF\,,
\\
&&\gamma_g=\frac{11\CA-4\TF\, \NF}{6}\,,\phantom{aaaaaaaaaaaaa}
\gamma_q=\gamma_{\bar q}= \frac{3}{2}\CF\,,
\\
&&\gamma^\prime_g=\( \frac{67}{9}-\frac{2\pi^2}{3}\)\CA
-\frac{23}{9}\TF \,\NF\,,\;\;\;\;\;\;
\gamma^\prime_q=\gamma^\prime_{\bar{q}}=
\(\frac{13}{2}-\frac{2\pi^2}{3}\)\CF\,.
\phantom{aaaa}
\end{eqnarray}
We stress that now the index $i$ in the sum of
eq.~(\ref{eq:Q_def}) runs only over the massless coloured final-state
partons. In fact, the contributions in square bracket arise from collinear
(rather than soft) final-state singularities, and thus apply only to massless
partons. The last line arises from initial-state collinear singularities.

The parameters $\delta_0$ and $\xi_c$ are arbitrary. In the framework of the
\POWHEGBOX, we have set $\delta_0 = 2$ and $\xi_c = 1$. An analogous
parameter for initial-state collinear singularities, $\delta_I$, that appears
in the collinear remnants, is also set to 2. These parameters are hardwired
in the code, since there is no reason to change them.

The soft-virtual contribution of eq.~(\ref{eq:soft_virt}) is implemented as
follows.  The \POWHEGBOX{} has access to the user-provided Born cross
section, including colour correlations. It thus builds automatically the
$\mathcal{Q}$, $\mathcal{I}_{i j}$ and $\mathcal{I}_i$ contributions of
eq.~(\ref{eq:soft_virt}) to the soft-virtual cross section. The corresponding
code is found in the \tmtexttt{btildevirt} subroutine, in the
\tmtexttt{sigsoftvirt.f} file.  As already stated in
section~\ref{sec:virtual}, the scale $Q$ is chosen equal to the
renormalization scale $\mur$. The user-provided virtual cross section should
then have $Q = \mur$.  The subroutine \tmtexttt{btildevirt} fills the array
\tmtexttt{resvirt} with the contribution of the soft-virtual cross section
for all possible underlying Born configurations.

\subsection{The collinear remnants}
The collinear remnants $G_{\oplus}^{\alpha_{\oplus}}$ and
$G_{\ominus}^{\alpha_{\ominus}}$ of eq.~(\ref{eq:Bbar}) are the finite
leftover from the subtraction of collinear singularities. Their general form,
in the FKS framework, is given in eq.~(2.102) of
ref.~\cite{Frixione:2007vw}. They are implemented in the \POWHEGBOX{} in the
$\overline{\rm MS}$ scheme.\footnote{Since currently the DIS factorization
  schemes are no longer used, we do not implement them.}  Their
implementation in the \tmtexttt{btilde} function has to properly handle the
distributions in the $z$ variable. This is done using the identities
\begin{eqnarray}
  \frac{1}{(1 - z)_{\xi_c}} & = & \frac{1}{(1 - z)_{1 - x}} + \log \frac{1 -
  x}{\xi_c} \,\delta (1 - z)\,, \\
  \left( \frac{\log (1 - z)}{1 - z} \right)_{\!\! \xi_c} & = & \left(
  \frac{\log (1 - z)}{1 - z} \right)_{1 - x} + \frac{\log^2 (1 - x) - \log^2
    \xi_c}{2} \,\delta (1 - z) \,,
\end{eqnarray}
where $x$ stands for either $x_{\oplus}$ or $x_{\ominus}$. The $z$
integration extends from $x$ to 1, and thus it is performed using the rules
\begin{eqnarray}
  \int_x^1\!\!  dz\, f (z) \frac{1}{(1 - z)_{\xi_c}} \!& = & \!\! \int_x^1
  \!\! d z\, \frac{f (z) - f (1)}{1 - z} + \log \frac{1 - x}{\xi_c} \,f (1),
  \\
\int_x^1 \!\!\! dz\, f (z) \!\left( \frac{\log (1 - z)}{1 - z} \right)_{\!\!
  \xi_c} \!\! \! \!& = & \!\!\!\int_x^1\!\!\! d z\left[ f (z) - f (1) \right]
\!\frac{\log (1 - z)}{1 - z} + \frac{\log^2 (1 - x) - \log^2 \xi_c}{2} f(1),
\phantom{aaa}
\end{eqnarray}
where we have always set $\xi_c=1$ in the code.  In this case too, a loop
over all underlying Born configurations is performed, and, for each of them,
all singular remnant contributions appropriate to the flavours of the
corresponding incoming legs are computed. The \tmtexttt{btildecoll} function
requires an extra integration variable, given as its first argument, to
generate a value for $z$. Within the \tmtexttt{btildecoll} function, if the
\tmtexttt{flg\_nlotest} flag is set, a call to the analysis routines is
performed to output the contribution of the collinear remnants to the
user-defined kinematic plots.

\subsection{The real contribution to \tmtexttt{btilde}}
The real contribution to the \tmtexttt{btilde} function has a certain
complexity, mostly due to the handling of the distributions in the $\xi$ and
$y$ variables. The function \tmtexttt{btilde} simply calls the function
\tmtexttt{btildereal} to get the contribution of the real cross section for
each underlying Born configuration. The complex task of evaluating the real
integrand is carried out in the subroutine \tmtexttt{btildereal}. In this
subroutine, there is a loop over all possible emitters, that envelopes its
whole body. All contributions are then evaluated for a given emitter. With
this choice we avoid repeating continuously complex phase-space
evaluations. For each emitter, for given underlying Born and radiation
variables, there is only one real kinematics to consider.

If the emitter is a final-state parton, the real phase space is generated by
a call to the subroutine \tmtexttt{gen\_real\_phsp\_fsr}. For initial-state
emission, a call to \tmtexttt{gen\_real\_phsp\_isr} is done.  The program
then calls the subroutine \tmtexttt{sigreal\_btl}, that fills its array
argument with the contributions of all regions (i.e.~\tmtexttt{alr}) that
have as emitter the current emitter \tmtexttt{kn\_emitter}.  This subroutine
returns an array whose elements are $R_{\alpha} (1 - y^i) \xi^2$, with $i=1$
in the FSR case and $i=2$ in the ISR case, rather than $R_{\alpha}$ alone
($R_\alpha$ is defined in the notation of~\cite{Frixione:2007vw}). The
quantity $R_{\alpha} (1 - y^i) \xi^2$ has well defined soft, collinear and
soft-collinear limits, that are obtained by a call to the subroutines
\tmtexttt{soft}, \tmtexttt{collfsr} (for final-state radiation) or
\tmtexttt{collisrp} and \tmtexttt{collisrm} (for initial-state $\oplus$ and
$\ominus$ collinear regions), \tmtexttt{softcollfsr} for soft-collinear limit
in final-state radiation, and \tmtexttt{softcollisrp} and
\tmtexttt{softcollisrm} for soft-collinear initial-state radiation. The
handling of the $\xi$ and $y$ distributions require some care, and we
describe it here in some detail.

We begin by looking at the final-state radiation case.  The integral that we
would like to perform has the form
\begin{equation}
\label{eq:realdistr2}
  \bar{B}_{\tmop{real}} = \int d \tmmathbf{\Phi}_n \int_0^{2 \pi}\! d \phi
  \int_{- 1}^1\! d y \int_0^{X(y)}\!\! d \xi \,\frac{J (\xi, y, \phi)}{\xi}
  \lq (1 - y) \,\xi^2 R_{\alpha}\rq \left( \frac{1}{\xi} \right)_+
  \(\frac{1}{1 - y}\)_+\,,
\end{equation}
the Jacobian being given by formula~(5.40) in ref.~\cite{Frixione:2007vw}. We
have indicated explicitly the dependence of the Jacobian upon the radiation
variables, but one should keep in mind that it also depends upon the
underlying Born variables. We have assumed, in all generality, that the upper
limit in the $\xi$ integration in eq.~(\ref{eq:realdistr2}) may depend upon
$y$, and we have denoted it as $X(y)$.  This is in fact not the case in our
choice of the final-state radiation kinematics, but it is the case for
initial-state radiation~(ISR).  The \tmtexttt{gen\_real\_phsp\_fsr}
subroutine returns the Jacobian for the integration of the radiation
variables divided by $\xi$, in the variable
\tmtexttt{}\tmtexttt{jac\_over\_csi}.

In eq.~(\ref{eq:realdistr2}) we introduce a new rescaled variable
$\tilde{\xi}$, whose upper bound does not depend upon $y$
\begin{equation}
\label{eq:csitildedef}
  \xi = X(y) \tilde{\xi} \,. 
\end{equation}
We can easily show that
\begin{equation}
  \int_0^{X(y)} \! d \xi \left( \frac{1}{\xi} \right)_+ \!  F (\xi) =
  \int_{0^{}}^1 d \tilde{\xi}  \left[ \left( \frac{1}{\tilde{\xi}} \right)_+ +
  \log X(y)  \delta ( \tilde{\xi}) \right]\! F (\xi)\,.
\end{equation}
We now rewrite eq.~(\ref{eq:realdistr2}) as
\begin{eqnarray}
  \bar{B}_{\tmop{real}} & = & \int d \tmmathbf{\Phi}_n \int_0^{2 \pi} d \phi
  \int_{- 1}^1 d y \(\frac{1}{1 - y}\)_+ \bigg[ \int_0^1 d \tilde{\xi} \,
    \frac{J (\xi, y, \phi)}{\xi} \lq (1 - y) \,\xi^2 R_{\alpha} \rq \left(
    \frac{1}{\tilde{\xi}} \right)_+  \nonumber\\
  & + & \log X(y) \, \lim_{\xi \rightarrow 0} \left( \frac{J (\xi, y,
  \phi)}{\xi} \, \lq (1 - y) \,\xi^2 R_{\alpha}\rq \right)\bigg], 
\end{eqnarray}
where we should not forget that $\xi$ is now also a function of $y$ through
eq.~(\ref{eq:csitildedef}). Defining
\begin{equation}
\label{eq:small_f}
  f (\xi, y) = \frac{J (\xi,y,\phi)}{\xi}\, \lq (1 - y)\, \xi^2 R_{\alpha}\rq,
\end{equation}
we get
\begin{eqnarray}
\label{eq:btildereal}
  \bar{B}_{\tmop{real}} & = & \int d \tmmathbf{\Phi}_n \int_0^{2 \pi} d \phi
  \int_{- 1}^1 \frac{d y}{1 - y} \Bigg\{ \int_0^1 d \tilde{\xi}  \left[
  \frac{f ( \tilde{\xi} X(y), y) - f (0, y)}{\tilde{\xi}} - \frac{f (
  \tilde{\xi} X(1), 1) - f (0, 1)}{\tilde{\xi}} \right] 
   \nonumber\\
  & + &  \big[ \log X(y) \,f (0, y) - \log X(1) \, f (0, 1) \big]
  \Bigg\} .  
\end{eqnarray}
We now see that both $(1 - y) \,\xi^2 R_{\alpha}$ and $J (\xi, y, \phi) /
\xi$ in eq.~(\ref{eq:small_f}) should be computed also with $\xi = 0$ (the
soft limit), $y = 1$ at fixed $\tilde{\xi}$ (the collinear limit) and both
$\xi = 0$ and $y = 1$ (the soft-collinear limit).

The case of initial-state radiation is handled similarly, except that now our
starting formula is
\begin{eqnarray}
 \label{eq:realdistr1}
  \bar{B}_{\tmop{real}} &=& \int\! d \tmmathbf{\Phi}_n \int_0^{2 \pi}\!\! d
  \phi \int_{- 1}^1 \!d y \int_0^{X(y)} \!\! d \xi \,\frac{J (\xi, y,
    \phi)}{\xi} \lq\(1 - y^2\) \xi^2 R_{\alpha}\rq \left( \frac{1}{\xi}
  \right)_+ \nonumber\\
&\times&\frac{1}{2} \left[ \(\frac{1}{1 - y}\)_+ + \(\frac{1}{1 + y}\)_+
    \right],
\end{eqnarray}
and one proceeds as before, by treating the $y = 1$ and $y = - 1$ regions
independently.

\subsection{The \tmtexttt{btildereal} subroutine}
We now give a more detailed description of the way \tmtexttt{btildereal} is
implemented. First of all, the generation of the radiation phase space is
performed according to the description given in section~5.1.1 of
ref.~\cite{Frixione:2007vw} for initial-state radiation, and in section~5.2.1
for final-state radiation. The subroutines \tmtexttt{gen\_real\_phsp\_fsr}
and \tmtexttt{gen\_real\_phsp\_isr} generate the phase space as a function of
the underlying Born kinematics, and of three real variables
\tmtexttt{xrad(3)}, that assume random values between zero and one. Since we
use eq.~(\ref{eq:btildereal}), a common Jacobian factor \tmtexttt{xjac} for
the transformation \tmtexttt{xrad(3)} into $\tilde{\xi}$, $y$ and $\phi$ is
also provided. The program also sets the variables \tmtexttt{jac\_over\_csi},
\tmtexttt{jac\_over\_csi\_coll} and \tmtexttt{jac\_over\_csi\_soft} to $J
(\xi, y, \phi) / \xi$, to its collinear limit and to its soft limit
respectively, and multiplies them by \tmtexttt{xjac}.

In \tmtexttt{btildereal}, these limits are obtained through the calls to
\tmtexttt{soft}, \tmtexttt{collfsr} and \tmtexttt{softcollfsr}, and the
limits for $J (\xi, y, \phi) / \xi$ are provided by the phase space
subroutines \tmtexttt{gen\_real\_phsp\_fsr}, in the variables
\tmtexttt{jac\_over\_csi\_coll} and \tmtexttt{jac\_over\_csi\_soft}. Notice
also that, while the soft limit of $J (\xi, y, \phi) / \xi$ is $y$
independent, this may not be the case for $(1 - y) \,\xi^2 R_{\alpha}$ . The
computed values of the real contribution, the soft, collinear and
soft-collinear counterterms are divided by $(1 - y) \,\tilde{\xi}$ (as in
eq.~(\ref{eq:btildereal})) and accumulated with the appropriate sign in the
array \tmtexttt{resreal}, indexed by the underlying Born index of the current
\tmtexttt{alr}. In the case of final-state radiation, the upper limit for
$\xi$ does not depend upon $y$, being given by formula~(5.49) of
ref.~\cite{Frixione:2007vw}, and is set by the phase space program
\tmtexttt{gen\_real\_phsp\_fsr} in the common variable \tmtexttt{kn\_csimax}.
Its value is taken from the array \tmtexttt{kn\_csimax\_arr}, indexed by
\tmtexttt{kn\_emitter}, that is filled by the routine
\tmtexttt{gen\_born\_phsp} when the underlying Born phase space is generated
in \tmtexttt{btilde}. In case of initial-state radiation,
\tmtexttt{kn\_csimax} is $y$ dependent, and is computed by an appropriate
routine when the real phase space is generated.  The last two terms in
eq.~(\ref{eq:btildereal}), $\tilde{\xi}$-independent, are also accumulated in
the \tmtexttt{resreal} array.

After the appropriate calls to the routine that generates the real
contributions and its various limits, the program loops through all real
regions (i.e.\ \tmtexttt{alr}), and accumulates the real contributions
according to eq.~(\ref{eq:btildereal}) or its initial-state radiation
version, in an array indexed by the index of the underlying Born of the
current \tmtexttt{alr}. This was set in the combinatoric routines, in the
array \tmtexttt{flst\_alr2born}. The accumulated values include the
underlying Born Jacobian, the real contribution or one of its various limits,
\tmtexttt{jac\_over\_csi} or one of its limits, and the remaining factor of
$1 / (\xi (1 - y))$ for FSR, or $1 / (2 \xi (1 \pm y))$ in the ISR case. The
sign of each contribution can be read out from
eq.~(\ref{eq:btildereal}). Besides filling the output array
\tmtexttt{resreal}, if the flag \tmtexttt{flg\_nlotest} is set, the result is
also output to NLO analysis routines, that perform a parton-level NLO
calculation of user-defined distributions. The analysis driver for the NLO
output, i.e.\ the subroutine \tmtexttt{analysis\_driver}, described in
section~\ref{sec:analysis}, is then called with the flag set to 1 for the
true real contribution, and 0 for all remaining terms.


\subsection{The subroutine \tmtexttt{sigreal\_btl}}
The subroutine \tmtexttt{sigreal\_btl} fills its output array argument,
indexed by the \tmtexttt{alr}, with the real contributions that have as
emitter \tmtexttt{kn\_emitter}.  The real contribution should also be
multiplied by $\xi^2 (1 - y)$ for FSR, or $\xi^2 (1 - y^2)$ for ISR, and,
furthermore, should also be multiplied by the $S^{\alpha}$ functions,
described in section~2.4 of ref.~\cite{Frixione:2007vw}. Two flags control
the behaviour of this function: \tmtexttt{flg\_smartsig} and
\tmtexttt{flg\_withdamp}. An explanation of the role of
\tmtexttt{flg\_smartsig} is given in section~\ref{sec:flg_smartsig}.  The
\tmtexttt{flg\_withdamp} flag will instead be explained in section~\ref
	 {sec:damprem}.

The code is better understood if one assumes, to begin with, that these flags
are set to false. In this case, the program loops over all \tmtexttt{alr}.
For those that have emitter equal to the current emitter, it calls the
subroutine \tmtexttt{realgr}, passing as argument the list of flavours of the
current configuration, and the real momenta.  The function is supposed to
return, in its last argument, the value of $R$, the real matrix element
squared. Next, each contribution should be multiplied by its $S^{\alpha}$
factor. In the framework of the \POWHEGBOX{}, we define the
$S^{\alpha}$ factor in the following way. We consider the flavour structure
of the $\alpha$ region under consideration, and call it $f^{\alpha}$. We call
$R_{f^{\alpha}}$ the corresponding real contribution. $R_{f^{\alpha}}$ can
have several singular regions (see, for example, the list of pairs of indexes
in the last column of table~\ref{tab:example2}). Each such region is
characterized by a pair of indexes in the legs of the real process, of the
form $(i, j)$. These can be the indexes of two final-state lines becoming
collinear, or of an initial- and final-state line becoming
collinear. According to the \POWHEG{} conventions, one can also set $i = 0$,
meaning that there are initial-state collinear singularities in both
directions (gluon emission from initial-state partons), and they share the
same underlying Born. We call $\mathcal{I}_{\alpha}$ the array of all
singular regions, i.e.\ $\mathcal{I}_{\alpha}$ is an array of pairs of
indexes. In particular, the emitter associated with the $\alpha$ region,
together with the last parton (i.e.\ the \tmtexttt{nlegreal} parton) form a
pair that belongs to $ \mathcal{I}_{\alpha}$. Let us call it the $(k, n)$
pair. Then, $S^{\alpha}$ is given by
\begin{equation}
  S^{\alpha} = \frac{1}{d_{k n}} \(\sum_{(i, j) \in \mathcal{I}_{\alpha}}
  \frac{1}{d_{i j}} \)^{-1},
\end{equation}
where $d_{i j}$ are appropriate kinematic functions that vanish when lines
$i$ and $j$ become collinear. The choice of the $d_{i j}$ function
implemented in the \POWHEGBOX{} can be found in the
\tmtexttt{compdij} subroutine, in the \tmtexttt{gen\_real\_phsp.f} file. When
the phase space is generated, \tmtexttt{compdij} is called, and the array
\tmtexttt{kn\_dijterm} is filled.  It is an ordered array, i.e.~one always
assumes $i < j$. For initial-state singularities it is given by the
expressions
\begin{eqnarray}
  d_{0 j} & = & \lq E_j^2 \(1 - y_j^2\)\rq^{p_1}, \\
  d_{1 j} & = & \lq 2\,E_j^2 \(1 - y_j\)\rq^{p_1}, \\
  d_{2 j} & = & \lq 2\, E_j^2 \(1 + y_j\)\rq^{p_1}, 
\end{eqnarray}
where $y_j$ is the cosine of the emission direction of parton $j$, in the
real CM frame, relative to the positive collision direction.  Notice that we
assume, by convention, that $k = 0$ means that there are collinear
singularities from both the positive and negative direction with the same
underlying Born configuration, as is the case when a gluon is emitted. The
positive and negative collinear directions need to be considered separately
if the corresponding underlying Born differs.  The parameter $p_1$
corresponds to the parameter \tmtexttt{par\_diexp}, defined in the
\tmtexttt{pwhg\_kn.h} include file. Its default value is 1. For FSR regions
we define
\begin{equation}
  d_{i j} = \lq 2 \, \(k_i\cdot k_j\) \, \frac{E_i\,
    E_j}{\(E_i+E_j\)^2}\rq^{p_2}\,, 
\end{equation}
where $p_2$ corresponds to \tmtexttt{par\_dijexp}, and is set to 1 by
default.

The \tmtexttt{sigreal\_btl} subroutine builds the $S^{\alpha}$ factor for
each non vanishing \tmtexttt{alr}. At the combinatoric stage, for each
\tmtexttt{alr}, a list of the singular regions associated with its flavour
structure was built.  This list was stored in the \tmtexttt{flst\_allreg}
array, and is used to find the indexes $i j$ for all the singular regions of
the given \tmtexttt{alr} (see table~\ref{tab:example2} for an example).

One extra factor is supplied for final-state singularities if both the
emitter and radiated partons are gluons. One multiplies the result by
\begin{equation}
\label{eq:softemitterfac}
  \frac{2 E_{\tmop{em}}}{E_{\tmop{em}} + E_{\rm r}},
\end{equation}
where $E_{\tmop{em}}$ and $E_{\rm r}$ are the energy of the emitter and of
the radiated parton, evaluated in the partonic CM. This does not change the
cross section, because of the symmetry in the exchange of the two gluons, but
guarantees that only when the radiated gluon becomes soft we can have a soft
singularity. As a final step, the multiplicity of the current \tmtexttt{alr}
and the $\xi^2(1 - y^2)$ (for ISR) or $\xi^2 (1 - y)$ (for FSR) factors are
included.

\subsection{The \tmtexttt{flg\_smartsig} flag}
\label{sec:flg_smartsig} 
In several processes, organizing the program in the way described in the
previous section would lead to several calls to the same matrix elements,
with a consequent waste of computing time. In the process of $W$ production,
for example, the matrix element is the same whether it is a $u \bar{d}$ or a
$c \bar{s}$ collision. Or it may differ only by a Cabibbo-Kobayashi-Maskawa
matrix element factor. If the flag \tmtexttt{flg\_smartsig} is set to true,
upon the first call to the \tmtexttt{sigreal\_btl} subroutine, the routine
finds matrix elements that differ only by a constant factor, and builds
appropriate array of pointers and proportionality constants. Upon subsequent
entries in the program, multiple calls to proportional matrix elements will
thus be avoided. All subprograms that invoke user routines for matrix element
calculations are affected by the setting of \tmtexttt{flg\_smartsig}, and
implement the same mechanism for avoiding useless calls to user routines. By
closely examining the code, the reader can find out how this works. The
output of the program, however, should be independent on the setting of this
flag. Only speed will be affected. In fact, the random number generator, used
to set up the random kinematics to check matrix elements for proportionality,
is reset to its original value after the equivalent matrix elements are
found.

In order to understand how the programs operate when the
\tmtexttt{flg\_smartsig} flag is turned on (i.e.~is set to true) it is better
to examine the \tmtexttt{allborn} routine.  Upon the first call to
\tmtexttt{allborn}, the current random number is saved, and then the Born
cross section for all flavour components is computed, for several value of
randomly chosen external momenta. An integer array \tmtexttt{equivto} is set
up, its value being -1 by default. If the Born contribution for the
$j^{\tmop{th}}$ Born flavour configuration is found to be proportional to a
previous $k^{\tmop{th}}$ Born flavour configuration (with $k < j$), the value
of \tmtexttt{equivto(j)} is set equal to $k$, and the array of real numbers
\tmtexttt{equivcoeff(j)} is set to the proportionality constant. Upon
subsequent calls to \tmtexttt{allborn}, this information is used to avoid
further calls to \tmtexttt{setborn}, whenever possible. Notice the use of
\tmtexttt{randomsave} and \tmtexttt{randomrestore}. By enclosing a set of
instructions between a \tmtexttt{randomsave} and a \tmtexttt{randomrestore}
call, we make sure that the random number sequence is not altered by the
inserted instructions.

If the flag \tmtexttt{flg\_smartsig} is set to false, all calls to the matrix
element routines are performed, but, thanks to the saving and restoring of
the random number sequence, the output of the program should be independent
of it. In other words, using \tmtexttt{flg\_smartsig=true} should only
accelerate the program, without altering the output. This feature can be used
to check that nothing weird has happened in the setup phase of the
\tmtexttt{equivto} and \tmtexttt{equivcoef} arrays.

\subsection{The soft, collinear and soft-collinear limit functions}
These limit functions could be obtained, in principle, by numerical methods,
using the full real contribution. We have, however, preferred to compute them
using the factorization formulae for collinear singularities, and the eikonal
formulae for soft emission, to avoid numerical instabilities. Furthermore,
the real contributions, and all the manipulations performed by the
combinatoric package can be tested for consistency (see
appendix~\ref{app:check}).

These soft, collinear and soft-collinear routines are
collected in the file \tmtexttt{sigcollsoft.f}. The subroutines relevant for
the computation of the \tmtexttt{btilde} function are reported in
table~\ref{tab:softcoll}.
\begin{table}[tbh]
\begin{center}
  \begin{tabular}{|c|c|}
    \hline
    \tmtexttt{soft} & soft limit\\
    \hline
    \tmtexttt{collfsr} & collinear limit for FSR\\
    \hline
    \tmtexttt{softcollfsr} & soft-collinear limit for FSR\\
    \hline
    \tmtexttt{collisrp} & collinear in the $\oplus$ direction\\
    \hline
    \tmtexttt{softcollisrp}  & soft-collinear limit in the $\oplus$
    direction\\
    \hline
    \tmtexttt{collisrm} & collinear in the $\ominus$ direction\\
    \hline
    \tmtexttt{softcollisrm} & soft-collinear limit in the $\ominus$
    direction\\
    \hline
  \end{tabular}
  \caption{\label{tab:softcoll} Subroutines for the soft and collinear limits
    of the real contributions used in the computation of \tmtexttt{btilde}.}
\end{center}
\end{table}  
The basic formulae for the collinear limits are collected in
appendix~\ref{app:colllim}. Here we illustrate the code of the
\tmtexttt{collfsr} routine.  This routine in turns calls
\tmtexttt{collfsrnopdf}, and provides the luminosity factor to its output.
Another ingredient that is necessary to build the collinear limit functions
is the direction of the transverse momentum $\hat{k}_{\sss\rm T}$ of the
radiated parton with respect to the emitter in the collinear limit (see
appendix \ref{app:colllim}), defined in the partonic CM frame.  This is a
function of the emitter direction and of the azimuthal angle $\phi$. The
origin of the azimuth. i.e.~the plane along which $\phi = 0$ (or $\pi$)
should be consistent with what \tmtexttt{gen\_real\_phsp\_fsr} does in the
collinear limit. A change of $\phi \rightarrow \phi + \pi$ is instead
irrelevant. Our convention for the origin of $\phi$ is to take a plane
containing $\bar{k}_{\tmop{em}}$ (the momentum, in the CM frame, of the
parton that will undergo the splitting in the underlying Born, see
ref.~\cite{Frixione:2007vw}), and the third axis. The subroutine
\tmtexttt{buildkperp}, called from the \tmtexttt{collfsrnopdf} routine,
constructs the 4-vector \tmtexttt{kperp(0:3)}. This vector is normalized
arbitrarily, and has zero time component.  Its only requirement is that it
should be parallel to $\hat{k}_{\sss\rm T}$. Its modulus squared is also
returned by the \tmtexttt{buildkperp} routine.  For each \tmtexttt{alr}
sharing the current emitter, the subroutine \tmtexttt{collfsralr} is called,
with \tmtexttt{kperp} also passed as an argument.  The values of $\xi$ and
$x$, defined as
\begin{equation}
  x = \frac{k^0}{\bar{k}^0_{\tmop{em}}},\qquad\quad \xi = \frac{2
  k^0}{\sqrt{s}}\,,
\end{equation}
and $x / \xi$ are all passed to the subroutine, that is meant to work also if
$\xi$ and $x$ vanish, with their ratio remaining finite.  The subroutine
\tmtexttt{collfsralr} implements the formulae given in
appendix~\ref{app:colllim} in a straightforward way, multiplying them by
$\xi^2 (1 - y)$, and taking care to use the $x / \xi$ variable when
necessary, in such a way that one never divides by $x$ or $\xi$. There is
only one caveat to keep in mind.  The collinear approximation is meant to
reproduce an $R^{\alpha}$ contribution. In the collinear limit of the
$\alpha$ region, this coincides with $R$, since the $S^{\alpha}$ factor
becomes 1 in this limit. We should remember, however, that, in case the
emitter and radiated partons are both gluons, we have also supplied a factor
$2 E_{\tmop{em}} / (E_{\tmop{em}} + E_{\rm r})$, which becomes $2 (1 - x)$ in
the collinear limit (see eq.~(\ref{eq:softemitterfac})). In addition, in this
case, we supply a factor of $1 / 2$ to account for the two identical partons
in the final state. Thus, an extra factor of $(1 - x)$ is supplied.

The case of initial-state collinear singularities is handled similarly. In
this case we simply have $x = 1 - \xi$, and, as before, one should evaluate
all contributions taking care of never dividing by $1 - x$. The routine
\tmtexttt{collisralr} implements the formulae in appendix~\ref{app:colllim}.
It carries an integer argument $i$ that corresponds to 1 for the collinear
$\oplus$ direction and 2 for the $\ominus$ direction.

In order to get the soft-collinear limits, the subroutines
\tmtexttt{softcollfsr}, \tmtexttt{softcollisrp} and \tmtexttt{softcollisrm}
simply call the corresponding collinear subroutines setting temporarily
\tmtexttt{kn\_csi} equal to zero.

The soft limit is obtained with the subroutine \tmtexttt{soft}, that in turn
calls \tmtexttt{softalr} to get the soft contribution of a single
\tmtexttt{alr}. It implements formula~(\ref{eq:softlim}) (for $\epsilon
= 0$), multiplied by $\xi^2 (1 - y)$ for an FSR region, or $\xi^2 (1 - y^2)$
for an ISR region. In formula~(\ref{eq:softlim}) there are two powers of the
soft momentum $k$ in the denominator. We then factorize $k^0$ in front of the
four-vector $k$ and define $k = k^0 \hat{k}$, so that ($\xi=2k^0/\sqrt{s}$)
\begin{equation}
  \left( \frac{\xi}{k^0} \right)^2 = \frac{4}{s}
\end{equation}
is finite.  The vector $\hat{k}$ carries the information of the direction of
the radiated parton. In practice, we thus replace $k\,\to\,\hat{k}$ in
eq.~(\ref{eq:softlim}), and supply the factor $4 (1 - y) / s$ or $4 (1 - y^2)
/ s$. The vector $\hat{k}$ should equal $k / k^0$ in the limit $\xi
\rightarrow 0$, keeping $y$ and $\phi$ fixed, for the given underlying Born
kinematics. It is computed in the subroutine \tmtexttt{gen\_real\_phsp\_fsr}
and \tmtexttt{gen\_real\_phsp\_isr}, with a call to the subroutine
\tmtexttt{setsoftvec}. Furthermore, we should remember that the functions
$S^{\alpha}$ have non-trivial soft limits. They should be computed in the
soft limit and multiplied by the result. For this purpose, the routine
\tmtexttt{compdijsoft}, in the \tmtexttt{gen\_real\_phsp.f} file, computes
the soft limit of the $d_{i j}$ functions, and stores them in the array
\tmtexttt{kn\_dijterm\_soft}. This array has a single index, since the second
one is the index of the soft parton. There is no need to consider the other
$d_{i j}$ terms (those not involving the soft parton) since in the soft limit
they are finite, and do not contribute to $S^{\alpha}$. Observe that
\tmtexttt{compdijsoft} assumes that the $d_{i j}$ terms are homogeneous in
$k^0$, which is the case if the two parameters \tmtexttt{par\_diexp} and
\tmtexttt{par\_dijexp} are equal. If they differ, the fastest vanishing one
(in the $k^0 \rightarrow 0$ limit) will dominate $S^{\alpha}$. It is in
principle possible to experiment with settings that have different
\tmtexttt{par\_diexp} and \tmtexttt{par\_dijexp}, provided that the slowest
vanishing ones are excluded in some way from the list.  At the moment, this
possibility has not been investigated.  As a last point, special treatment
was required for the FSR collinear limit of two outgoing gluons.  Here, in
fact, no action should be taken: one should provide a factor $(1 - x)$, that
equals one in the soft limit.

\section{Tuning the real cross section in \POWHEG{}}
\label{sec:damprem}
In \POWHEG{} it is possible to tune the contribution to the real cross
section that is treated with the Monte Carlo shower technique.  This was
pointed out first in ref.~\cite{Nason:2004rx}, where the \POWHEG{} method was
formulated, and it was first implemented in ref.~\cite{Alioli:2008tz}.  In
\POWHEG{} there is the possibility to separate the real cross section, in a
given singular region $\alpha$, as follows
\begin{equation}
\label{eq:singplusreg}
  R^{\alpha} = R^{\alpha}_s + R^{\alpha}_f, 
\end{equation}
where $R^{\alpha}_f$ has no singularities and only $R^{\alpha}_s$ is singular
in the corresponding region. In practice, the separation may be achieved, for
example, using a function of the transverse momentum of the radiation $0
\leqslant F\! \left( k_T^2 \right) \leqslant 1$, that approaches 1 when its
argument vanishes, and define
\begin{eqnarray}
  R^{\alpha}_s & = & R^{\alpha} F\! \left( k_T^2 \right), \\
\label{eq:Rf_term}
  R^{\alpha}_f & = & R^{\alpha}  \lq 1 - F\! \left( k_T^2 \right) \rq . 
\end{eqnarray}
One carries out the whole \POWHEG-style generation using $R^{\alpha}_s$
rather than $R^{\alpha}$. The contribution $R^{\alpha_{}}_f$, being finite,
is generated with standard techniques, and fed into a shower Monte Carlo
as is.

More generally $F$ can be chosen as a general function of the kinematic
variables, provided it approaches 1 in the singular region. This turns out to
be useful in all cases when the ratio $R/B$ in
the \POWHEG{} Sudakov exponent becomes too large with respect to its
corresponding collinear or soft approximation (see for example
ref.~\cite{Alioli:2008gx}). In this case, radiation generation becomes highly
inefficient. A general solution to this problem (which has already been
implemented in refs.~\cite{Nason:2009ai} and~\cite{POWHEG_Zjet}) is to chose
the function $F$ in the following way: if the real squared amplitude (no
parton distribution functions included), in a particular singular region, is
greater than five times its soft and collinear approximation,
then $F$ is set to zero, otherwise is set to one. We also stress that this
procedure remedies automatically to the Born zeros problem examined in
ref.~\cite{Alioli:2008gx}.

This feature is implemented in the \POWHEGBOX.  By setting the flag
\tmtexttt{flg\_withdamp} to true, this behaviour is turned on.  When
computing the \tmtexttt{btilde} function, the real contribution will always
be multiplied by a damping factor, supplied by the routine
\tmtexttt{bornzerodamp}. The damping factor is not necessary in the soft and
collinear counterterm contributions, since, in these cases, we will certainly
have $F = 1$.  The routine \tmtexttt{bornzerodamp} takes as argument the
$\alpha$-region index (i.e.~the \tmtexttt{alr}), the value of
$\mathcal{R}^{\alpha}$ (i.e.~the real cross section \tmtextit{without} the
parton distribution function factors) and the value of its collinear and soft
limits (also without pdf factors). It returns the damping factor as its last
argument. The presence of the collinear and soft limits of
$\mathcal{R}^{\alpha}$ in the arguments of the subroutine, allows the user to
set a damping factor that depends upon the distance of the real contribution
from its collinear or soft approximation, as stated previously. This routine
can be easily modified by the user: for example, the sharp theta function
adopted in the \POWHEGBOX{} can be replaced by a smoother function, the
factor of five can be changed, and so on.

One can see the effects of setting the \tmtexttt{flg\_withdamp} flag in the
\tmtexttt{sigreal\_btl} subroutine. The soft and collinear limits of the real
contribution are obtained with calls to the subroutines \tmtexttt{collbtl}
and \tmtexttt{softbtl} (the \tmtexttt{btl} ending standing for
\tmtexttt{btilde}), that make use of the subroutines already described
previously for the calculation of the soft and collinear limits.

If a damping factor is used, the leftover term $R^{\alpha}_f$ of
eq.~(\ref{eq:Rf_term}) needs to be handled independently.  The subroutine
\tmtexttt{sigremnants} deals with this term together with the real terms that
do not have any associated singular region, if there are any. It has the same
calling sequence of \tmtexttt{btilde}. It is meant to be integrated using the
\tmtexttt{mint} integration program, that allows for the possibility of
generating phase space kinematics distributed with a probability proportional
to the integrand, after a single integration.  Within \tmtexttt{sigremnant},
the contribution from the regular real graphs can be integrated with an
arbitrary phase-space parametrization, that we choose to be the initial-state
radiation parametrization, i.e.~the \tmtexttt{gen\_real\_phsp\_isr}
subroutine. Both the underlying Born configuration and the real phase space
are generated within \tmtexttt{sigremnant}. The regular contributions to the
real cross section are returned by the subroutine
\tmtexttt{sigreal\_reg}. Within \tmtexttt{sigreal\_reg}, by making use of the
list of regular real contributions (that is setup when the combinatorics is
carried out), the regular contributions to the cross section are
computed. The contribution of the $R^{\alpha}_f$ terms is more delicate. This
is computed with a loop through all possible emitter values using the global
variable \tmtexttt{kn\_emitter}.  The real phase space is set according to
it.  Then the subroutine \tmtexttt{sigreal\_damp\_rem} (where
\tmtexttt{damp\_rem} stands for damp remnants) is invoked. This subroutine is
very similar to the \tmtexttt{sigreal\_btl} subroutine. For all
\tmtexttt{alr} that share the current emitter, the corresponding $R^{\alpha}$
is computed, the damping factor \tmtexttt{dampfac} is computed, and the real
result is multiplied by \tmtexttt{(1-dampfac)} (in \tmtexttt{sigreal\_btl}
the result was instead multiplied by \tmtexttt{dampfac}).

Notice that \tmtexttt{sigreal\_damp\_rem} and \tmtexttt{sigreal\_btl} carry
out very similar tasks, the only difference being the presence of the factor
$(1-F)$ in the first, and $F$ in the second. This fact is exploited in the
\POWHEGBOX{} by implementing both of them via a call to a single subroutine
\tmtexttt{sigreal\_btl0}, that carries an extra integer argument. When the
extra argument is zero, the multiplication factor is set equal to $F$, and
when it is 1, it is set to $(1-F)$.

\section{The initialization phase}
\label{sec:bbinit}
The preparation of the grids for the generation of the events is carried out
in the subroutine \tmtexttt{bbinit}. Its most important task is to execute
the integration of the \tmtexttt{btilde} function, determine the fraction of
negative weights, compute the total cross section, and, if required, plot the
NLO distributions. At the first step, the subroutine
\tmtexttt{mint}~\cite{Nason:2007vt} is invoked with \tmtexttt{imode} set to
0. In this mode, \tmtexttt{mint} integrates the absolute value of
\tmtexttt{btilde}, and sets up the importance-sampling grid. Next,
\tmtexttt{mint} is invoked with \tmtexttt{imode} set to 1, and the flag
\tmtexttt{negflag} set to true. In this mode, \tmtexttt{mint} computes the
negative contribution to the \tmtexttt{btilde} function. No histograms for
the NLO results are generated up to now. At this stage, \tmtexttt{negflag} is
set to false, the \tmtexttt{flg\_nlotest} is set to true, and \tmtexttt{mint}
is invoked again on \tmtexttt{btilde} to compute the positive contribution to
the integral. At this stage, the NLO histograms are filled. We stress that
also negative weights, if present, will end up in the histograms, so that the
NLO histograms should exactly correspond to a standard NLO calculation. The
positive weight total cross section computed by \tmtexttt{mint} is combined
with the negative weight part, and stored in a variable
\tmtexttt{rad\_sigbtl}, defined in the header file
\tmtexttt{pwhg\_rad.h}. After this, the contribution to the cross section
from the real remnants is also computed. These are terms that arise either
because there are real contributions with no associated singular regions, or
because \tmtexttt{flg\_withdamp} is set to true (see
section~\ref{sec:damprem}). The remnant cross section calculation is performed
with an independent set of grids. Also the remnant contributions will end up
in the NLO histograms. The remnant cross section is stored in the variable
\tmtexttt{rad\_sigrm}, and the total in
\tmtexttt{rad\_sigtot=rad\_sigrm+rad\_sigbtl}.

When \tmtexttt{mint} is called with \tmtexttt{imode} equal to 1, the upper
bounding envelope of the integrated function is also computed, and stored in
an array. This upper bounding envelope will be used later for the generation
of unweighted events. The arrays \tmtexttt{xgrid}, \tmtexttt{ymax},
\tmtexttt{xmmm} are all necessary for the generation of the events, and they
can be saved in a file, so that the time consuming initialization phase does
not need to be repeated if one wishes to generate more events in the same
context.

The final important task of the \tmtexttt{bbinit} routine is the call to the
\tmtexttt{do\_maxrat} subroutine, that sets up the normalization of the upper
bounding function for radiation, thus preparing the system for the generation
of full events.  This will be described in
section~\ref{sec:norm_upper_bound}.  In \tmtexttt{bbinit} an initialization
call to the function \tmtexttt{gen}, that generates the underlying Born
configuration, is also performed.

\section{The generation of radiation}\label{sec:rad}
There are two components that contribute to the generation of radiation: one
arises from the $\bar{B}$ term and the other from the remnant. The total
cross section for the two contributions is stored in the global variables
\tmtexttt{rad\_sigbtl} and \tmtexttt{rad\_sigrm}. When radiation is
generated, one begins by picking one of the two cases with a probability
proportional to the respective cross section. In the \POWHEGBOX{}, the
generation of radiation is carried out in the subroutine
\tmtexttt{pwhgevent}, that begins precisely by performing this random
choice. We describe, in turn, the two components.

\subsection{Radiation from the $\bar{B}$ component}
This begins with the generation of an underlying Born configuration
distributed according to the $\bar{B}$ function.  Radiation is generated
using the \POWHEG{} Sudakov form factor (see eq.~(4.21) of
ref.~\cite{Frixione:2007vw})
\begin{equation}
  \Delta^{f_b} (\Phin, \pt) = \prod_{\alr \in \{\alr |f_b \}}
  \Delta^{f_b}_{\alr} (\Phin, \pt),
\end{equation}
where
\begin{equation}
\label{eq:pwhgsuda}
  \Delta^{f_b}_{\alr} (\Phin, \pt) = \exp \left\{ - \left[ \int d
  \Phi_{\tmop{rad}} \,\frac{R (\Phinpo)}{B^{f_b} (\Phin)} \,\theta\!\(
\kt\!\(\Phinpo\) - \pt\) \right]^{\bar{\bf \Phi}_n^{\alr} = \Phin}_{\alr}
  \right\} . 
\end{equation}
If $R$ has been separated into a regular and singular part, according to
eq.~(\ref{eq:singplusreg}), only the singular part will appear in the Sudakov
form factor. According to the notation of ref.~\cite{Frixione:2007vw}, the
square bracket with the $\alr$ suffix indicates that all quantities inside
the bracket should be taken relative to the $\alr$ region. So, the $n + 1$
phase space in eq.~(\ref{eq:pwhgsuda}) is given as a function of the
underlying Born phase space $\bar{\bf \Phi}_n^{\alr}$, taken at the point
$\Phin$, and of the radiation variables $\Phi_{\tmop{rad}}$, according to the
phase-space mapping defined for the $\alr$ region. In \POWHEG{}, the
individual Sudakov form factors for each $\alr$ are further assembled into a
product of form factors sharing the same underlying Born and the same
radiation region. As we have seen, each singular region is characterized by
an emitter and an emitted parton. Within \POWHEG{}, the emitted parton is
always the last one, while the emitter can be any light coloured parton in
the initial or final state. There is one single initial-state radiation
kinematics, independent of which incoming parton is emitting. The final-state
radiation kinematics depends instead upon the index of the emitter. We
introduce here the label \tmop{rr} to specify the radiation region
kinematics: \tmop{rr} = 1, if \tmtexttt{kn\_emitter}=0, 1 or 2, and \tmop{rr}
= \tmtexttt{kn\_emitter\,-\,flst\_lightparton\,+\,2}, if
\tmtexttt{kn\_emitter} $\ge$ \tmtexttt{flst\_lightparton}.  In fact, within
the \POWHEGBOX{} framework, the radiation kinematics is the same for
\tmtexttt{kn\_emitter}=0, 1 or 2.  We write
\begin{equation}
  \Delta^{f_b}(\Phin, \pt) = \prod_{\tmop{rr} \in \{\tmop{rr}
  |f_b \}} \Delta^{f_b}_{\tmop{rr}} (\Phin, \pt),
\end{equation}
where
\begin{equation}
  \label{eq:pwhrrgsuda}
  \Delta^{f_b}_{\tmop{rr}} (\Phin, \pt) = \exp \left\{ - \sum_{\alr \in
  \{\alr |f_b, \tmop{rr}\}} \left[ \int d \Phi_{\tmop{rad}} \,\frac{R
  (\Phinpo)}{B^{f_b} (\Phin)} \,\theta (\kt (\Phinpo) - \pt)
  \right]^{\bar{\bf \Phi}_n^{\alr} = \Phin}_{\alr} \right\},
\end{equation}
and the notation $\{\alr |f_b, \tmop{rr}\}$ indicates the ensemble of all
$\alr$ that share the same underlying Born $f_b$ and the same radiation
region kinematics $\tmop{rr}$. It makes then sense to define
\begin{equation}
  R^{\tmop{rr}} (\Phinpo) = \sum_{\alr \in \{\alr |f_b,
    \tmop{rr}\}} R^{\alr} (\Phinpo),
\end{equation}
since the phase space only depends upon the radiation region kinematics
$\tmop{rr}$, and
not on the specific $\alr$. With this definition we have
\begin{equation}
\label{eq:pwhrrgsudarr}
  \Delta^{f_b}_{\tmop{rr}} (\Phin, \pt) = \exp \left\{ - \left[ \int d
  \Phi_{\tmop{rad}}\, \frac{R^{\tmop{rr}} (\Phinpo)}{B^{f_b} (\Phin)}
  \, \theta (\kt (\Phinpo) - \pt) \right]^{\bar{\bf \Phi}_n^{\alr} =
  \Phin}_{\alr} \right\} . 
\end{equation}
In order to generate the radiation, the \POWHEGBOX{} uses the highest-bid
algorithm. For each $\tmop{rr}$, it generates a $\pt$ value with a
probability distribution equal to
\begin{equation}
P^{f_b}_{\tmop{rr}} (\pt) =
\frac{\partial}{\partial \pt} \Delta^{f_b}_{\tmop{rr}} (\Phin, \pt).
\end{equation}
The program then selects the highest $\pt$ value, and thus fixes the
corresponding \tmop{rr} region.  The $\alr$ value is picked in the ensemble
$\{\alr |f_b, \tmop{rr}\}$, with a probability proportional to the
corresponding $R_{\alr}$.

The individual $\pt$ values for each $\Delta^{f_b}_{\tmop{rr}}$ are generated
with the veto method. We define
\begin{equation}
  d \Phi_{\tmop{rad}} = J^{\tmop{rr}} \, d\xi \, d y \, d \phi,
\end{equation}
where $J^{\tmop{rr}}$ is the Jacobian of the $\Phi_{\tmop{rad}}$ phase space,
when written as function the three radiation variables: $\xi$, $y$ and
$\phi$. We then introduce a suitable upper bounding function $U^{\tmop{rr}}
(\xi, y)$, and determine its normalization $\Nrr$ by requiring
\begin{equation}
  \frac{J^{\tmop{rr}} \, R^{\tmop{rr}} (\Phinpo)}{B^{f_b} (\Phin)}
  \leqslant \Nrr \, U^{\tmop{rr}} (\xi, y)\,,
  \label{eq:normubound}
\end{equation}
for all $\Phin$ and $\Phi_{\tmop{rad}}$. In order to generate the radiation,
one first generates a $\pt$ value according to the probability distribution
\begin{equation}
 \label{eq:sudubound}
  P^U_{\tmop{rr}} (\pt) = \frac{\partial}{\partial \pt} \Delta^U_{\tmop{rr}}
  (\pt), 
\end{equation}
where
\begin{equation}
\Delta^U_{\tmop{rr}} (\pt) = \exp \left[ -\Nrr \int
  d \xi \, d y \, d \phi\, U^{\tmop{rr}} (\xi, y) \,\theta (\kt (\Phinpo) - \pt)
  \right],
\end{equation}
and then generates the corresponding values for the radiation variables
$\xi$, $y$ and $\phi$, distributed with a probability proportional to
$U^{\tmop{rr}}$. At this point one accepts the event with a probability
\begin{equation}
  \label{eq:acceptprob}
  \frac{1}{\Nrr\, U^{\tmop{rr}} (\xi,y)}
  \frac{J^{\tmop{rr}}\, R^{\tmop{rr}} (\Phinpo)}{B^{f_b} (\Phin)}\, .
\end{equation}
If the event is rejected, one goes back to the beginning, and generates a new
$\pt'$ value, smaller than the current one, using the probability
\begin{equation}
\label{eq:vetoedprob}
  P^U_{\tmop{rr}} (\pt', \pt) = \frac{\partial}{\partial \pt'} \,
  \frac{\Delta^U_{\tmop{rr}} (\pt')}{\Delta^U_{\tmop{rr}} (\pt)}\,, \qquad \pt'
  \leqslant \pt\, . 
\end{equation}
Ideally, the function $U^{\tmop{rr}}$ should be chosen in such a way that the
equation $r = \Delta^U_{\tmop{rr}} (\pt)$ can be easily solved for $\pt$, and
a set of radiation variables, distributed according to $U^{\tmop{rr}}$, are
easily generated. In practice we only require this second feature, and solve
for the equation $r = \Delta^U_{\tmop{rr}} (\pt)$ numerically (in this way,
if $r$ is a uniform random number between 0 and 1, the corresponding $\pt$ is
distributed according to eq.~(\ref{eq:sudubound})).  Several choices are
possible. In appendixes~\ref{app:ubfsr} and~\ref{app:ubisr} we describe the
functions used in the \POWHEGBOX.

In the \POWHEGBOX, a straightforward variant of the veto method is used
several times. Suppose that we know a function $F (\Phi_{\tmop{rad}})$ such
that
\begin{equation}
\label{eq:F_rad}
  \frac{J^{\tmop{rr}}\,R^{\tmop{rr}} (\Phinpo)}{B^{f_b} (\Phin)} \leqslant F
  (\Phi_{\tmop{rad}}) \leqslant \Nrr \, U^{\tmop{rr}}
  (\Phi_{\tmop{rad}})\, .
\end{equation}
Then we can first use the veto method accepting events with a probability
\begin{equation}
  \frac{F (\Phi_{\tmop{rad}})}{\Nrr\,  U^{\tmop{rr}}
    (\Phi_{\tmop{rad}})} .
\end{equation}
If the event is accepted, we go through a second veto, accepting the event with
a probability
\begin{equation}
  \frac{1}{F (\Phi_{\tmop{rad}})}  \frac{J^{\tmop{rr}} R^{\tmop{rr}} (\Phi_{n
  + 1})}{B^{f_b} (\Phin)} .
\end{equation}
This is obviously the same as accepting according to the probability of
eq.~(\ref{eq:acceptprob}), but it has the advantage that, in many instances,
when a veto is imposed, one only evaluates the function $F$, without the need
to compute the real and Born contributions. This method is also applied in
the \POWHEGBOX{} by replacing eq.~(\ref{eq:normubound}) with
\begin{equation}
  \label{eq:normuboundnew}
  \frac{J^{\tmop{rr}} \, R^{\tmop{rr}} (\Phinpo)}{B^{f_b} (\Phin)} \leqslant
  \Nrr (\xi, y) \, U^{\tmop{rr}} (\Phi_{\tmop{rad}})
  \leqslant \Nrr\, U^{\tmop{rr}} (\Phi_{\tmop{rad}})\,,
\end{equation}
where $\Nrr (\xi, y)$ is a stepwise function in the $\xi$ and $y$ radiation
variables, and where
\begin{equation}
  \Nrr = \max_{\xi, y} \Nrr (\xi, y) \,.
\end{equation}
One determines the step function $\Nrr(\xi, y)$ so to have the smallest
values that satisfy the first bound of eq.~(\ref{eq:normuboundnew}). Then,
the method described above is used, where, according to eq.~(\ref{eq:F_rad}),
\begin{equation}
  F (\Phi_{\tmop{rad}}) = \Nrr (\xi, y)\, U^{\tmop{rr}}
  (\Phi_{\tmop{rad}})\,,
\end{equation}
so that events are first accepted with a probability $\Nrr (\xi, y) / \Nrr$.

Within the \POWHEGBOX, the generation of the underlying Born kinematics is
performed by the routines \tmtexttt{gen\_btilde}, that invokes \tmtexttt{gen}
with the appropriate arguments.  After that, the subroutine
\tmtexttt{gen\_uborn\_idx} is called and it generates the underlying Born
flavour configuration. The purpose of this call is to pick a random $f_b$
configuration, with a probability proportional to its contribution to the
$\tilde{B}$ value at the given kinematic point.  By inspecting the
\tmtexttt{gen\_uborn\_idx} subroutine, we see how this task is performed. We
recall that when \tmtexttt{gen} returns, the last call to the
\tmtexttt{btilde} function has been performed at the generated Born
kinematics configuration. The contribution of each flavour component of the
$\widetilde{B_{}}$ cross section is stored in the array
\tmtexttt{rad\_btilde\_arr}. In \tmtexttt{gen\_uborn\_idx} a generic utility
subroutine \tmtexttt{pick\_random} is invoked with arguments
\tmtexttt{flst\_nborn}, \tmtexttt{rad\_btilde\_arr} and
\tmtexttt{rad\_ubornidx}. The \tmtexttt{pick\_random} subroutine returns the
value \tmtexttt{rad\_ubornidx} with a probability proportional to
\tmtexttt{rad\_btilde\_arr(rad\_ubornidx)}. The variable
\tmtexttt{rad\_ubornidx} represents the index of the currently generated
underlying Born configuration. There are several other \tmtexttt{rad\_}
prefixed global variables that need to be set, in order to perform the
generation of radiation from the current underlying Born. First of all, a
list of all \tmtexttt{alr}, that share the current underlying Born structure,
should be constructed. This is done by filling the array
\tmtexttt{rad\_alr\_list}, of length \tmtexttt{rad\_alr\_nlist}, using
\tmtexttt{flst\_born2alr}, that was constructed at the combinatoric stage of
the program.

The variable denoting the singular region $\tmop{rr}$ is represented by the
global variable \tmtexttt{rad\_kinreg} in the \POWHEGBOX. As already stated,
it takes the value 1, for initial-state radiation, and the value
\tmtexttt{rad\_kinreg\,=\,kn\_emitter\,-\,flst\_lightparton\,+\,2} for
final-state radiation.  Not all values of \tmtexttt{rad\_kinreg} may be
associated with an active radiation region for the given underlying Born. A
logical array \tmtexttt{rad\_kinreg\_on} is set up, with its entries indexed
by the \tmtexttt{rad\_kinreg} values. The entries set to true correspond to
active radiation regions. The array \tmtexttt{rad\_kinreg\_on} is set in the
subroutine \tmtexttt{gen\_uborn\_idx}.
\begin{table}[tbh]
\begin{center}
  \begin{tabular}{|c|c|}
    \hline
    \tmtexttt{rad\_ubornidx} &  index in the current underlying Born flavour
    structure\\
    \hline
    \tmtexttt{rad\_alr\_list} & list of \tmtexttt{alr}'s that share the
    current underlying Born\\
    \hline
    \tmtexttt{rad\_kinreg\_on} & marks the active singular regions for the
    current underlying Born\\
    \hline
    \tmtexttt{rad\_kinreg} & current singular region\\
    \hline
  \end{tabular}
  \caption{Global variables that characterize the generation of radiation for
    the given underlying Born configuration.\label{tab:radlists}}
\end{center}
\end{table} 
In table~\ref{tab:radlists} we summarize the global variables relevant to the
generation of radiation. We do need all these variables, because we typically
need to consider the \tmtexttt{alr} that share the current underlying Born
\tmtextit{and} the current kinematic region. This is done by going through
the \tmtexttt{rad\_alr\_list}, and considering only the \tmtexttt{alr}'s
whose emitter is compatible with the current \tmtexttt{rad\_kinreg} value.

\subsubsection{Normalization of the upper bounding function}
\label{sec:norm_upper_bound}
Before the radiation is generated, the normalization of the upper bounding
functions should be computed. This task is carried out by the subroutine
\tmtextup{\tmtexttt{do\_maxrat}}, which, in turn, is invoked in the
\tmtexttt{bbinit} subroutine. The normalizations $\Nrr (\xi, y)$ and $\Nrr$
are stored in the arrays

    \tmtexttt{rad\_csiynorms(rad\_ncsinorms,\,rad\_nynorms,\,rad\_nkinreg,\,flst\_nborn)},\\
\indent    \tmtexttt{rad\_norms(rad\_nkinreg,\,flst\_nborn)}.\\
By inspecting the \tmtexttt{do\_maxrat} routine, one can see that there is a
mechanism (that is better understood by studying the code) to store and
retrieve previously computed values for these arrays. The core of the
\tmtexttt{do\_maxrat} routine is a loop repeated \tmtexttt{nubound} times,
where \tmtexttt{nubound} is a parameter set in the \POWHEGBOX{} data file.
Within this loop, \tmtexttt{gen\_btilde} and \tmtexttt{gen\_uborn\_idx} are
called in sequence.  After that, radiation kinematic variables are set up
randomly. The program then loops over all valid radiation regions
(i.e.~\tmtexttt{rad\_kinreg} values). For the ISR radiation region, the
initial-state radiation phase space is generated, with a call to
\tmtexttt{gen\_real\_phsp\_isr\_rad0}, and for the final state a call to
\tmtexttt{gen\_real\_phsp\_fsr\_rad0} is performed.  The task of effectively
increasing the norms is performed in the routine \tmtexttt{inc\_norms}. The
phase space generation routines are slight variants of the phase space
routines previously encountered. They perform a similar task, but are
dependent upon the \tmtexttt{rad\_kinreg} setting (rather than the
\tmtexttt{kn\_emitter} value) and furthermore they compute the kinematics
starting from the values of \tmtexttt{kn\_csitilde}, \tmtexttt{kn\_y} and
\tmtexttt{kn\_azi} (details are found in the \tmtexttt{gen\_real\_phsp.f}
code). The \tmtexttt{inc\_norms} subroutine first sets factorization and
renormalization scales for radiation (the \tmtexttt{set\_rad\_scales} call),
then computes the Born and real cross section. The real cross section is
multiplied by the Jacobian $J^{\tmop{rr}}$. The upper bounding function
$U^{\tmop{rr}}$ is returned by the function \tmtexttt{pwhg\_upperb\_rad()},
and the ratio
\begin{equation}
\label{eq:normurat}
  \frac{J^{\tmop{rr}} \, R^{\tmop{rr}} (\Phinpo)}{B^{f_b} (\Phin)\,
    U^{\tmop{rr}} (\xi, y)} 
\end{equation}
is formed. Its maximum gives the $N^{\tmop{rr}}_{f_b} (\xi, y)$
normalization.  Notice that the subroutines \tmtexttt{sigborn\_rad} and
\tmtexttt{sigreal\_rad} return respectively the value of the Born cross
section for the current underlying Born (i.e.~for the underlying Born indexed
by \tmtexttt{rad\_ubornidx}), and the $R^{\tmop{rr}}$ real cross section.
Thus, the subroutine \tmtexttt{sigreal\_rad} is similar to
\tmtexttt{sigreal\_btl}, but it only computes the cross section contributions
that share the underlying Born stored in \tmtexttt{rad\_ubornidx} and the
singular region stored in \tmtexttt{rad\_kinreg}.

The first two indexes, $\xi_i$ and $y_i$, in the array
\tmtexttt{rad\_csiynorms} represent the step number of the stepwise function
$\Nrr(\xi, y)$ (i.e.~they are integer functions of $\xi$ and $y$
respectively). Their form can be found in the code, but may be subject to
future modifications.

For the purpose of tuning the choice of the upper bounding function, each
evaluation of formula~(\ref{eq:normurat}) is histogrammed, and the histogram
is printed in \TOPDRAWER{} format at the end of the upper-bound evaluation,
in the file \tmtexttt{pwghistnorms.top}. The efficiency in the generation of
radiation will depend upon the shape of this histogram. It is roughly
estimated by the ratio of the average value of formula~(\ref{eq:normurat})
over its maximum. Highest efficiencies are achieved if the histogram goes
sharply to zero near the maximum value of the abscissa. Lowest efficiencies
are characterized by histograms with long tiny tails.

\subsubsection{The \tmtexttt{gen\_radiation} routine}
This routine is invoked from \tmtexttt{pwhgevent}, after the call to
\tmtexttt{gen\_btilde} and \tmtexttt{gen\_uborn\_idx}. It loops through the
valid radiation regions (i.e.~the allowed \tmtexttt{rad\_kinreg} values) and
it calls either the \tmtexttt{gen\_rad\_isr} or the \tmtexttt{gen\_rad\_fsr}
routines, that generate and store in the global variables \tmtexttt{kn\_csi},
\tmtexttt{kn\_y} and \tmtexttt{kn\_azi} a set of kinematics radiation
variables. It also returns, in its argument, the value of the radiation
transverse momentum squared, \tmtexttt{t}, that is defined in the function
\tmtexttt{pwhg\_pt2}. If \tmtexttt{t} is the largest generated so far, the
kinematics radiation variables and the \tmtexttt{rad\_kinreg} value are saved
in local variables, because they are the candidate for the highest bid method
(discussed in appendix~B of ref.~{\cite{Frixione:2007vw}}). At the end of the
loop, if no call has generated any radiation, the routine exits, after
setting \tmtexttt{kn\_csi} to zero, which signals the generation of a
Born-like event. If radiation was generated, the saved values of the
radiation variables are restored in global variables, and the appropriate
phase-space generation routine is invoked. The \tmtexttt{sigreal\_rad}
routine is invoked again, followed by a \tmtexttt{gen\_real\_idx}
call. Besides returning $R^{\tmop{rr}}$, \tmtexttt{sig\_real\_rad} also
stores each cross section contribution, so that, after the radiation
kinematics is generated, the corresponding flavour structure can also be
generated with a probability proportional to each cross section
contribution. This is what the \tmtexttt{gen\_real\_idx} call does. The index
(i.e.~the \tmtexttt{alr}) of the corresponding real flavour structure is
stored in the variable \tmtexttt{rad\_realalr}.

The subroutine \tmtexttt{sigreal\_rad}, as stated earlier, is similar to
\tmtexttt{sigreal\_btl}, but it only computes the cross section contributions
that share the underlying Born stored in \tmtexttt{rad\_ubornidx} and the
singular region stored in \tmtexttt{rad\_kinreg}. It also takes care to avoid
generating gluon splittings into heavy quark pairs below threshold. More
precisely, if the radiation $\kt^2$, returned by the function
\tmtexttt{pwhg\_pt2()}, is below \tmtexttt{rad\_charmthr2} for $g \rightarrow
c \bar{c}$ or below \tmtexttt{rad\_bottomthr2} for $g \rightarrow b \bar{b}$,
the corresponding result is set to zero.

\subsubsection{The \tmtexttt{gen\_rad\_isr} and \tmtexttt{gen\_rad\_fsr}
routines} 
These routines generate a $\pt$ value according to the Sudakov form factors
in eq.~(\ref{eq:pwhrrgsudarr}).  They make essential use of the function
\tmtexttt{pt2solve(pt2,i)}, that represents the function $\log \tilde{\Delta}
(\pt^{\tmop{old}}) / \tilde{\Delta} (\pt)$.  The expression of $\log
\tilde{\Delta} (\pt)$ is as given in eq.~(\ref{eq:logdeltisr}), in the case
of initial-state radiation, or as given in eqs.~(\ref{eq:ubsolution1fsr})
or~(\ref{eq:ubsolution2fsr}) (in this last case, depending upon the form of
the upper-bounding function used, controlled by the value of the integer
variable \tmtexttt{iupperfsr}). The value of $\pt^{\tmop{old}}$ corresponds
to the last vetoed $\pt$.  The value $\log \tilde{\Delta} (\pt^{\tmop{old}})$
is represented by the variable \tmtexttt{xlr}. Thus, solving
\tmtexttt{pt2solve} for zeros represents the first step of each iteration of
the veto procedure.

We examine now in detail the case of final-state radiation, with
\tmtexttt{iupperfsr=1}, that also illustrates how the other cases work.
Looking at the function \tmtexttt{pwhg\_upperb\_rad}, we see that, for
\tmtexttt{iupperfsr=1}, the upper bounding function (up to the normalization
factor) has the form
\begin{equation}
\label{eq:Urr}
  \frac{U^{\tmop{rr}}}{\Nrr (\xi, y)} = \frac{\as^{\tmop{PW}}}{\xi (1 -
    y)}\,,
\end{equation}
where $\as^{\tmop{PW}}$ is the variable-flavour two loop expression for the
strong coupling constant used for radiation generation throughout the
\POWHEGBOX{} program. It is set by a call to \tmtexttt{set\_rad\_scales} (the
choice of scales are discussed in detail in appendix~\ref{app:scales}). In
appendix~\ref{app:ubfsr}, it is shown how to deal with this form of the upper
bounding function for the case of one loop, constant flavour $\as$, and for
constant $N^{\tmop{rr}}$ (i.e.\ not dependent upon $\xi$ and $y$). We thus
begin by considering the upper bounding function
\begin{equation}
\label{eq:Utilderr}
  \tilde{U}^{\tmop{rr}} = N^{\tmop{rr}}_{f_b} \,
  \frac{\as^{\tmop{rad}}}{\xi (1 - y)}\,,
\end{equation}
with $\as^{\tmop{rad}}$ given by the one-loop expression
(see eq.~(\ref{eq:alphagenrad}))
\begin{equation}
  \as^{\tmop{rad}}(\mu) = \frac{1}{b^{\tmop{rad}}_0 \log
    \frac{\mu^2}{\Lambda_{\tmop{rad}}^2}}\, .
\end{equation}
We must choose $b_0^{\tmop{rad}}$ and $\Lambda_{\tmop{rad}}$ in such a way
that
\begin{equation}
  \as^{\tmop{rad}} (\mu) \geqslant \as^{\tmop{PW}} (\mu)\,,
\end{equation}
in all the range $\mu > \pt^{\min}$, where $\pt^{\min}$ is the minimum
allowed $\pt$ for radiation. If this inequality is fulfilled, we will have
$\tilde{U}^{\tmop{rr}} \geqslant U^{\tmop{rr}}$ in all the relevant
range. The value of $b_0^{\tmop{rad}}$ is taken equal to
\begin{equation}
  b_0^{\tmop{rad}} = \frac{33 - 2 \times 5}{12 \pi}\,,
\end{equation}
while $\Lambda_{\tmop{rad}}$ is computed and stored in the global variable
\tmtexttt{rad\_lamll} by a call to the subroutine
\tmtexttt{init\_rad\_lambda} at initialization stage, from the routine
\tmtexttt{init\_phys}.

The routine \tmtexttt{gen\_rad\_isr} proceeds by initializing the variable
\tmtexttt{unorm} to the value of $\Nrr$, stored in the \tmtexttt{rad\_norms}
array. The variable \tmtexttt{unorm} is also made available, via a common
block, to the function \tmtexttt{pt2solve}. In the same way, the value of
\tmtexttt{kt2max}, appropriate to the current kinematics and radiation
region, is computed and made available to the \tmtexttt{pt2solve} routine,
together with the value of $\Lambda_{\tmop{rad}}$ and the number of flavour
(i.e.~5) to be used in $b_0^{\tmop{rad}}$. At this stage the function
\tmtexttt{pt2solve} is in the condition to operate properly. Its return
value, for \tmtexttt{iupperfsr=1}, corresponds to
formula~(\ref{eq:ubsolution1fsr}). The veto loop is started, with the
variable \tmtexttt{xlr} set to the log of a uniform random number $0 < r <
1$. The zero of the \tmtexttt{pt2solve} function is found (using the
\tmtexttt{dzero} \tmtexttt{CERNLIB} routine), which thus corresponds to a
$\pt$ value that solves the equation
\begin{equation}
  \log (r) - \log \Delta^{( \tilde{U}^{\tmop{rr}})} (\pt) = 0.
\end{equation}
If $\pt^2$ (denoted by \tmtexttt{t} in the \POWHEGBOX{}) is below the allowed
minimum value, a negative \tmtexttt{t} is returned to signal the generation
of an event with no radiation (i.e.~with Born-like kinematics). Otherwise a
sequence of vetoes is applied. First, the event is accepted with a
probability
\begin{equation}
  \frac{\as^{\tmop{PW}} (\pt^2)}{\as^{\tmop{rad}} (\pt^2)},
\end{equation}
and vetoed otherwise. After this veto is passed, the distribution of
eq.~(\ref{eq:Utilderr}) has been corrected for the use of $\as^{\tmop{rad}}$,
instead of the correct one, $\as^{\tmop{PW}}$.  At this stage, $\xi$ is
generated (at fixed $\pt$): its probability distribution is uniform in its
logarithm, as can be evinced from eq.~(\ref{eq:ubsolution1fsr}). The value of
$y$ is obtained by solving for $y$ the $\kt^2=\pt^2$ definition for FSR,
i.e.~eq.~(\ref{eq:k2fsr}). At this stage we can compute a further veto,
accepting the event with a probability
\begin{equation}
  \frac{\Nrr (\xi, y)}{\Nrr}\,,
\end{equation}
which is the number returned by the subroutine \tmtexttt{uboundfct}. After
this veto, $\xi$ and $y$ have been generated with probability
\begin{equation}
  \exp \left[ - \int U^{\tmop{rr}} \,
   \theta (\kt - \pt)\, d \xi\, d y\, d
  \phi \right]  2 \pi\ U^{\tmop{rr}}\, d\xi\,dy \,.
\end{equation}
At this stage, the Born cross section is computed with a call to
\tmtexttt{sigborn\_rad}, a uniform azimuth for radiation is also generated
and also \tmtexttt{sigreal\_rad} is called to compute the real cross
section. One now vetoes again accepting the event with a probability
\begin{equation}  
\frac{J^{\tmop{rr}}R^{\tmop{rr}}}{B_{f_b}}\times \frac{1}{U^{\tmop{rr}}}\,,
\end{equation}
and, after this, the $\xi$, $y$ and $\phi$ variables have been generated
according to the probability
\begin{equation}
  \exp \left[ - \int \frac{R^{\tmop{rr}}}{B_{f_b}} \, \theta (\kt - \pt)\, d
  \Phi_{\rm rad} \right] \frac{R^{\tmop{rr}}}{B_{f_b}} \, d \Phi_{\rm rad}\,,
\end{equation}
which is the desired result.

\subsection{Remnant radiation}
Within the subroutine \tmtexttt{pwhgevent}, the generation of an event with
the remnant component of the cross section is carried out as follows. First
the subroutine \tmtexttt{gen\_sigremnant} is invoked. This subroutine uses
the routine \tmtexttt{gen} to generate a point in the full phase space,
distributed with a probability proportional to the \tmtexttt{sigremnant}
cross section, using the grids previously prepared, as described in
section~\ref{sec:bbinit}. The phase space point remains stored in the
kinematics global variables. After that, the \tmtexttt{gen\_remnant}
subroutine is invoked. This subroutine generates the flavour structure of the
current event with the appropriate probability. This is possible because the
subroutine \tmtexttt{gen\_remnant} stores in the global arrays
\tmtexttt{rad\_damp\_rem\_arr} and \tmtexttt{rad\_reg\_arr} (defined in
\tmtexttt{pwhg\_rad.h}) each contribution to the cross section for the last
kinematics point computed. The \tmtexttt{gen\_remnant} subroutine picks a
contribution with a probability proportional to the values stored in these
arrays. If the contribution is a remnant (described in
section~\ref{sec:damprem}), its index is stored in \tmtexttt{rad\_realalr},
and the corresponding underlying Born and emitter is found. The radiation
phase space is thus generated again with this value of the emitter, and the
same values for the three parameters used to parametrize the radiation phase
space in \tmtexttt{sigremnant}. These parameters are stored by the
\tmtexttt{sigremnant} subroutine in the global array
\tmtexttt{rad\_xradremn}.  The recalculation of the radiation phase space is
necessary, since only in the case when \tmtexttt{gen\_remnant} picks the last
contribution computed, the phase space would already have the appropriate
settings. The \tmtexttt{gen\_remnant} subroutine also returns in its integer
argument the value 1 for remnant contributions or the value 2 for regular
contributions. If the contribution is from a regular part, its index is
retrieved and stored in \tmtexttt{rad\_realreg}, and the ISR phase space is
used, since this is the one we have chosen to use for all regular contributions.

\subsection{Completion of the event}
For simplicity, in the \POWHEGBOX{}, one always assumes that there is an
azimuthal symmetry, so that, in the generation of the Born phase space, one
can always require that some reference particle in the final state lies on
the $x z$ (or $y z$) plane, where $z$ is the direction of the beam axis. At
the end of the event generation, a random azimuthal rotation of the whole
event is performed. This is done within the \tmtexttt{pwhgevent} routine,
through a call to the subroutine \tmtexttt{add\_azimuth}.

Besides setting up the kinematics and the flavour structure, in order to pass
the event to the Les Houches Interface for User Processes~\cite{Boos:2001cv}
(LHIUP from now on), we must also decide up to which scale the subsequent
(SMC generated) shower should start. In case of a \tmtexttt{btilde} generated
event, this scale should coincide with the radiation transverse momentum. In
case of remnant or regular contribution, this choice is to some extent
ambiguous. In order to maintain some continuity of the remnant events with
the \tmtexttt{btilde} events, we also set this scale to the radiation
transverse momentum. For regular contributions, this value is better decided
on the basis of the specific process, and an appropriate function
\tmtexttt{pt2max\_regular} should be provided by the user, in the file
\tmtexttt{pt2maxreg.f}. The global variable \tmtexttt{rad\_pt2max} is set to
the maximum $\pt$ for the subsequent shower. It will be used in the LHIUP
interface to set the variable \tmtexttt{SCALUP}.

\section{The Les Houches interface for user processes}\label{sec:leshouches}
At last, the generated event is put on the LHIUP interface. The scale for
subsequent radiation is setup, and colours are assigned to the incoming and
outgoing partons. For $\bar{B}$ generated and remnant events, this task is
carried out by the subroutine \tmtexttt{gen\_leshouches}. For regular
remnants, a special routine, \tmtexttt{gen\_leshouches\_reg}, does the job
and should be provided by the user in the file
\tmtexttt{LesHouchesreg.f}. The different treatment in the two cases is due
to the fact that, in the case of $\bar{B}$ and remnant events, we have a
standard method to assign colour, that is correct in the singular region. For
regular contributions, instead, other methods should be used, like resorting,
for example, to the planar limit of the cross section formulae.

\begin{figure}[tbh]
\begin{center}
  \epsfig{file=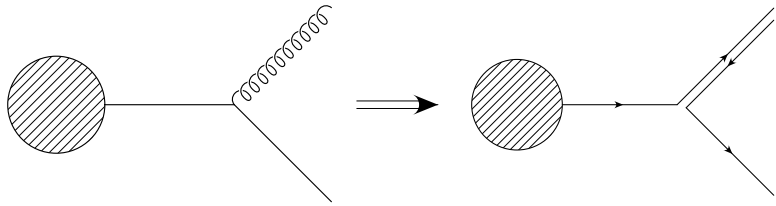,width=0.68\textwidth}
\end{center}
  \caption{\label{fig:qgcolor} Colour assignment for a singular region
    corresponding to a quark radiating a collinear gluon in the final state.}
\end{figure}

\begin{figure}[tbh]
\begin{center}
  \epsfig{file=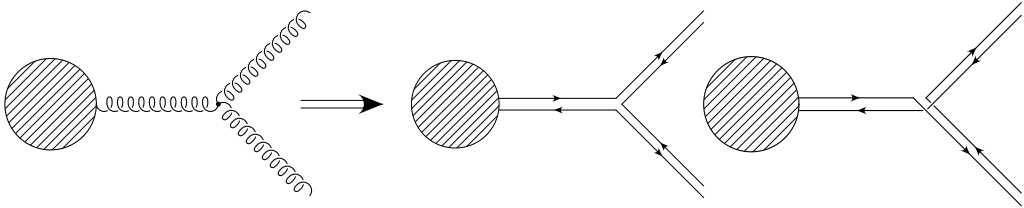,width=0.9\textwidth}
\end{center}
  \caption{\label{fig:ggcolor} The two alternative colour assignments for a
    singular region corresponding to a gluon radiating a collinear gluon in
    the final state.}
\end{figure}

In general, there is much room for improvement in the technique used for
colour assignment~\cite{Frixione:2007vw}. We do not consider this a crucial
problem at the present stage. However, if the need of a better colour
treatment will emerge, it is clear that the user should provide more colour
information. We thus limit ourselves, in the present case, to an approximate
colour implementation that is general enough to be process independent. What
we do is to assign colour on the basis of the underlying Born configuration
first. Then, depending upon the region we are considering, we assign the
colour of the real emitter and radiated parton as if a collinear splitting
process had really taken place. In the planar limit, this yields a unique
colour prescription for the emitter and the radiated parton, except for the
case of a gluon splitting into two gluons, that yields two possible colour
assignments with equal probabilities. In figures~\ref{fig:qgcolor}
and~\ref{fig:ggcolor} two particular cases are illustrated.

The routine \tmtexttt{gen\_leshouches} begins with a call to
\tmtexttt{born\_lh}, that sets up a few event-specific LHIUP parameters, like
the flavour, status and mothers of the underlying Born incoming and outgoing
particles. In order to understand this part of the code, the reader should
refer to the specifications of the LHIUP~\cite{Boos:2001cv}.  The
\tmtexttt{born\_lh} subroutine sets up also the colours of the underlying
Born process, by calling the \tmtexttt{borncolour\_lh} subroutine. This
subroutine is process dependent, and must be provided by the user. It should
return, in the LHIUP, a planar colour connection, with a probability
proportional to its Born contribution in the planar limit. For the most
simple processes, like for example $Z$ production, there is one single planar
colour connection, i.e.~the incoming quark and antiquark should have
complementary colours. For more complicated processes, more colour structures
can arise. In the planar limit, different colour structures do not interfere,
so it is possible to generate a single colour structure with a probability
equal to the corresponding contribution to the cross section. In some cases,
it may be useful to include also some colour-suppressed contributions. This
is the case, for example, in heavy flavour production, where the leading
colour configuration always leads to a heavy-flavour pair in an octet colour
state.  Singlet production may be more suited to the direct production of
bound states, and so, it may be better to include it. This can be done, as
long as different colour configurations do not interfere with each other.

In the case when $\xi = 0$, a Born event is produced, with a very low value
of the \tmtexttt{SCALUP} variable, so that no further radiation is generated
by the SMC. Within the \tmtexttt{gen\_leshouches} subroutine, this is
achieved by calling \tmtexttt{born\_lh}, by copying the \tmtexttt{kn\_pborn}
momenta on the LHIUP (a task carried out by the subroutine
\tmtexttt{momenta\_lh}) and by setting \tmtexttt{SCALUP} to the minimum
radiation transverse momentum. If $\xi \neq 0$, radiation has taken
place. The routine \tmtexttt{born\_lh} is called first, and one more parton
is added with the flavour of the radiated parton; the leg corresponding to
the emitter in the real graph is assigned the correct flavour. The colours
are assigned on the basis of the Born colour already stored in the LHIUP.
The subroutine \tmtexttt{setcolour\_rad} is used to perform this task for
both initial- and final-state radiation.

A final task of the LHIUP subroutine is to put on the interface the
intermediate resonances. The LHIUP specifies how resonances should be put in
the interface. This information should be made available to the shower
program, since resonance masses must be preserved by the shower. This is
achieved by calling the user routine \tmtexttt{resonances\_lh}, which calls
the \tmtexttt{add\_resonance} routine for each particle id of intermediate
resonances, specifying also its decay products.

\section{Conclusions}
\label{sec:conc}
In this work, we have introduced and documented the \POWHEGBOX{}, a computer
framework for the construction of a \POWHEG{} implementation of any given NLO
process. The \POWHEGBOX{} code is available at the
\url{http://moby.mib.infn.it/~nason/POWHEG/.}

In the \POWHEGBOX{} package, the user can found three directories:
\tmtexttt{W}, \tmtexttt{Z} and \tmtexttt{VBF\_H}.  They contain the code for
$W$, $Z$ and Higgs boson production in vector boson fusion, and can serve as
template for any further process that a user may want to implement.

We would like to emphasize that the \POWHEGBOX{} is a tool to develop
programs.  It is not something ready to run out of the box. Thus, even in
order to compile the examples, the user should examine the
\tmtexttt{Makefile}, and make sure that the pdf and jet libraries are
available in the system and are linked with the correct path.  The copy of
the code present in the repository is the SVN current version, marked with
its version number. From time to time, we will make new SVN versions
available as we implement new processes ourselves, or following important
code improvements.

A byproduct of our work is the implementation of the NLO corrections for an
arbitrary hadronic collision process, within the Frixione, Kunszt and
Signer~(FKS) subtraction scheme. The authors of ref.~\cite{Frederix:2009yq}
have also proposed a general FKS implementation for $e^+ e^-$ processes, and
they are currently extending it to hadronic
collisions~\cite{Frederix-private}.  We stress that in our approach, however,
the role of the NLO calculation is only meant to test the consistency of the
implementation, and no effort was made to improve its efficiency.

\section{Acknowledgments}
The work of S.A. has been supported in part by the Deutsche
Forschungsgemeinschaft in SFB/TR 9.

\appendix

\section{Soft contributions}
\label{app:soft}
In this appendix we document the calculation of the soft contribution in the
FKS subtraction framework. The real cross section in the soft approximation
is given by
\begin{equation}
  \mathcal{R} = 4 \pi \as \mur^{2 \epsilon} \left[ \sum_{i \neq j}
  \mathcal{B}_{i j} \frac{k_i \cdot k_j}{(k_i \cdot k) (k_j \cdot k)} -
  \mathcal{B} \sum_i \frac{k_i^2}{(k_i \cdot k)^2} \,C_i \right] +
  \mathcal{R}^f,
  \label{eq:softlim}
\end{equation}
where $\mathcal{B}_{i j}$ is the colour correlated Born cross section of
eq.~(\ref{eq:colourcorr}), and $C_i$ is the Casimir invariant for the
$i^{\tmop{th}}$ leg.  $\mathcal{R}^f$ has no singularities as the momentum of
the radiated parton, $k$, goes to zero.

\subsection{Soft phase space}
The phase space in the soft limit always factorizes as
\begin{equation}
  d \Phi^{n + 1} = d \Phi^n \frac{d^{d - 1} k}{2 k_0 (2 \pi)^{d - 1}} .
\end{equation}
We write now
\begin{equation}
\label{eq:dl1}
  d^{d - 1} k = d k_1\, d k_2\, d^{d - 3} k_{\perp} = d k_1 \,d k_2\, d
  k_{\perp} k_{\perp}^{- 2 \epsilon} \,\Omega^{1 - 2 \epsilon}, 
\end{equation}
where we have set $d = 4 - 2 \epsilon$, and $\Omega^\alpha$ is the solid
angle in $\alpha$ dimension
\begin{equation}
  \Omega^\alpha = \frac{\alpha\, \pi^{\alpha / 2}}{\Gamma (1 + \alpha / 2)} =
  \frac{\pi^{\alpha / 2} 2^\alpha \Gamma \left( \frac{\alpha + 1}{2}
  \right)}{\sqrt{\pi} \Gamma (\alpha)} \quad \Longrightarrow \quad\Omega^{1 -
  2 \epsilon} = 2 \frac{(4 \pi)^{- \epsilon} \Gamma \left( 1 - \epsilon
  \right)}{\Gamma (1 - 2 \epsilon)} .
\end{equation}
Turning eq.~(\ref{eq:dl1}) into polar coordinates we get
\begin{equation}
  \frac{d^{d - 1} k}{2 k^0 (2 \pi)^{d - 1}} = \frac{\pi^{\epsilon} \,\Gamma
  \left( 1 - \epsilon \right)}{\Gamma (1 - 2 \epsilon)} \frac{1}{(2 \pi)^3}
  k_0^{1 - 2 \epsilon} (\sin \theta \sin
  \phi)^{- 2 \epsilon} d k_0\, d \cos \theta \,d \phi  ,
\end{equation}
where we have defined
\begin{equation}
  k_1 = k_0\cos \theta,\qquad k_2 = k_0\sin \theta \cos \phi, \qquad k_{\perp}
  = k_0 \sin \theta \sin \phi.
\end{equation}
Since $k_{\perp}\ge 0$, this means that $0 \le \phi \le \pi$, and that only
even quantities can be integrated in this way. In other words, $k_{\perp}$
should not be confused with $k_3$ ($k_3$ is no longer available at this
stage). Inserting
\begin{equation}
  k_0 = \xi \frac{\sqrt{s}}{2},
\end{equation}
this becomes
\begin{equation}
  \frac{d^{d - 1} k}{2 k_0 (2 \pi)^{d - 1}} = \left[ \frac{(4 \pi)^{\epsilon}
  \Gamma \left( 1 - \epsilon \right)}{\Gamma (1 - 2 \epsilon)} \right] s^{-
  \epsilon} \frac{1}{(2 \pi)^3} \frac{s}{4} \,\xi^{1 - 2 \epsilon} (\sin \theta
  \sin \phi)^{- 2 \epsilon} d \xi \, d \cos \theta \, d \phi \, .
\end{equation}
By writing the $\mathcal{R}$ term as $\xi^{- 2} (\xi^2 \mathcal{R})$, we
notice that $(\xi^2 \mathcal{R})$ has a finite limit as $\xi \rightarrow
0$. The $\xi$ integration is performed by separating first
\begin{equation}
  \xi^{- 1 - 2 \epsilon} = - \frac{\xi_c^{- 2 \epsilon}}{2 \epsilon} \delta
  (\xi) + \left( \frac{1}{\xi} \right)_{\xi_c} - 2 \epsilon \left( \frac{\log
  \xi}{\xi} \right)_{\xi_c},
\end{equation}
where the $\delta(\xi)$ term yields the soft contribution. We thus have that
the integral of the soft-divergent part of $\mathcal{R}$ is given by
\begin{eqnarray}
  \mathcal{R}^s & = & - \frac{1}{2 \epsilon} \left[ \frac{(4 \pi)^{\epsilon}
  \Gamma \left( 1 - \epsilon \right)}{\Gamma (1 - 2 \epsilon)} \right] s^{-
  \epsilon} \xi_c^{- 2 \epsilon} \frac{1}{(2 \pi)^3} \int d \cos \theta\, d
  \phi (\sin \theta \sin \phi)^{- 2 \epsilon} 
  \nonumber\\ & & \times \frac{s
  \xi^2}{4} 4 \pi \as \mur^{2 \epsilon} \left[ \sum_{i \neq j}
  \mathcal{B}_{i j} \frac{k_i \cdot k_j}{(k_i \cdot k) (k_j \cdot k)} -
  \mathcal{B} \sum_i \frac{k_i^2}{(k_i \cdot k)^2} C_i \right],
\end{eqnarray}
where $\mathcal{R}^s$ is now independent upon $\xi$ (the dependence on $\xi$
of $k$ is canceled by the $\xi^2$ term in the numerator). Collecting the
normalization factor of eq.~(\ref{eq:Ndef}) in front, we get
\begin{eqnarray}
  \mathcal{R}^s & = & \mathcal{N} \left[ 1 - \frac{\pi^2}{6} \epsilon^2
  \right] \left( \frac{Q^2}{s \xi_c^2} \right)^{\epsilon} \left( \frac{- 1}{2
  \epsilon} \right) \frac{\as}{2 \pi} \int d \cos \theta \,\frac{d
  \phi}{\pi} (\sin \theta \sin \phi)^{- 2 \epsilon} 
  \nonumber\\ & &\times 
  \frac{s \xi^2 }{4} \left[ \sum_{i \neq j} \mathcal{B}_{i j} \frac{k_i \cdot
  k_j}{(k_i \cdot k) (k_j \cdot k)} - \mathcal{B} \sum_i \frac{k_i^2}{(k_i
  \cdot k)^2} \,C_i \right] .
\end{eqnarray}
We now define
\beqn
\label{eq:I_ij}
  \mathcal{I}_{i j} &=& \left[ 1 - \frac{\pi^2}{6} \epsilon^2 \right] \left(
  \frac{Q^2}{s \xi_c^2} \right)^{\epsilon} \left( -\frac{ 1}{2 \epsilon}
  \right) \int d \cos \theta \,\frac{d
  \phi}{\pi}  (\sin \theta \sin \phi)^{- 2 \epsilon} 
  \,\frac{s \xi^2 }{4} \frac{k_i \cdot k_j}{(k_i \cdot k) (k_j \cdot
    k)},\phantom{aaaa} 
\\
\label{eq:I_i}
 \mathcal{I}_{i} &=& \left[ 1 - \frac{\pi^2}{6} \epsilon^2 \right] \left(
   \frac{Q^2}{s \xi_c^2} \right)^{\epsilon} \left(- \frac{ 1}{2 \epsilon}
   \right) \int d \cos \theta\,\frac{d
  \phi}{\pi}  (\sin \theta \sin \phi)^{- 2 \epsilon} 
   \,\frac{s \xi^2 }{4} \frac{C_i k_i^2}{(k_i \cdot k)^2},
\eeqn
so that
\beq
 R^s = \mathcal{N} \frac{\as}{2 \pi} \left[ \sum_{i \neq j}
   \mathcal{I}_{i j} \mathcal{B}_{i j} - \mathcal{B} \sum_i \mathcal{I}_i
   \right] .
\eeq
We introduce then our basic integral
\begin{equation}
  I (p, q) = \int d \cos \theta\, \frac{d \phi}{\pi} (\sin \theta \sin \phi)^{-
  2 \epsilon}  \left[ \frac{s \xi^2}{4} \frac{p \cdot q}{p \cdot k\, q \cdot k}
  \right] = \frac{1}{\epsilon} I_d (p, q) + I_0 (p, q) + \epsilon I_{\epsilon}
  (p, q) .
\end{equation}
The expression of $I (p, q)$ will be substantially different for the case
when both $p$ and $q$ are massless, when one is massive and one massless, and
when both are massive.

\subsection{One massless and one massive particle}
Consider two momenta $p$ and $m$, with $p^2 = 0$ and $m^2 \neq 0$. We want to
evaluate $I (p, m)$. We separate out the collinear divergent component from
the eikonal factor
\begin{equation}
 \frac{p \cdot m}{p \cdot k\; m \cdot k} = \left[ \frac{p \cdot m}{p \cdot k\; m
  \cdot k} - \frac{n \cdot p}{p \cdot k\; n \cdot k} \right] + \frac{p \cdot
  n}{p \cdot k\; n \cdot k},
\end{equation}
where the square bracket term has no collinear singularities. Assuming $n$
along the time direction, we have:
\begin{equation}
  \frac{s \xi^2}{4}  \frac{n \cdot p}{p \cdot k\; n \cdot k} = \frac{1}{1 - \cos
  \theta},
\end{equation}
and
\begin{equation}
  \int d \cos \theta \,\frac{d \phi}{\pi} (\sin \theta \sin \phi)^{- 2
  \epsilon}\; \frac{1}{1 - \cos \theta} = -\frac{ 1}{\epsilon},
\end{equation}
so that
\begin{equation}
  I_d (p, m) = - 1 .
\end{equation}
The remaining integral has no collinear singularities so that we can write
\begin{equation}
  \int d \cos \theta\, \frac{d \phi}{\pi} (\sin \theta \sin \phi)^{- 2
  \epsilon} \,\frac{s \xi^2}{4} \left[ \frac{p \cdot m}{p \cdot k \, m \cdot k}
  - \frac{n \cdot p}{p \cdot k\, n \cdot k} \right] = I_0 (p, m) + \epsilon
  I_{\epsilon} (p, m)\,.
\end{equation}
Defining
\begin{equation}
  \hat{p} = \frac{p}{p_0}, \qquad \hat{m} = \frac{m}{m_0}, \qquad \beta =
  \frac{| \vec{m} |}{m_0},
\end{equation}
we have
\begin{eqnarray}
  I_d (p, m)\! & = & \! - 1, \\
  I_0 (p, m)\! & = & \!\log \frac{( \hat{p} \cdot \hat{m})^2}{\hat{m}^2}, \\
  I_{\epsilon} (p, m)\! & = &\! - 2 \left[ \frac{1}{4} \log^2  \frac{1 -
  \beta}{1 
  + \beta} + \log \frac{\hat{p} \cdot m}{1 + \beta} \log \frac{\hat{p} \cdot
  m}{1 - \beta} + \tmop{Li}_2 \left( 1 - \frac{\hat{p} \cdot m}{1 + \beta}
  \right) + \tmop{Li}_2 \left( 1 - \frac{\hat{p} \cdot m}{1 - \beta} \right)
  \right]\!. \nonumber\\ 
\end{eqnarray}
Thus, eq.~(\ref{eq:I_ij}), in the case where $k_1$ is massless and $k_2$ is
not, is given by
\begin{eqnarray}
  \mathcal{I}_{12} & = & \left[ 1 - \frac{\pi^2}{6} \epsilon^2 \right] \left(
  \frac{Q^2}{s \xi_c^2} \right)^{\epsilon} \left( -\frac{ 1}{2 \epsilon}
  \right) \int d \cos \theta \,\frac{d \phi}{\pi} (\sin \theta \sin \phi)^{- 2
  \epsilon}  \left[ \frac{s \xi^2}{4} \frac{k_1 \cdot k_2}{k_1 \cdot k\, k_2
  \cdot k} \right] \nonumber\\
  & = & \left[ 1 + \epsilon \log \frac{Q^2}{s \xi_c^2} + \left( \frac{1}{2}
  \log^2 \frac{Q^2}{s \xi_c^2} - \frac{\pi^2}{6} \right) \epsilon^2 \right]
  \left(- \frac{ 1}{2 \epsilon} \right) I (k_1, k_2) \nonumber\\
  & = & \frac{A}{\epsilon^2} + \frac{B}{\epsilon} + C \,,
\end{eqnarray}
where
\begin{eqnarray}
  A & = & \frac{1}{2} \,,\\
  B & = & \frac{1}{2} \log \frac{Q^2}{s \xi_c^2} - \frac{1}{2} I_0 (k_1,
  k_2)\,, 
  \\
  C & = & \frac{1}{2} \left[ \log^2 \frac{Q^2}{s \xi_c^2} - \frac{\pi^2}{6}
  \right] - \frac{1}{2} I_0 (k_1, k_2) \log \frac{Q^2}{s \xi_c^2} -
  \frac{1}{2} I_{\epsilon} (k_1, k_2) \,. 
\end{eqnarray}

\subsection{Two massless particles}
Using the identity
\begin{equation}
  \frac{k_1 \cdot k_2}{k_1 \cdot k\, k_2 \cdot k} = \frac{k_1 \cdot (k_1 +
  k_2)}{k_1 \cdot k \, (k_1 + k_2) \cdot k} + \frac{k_2 \cdot (k_1 + k_2)}{k_2
  \cdot k \,(k_1 + k_2) \cdot k},
\end{equation}
that holds if $k_1^2 = 0$ and $k_2^2 = 0$, we can immediately obtain the
expression of $I (k_1, k_2)$ for two massless momenta
\begin{equation}
  I (k_1, k_2) = I (k_1, k_1 + k_2) + I (k_2, k_1 + k_2) ,
\end{equation}
and use the results of the previous subsection for the two terms on the
right hand side.
We can write
\begin{eqnarray}
  \mathcal{I}_{12}  & = & \left[ 1 - \frac{\pi^2}{6} \epsilon^2 \right] \left(
  \frac{Q^2}{s \xi_c^2} \right)^{\epsilon} \left( -\frac{ 1}{2 \epsilon}
  \right) \int d \cos \theta\, \frac{d \phi}{\pi} (\sin \theta \sin \phi)^{- 2
  \epsilon}  \left[ \frac{s \xi^2}{4} \frac{k_1 \cdot k_2}{k_1 \cdot k\, k_2
  \cdot k} \right] \nonumber\\
  & = & \left[ 1 + \epsilon \log \frac{Q^2}{s \xi_c^2} + \left( \frac{1}{2}
  \log^2 \frac{Q^2}{s \xi_c^2} - \frac{\pi^2}{6} \right) \epsilon^2 \right]
  \left(- \frac{ 1}{2 \epsilon} \right) \left[ I (k_1, k_1 + k_2) + I (k_2,
  k_1 + k_2) \right] \nonumber\\
  & = & \frac{A}{\epsilon^2} + \frac{B}{\epsilon} + C 
\end{eqnarray}
with
\begin{eqnarray}
  A & = & 1 \\
  B & = & \log \frac{Q^2}{s \xi_c^2} - \frac{1}{2} \left[ I_0 (k_1, k_1 + k_2)
  + I_0 (k_2, k_1 + k_2) \right] \\
  C & = & \left[ \frac{1}{2} \log^2 \frac{Q^2}{s \xi_c^2} - \frac{\pi^2}{6}
  \right] - \frac{1}{2} \left[ I_0 (k_1, k_1 + k_2) + I_0 (k_2, k_1 + k_2)
  \right] \log \frac{Q^2}{s \xi_c^2}  \nonumber\\
  &  & - \frac{1}{2} \left[ I_{\epsilon} (k_1, k_1 + k_2) + I_{\epsilon}
  (k_2, k_1 + k_2) \right] . 
\end{eqnarray}

\subsection{Two massive particles}
In case both $k_1$ and $k_2$ are massive, we define
\beqn
  I (k_1, k_2) &=& I_0 (k_1, k_2) + \epsilon I_{\varepsilon} (k_1, k_2),
\\
  I_0 (k_1, k_2) &=& \int d \cos \theta \,\frac{d \phi}{\pi}  \left[ \frac{s
  \xi^2}{4} \frac{k_1 \cdot k_2}{k_1 \cdot k\; k_2 \cdot k} \right] ,
\\
  I_{\epsilon} (k_1, k_2) &=& - 2 \int d \cos \theta \,\frac{d \phi}{\pi} \log
  \(\sin \theta \sin \phi\) \left[ \frac{s \xi^2}{4} \frac{k_1 \cdot k_2}{k_1
  \cdot k \; k_2 \cdot k} \right] ,
\eeqn
and get (neglecting now $\epsilon^2$ terms)
\begin{eqnarray}
  \mathcal{I}_{12} & = & \left( \frac{Q^2}{s \xi_c^2} \right)^{\epsilon}
  \left(- \frac{ 1}{2 \epsilon} \right) \int d \cos \theta \,\frac{d \phi}{\pi}
  (\sin \theta \sin \phi)^{- 2 \epsilon}  \left[ \frac{s \xi^2}{4} \frac{k_1
  \cdot k_2}{k_1 \cdot k \; k_2 \cdot k} \right] \nonumber\\
  & = & \left[ 1 + \epsilon \log \frac{Q^2}{s \xi_c^2} \right] \frac{- I
  (k_1, k_2)}{2 \epsilon} = \frac{B}{\epsilon} + C \,,
\end{eqnarray}
with
\begin{eqnarray}
  B & = & - \frac{1}{2} I_0 (k_1, k_2)\,, \\
  C & = & - \frac{1}{2} I_0 (k_1, k_2) \log \frac{Q^2}{s \xi_c^2} -
  \frac{1}{2} I_{\epsilon} (k_1, k_2)\,, 
\end{eqnarray}
and
\begin{equation}
  I_0 (k_1, k_2) = \frac{1}{\beta} \log \frac{1 + \beta}{1 - \beta}\,, 
\qquad\quad
\beta =
  \sqrt{1 - \frac{k_1^2 k_2^2}{(k_1 \cdot k_2)^2}}\,.
\end{equation}
The expression for $I_{\epsilon}$ is defined by the equations below
\begin{eqnarray}
  a & = & \beta_1^2 + \beta_2^2 - 2 \,\vec{\beta}_1 \cdot \vec{\beta}_2, \\
  x_1 & = & \frac{\beta_1^2 - \vec{\beta}_1 \cdot \vec{\beta}_2}{a}, \\
  x_2 & = & \frac{\beta_2^2 - \vec{\beta}_1 \cdot \vec{\beta}_2}{a} = 1 - x_1,
  \nonumber\\
  b & = & \frac{\beta_1^2 \beta_2^2 - ( \vec{\beta}_1 \cdot
  \vec{\beta}_2)^2}{a}, \nonumber\\
  c & = & \sqrt{\frac{b}{4 a}}, \\
  z_+ & = & \frac{1 + \sqrt{1 - b}}{\sqrt{b}}, \\
  z_- & = & \frac{1 - \sqrt{1 - b}}{\sqrt{b}}, \\
  z_1 & = & \frac{\sqrt{x_1^2 + 4 c^2} - x_1}{2 c}, \\
  z_2 & = & \frac{\sqrt{x_2^2 + 4 c^2} + x_2}{2 c}, \\
  K (z) & = & - \frac{1}{2} \log^2  \frac{(z - z_-) (z_+ - z)}{(z_+ + z) (z_-
  + z)} - 2 \tmop{Li}_2 \left( \frac{2 z_- (z_+ - z)}{(z_+ - z_-) (z_- + z)}
  \right) \nonumber\\
  &  & - 2 \tmop{Li}_2 \left( - \frac{2 z_+ (z_- + z)}{(z_+ - z_-) (z_+ - z)}
  \right), \\
  I_{\epsilon} (k_1, k_2) & = & \lq K (z_2) - K (z_1)\rq \frac{}{} \frac{1 -
  \vec{\beta}_1 \cdot \vec{\beta}_2}{\sqrt{a (1 - b)}} \,, 
\end{eqnarray}
where
\beq
\vec{\beta}_1 = \frac{\vec{k}_1}{(k_1)_0}\,,\qquad
\vec{\beta}_2 = \frac{\vec{k}_2}{(k_2)_0}\,.
\end{equation}
The calculation of this integral is long and cumbersome. However, its
correctness can be easily checked numerically, once the analytic answer has
been obtained. Codes for checking the soft integrals are included in the
\tmtexttt{Notes} subdirectory of the \POWHEGBOX.

\subsubsection{Two massive particles with equal momenta}
In the particular case when $k_1 = k_2 = p$, $p^2 \ne0$,  we have
\begin{equation}
  I_0(p,p) = 2, \qquad I_{\epsilon}(p,p) = \frac{2}{\beta} \log \frac{1 +
    \beta}{1 - \beta}, \qquad \beta = \frac{| \vec{p} |}{p_0},
\end{equation}
so that
\begin{equation}
  \mathcal{R}^s_{12} = \frac{B}{\epsilon} + C,
\end{equation}
with
\begin{eqnarray}
  B & = & - 1\,, \\
  C & = & - \log \frac{Q^2}{s \xi_c^2} - \frac{1}{\beta} \log \frac{1 +
  \beta}{1 - \beta}\, . 
\end{eqnarray}

\section{Collinear limits}
\label{app:colllim}
We list here the relationship between real- and Born-level squared amplitude,
summed over the final-state and averaged over initial-state colours and spins,
and divided by the appropriate flux factor.

\subsection{Initial-state radiation}
We call $p$ the momentum of the incoming parton and $k$ the momentum of the
parton that enters the underlying Born process. Thus, $(p-k)$ is the momentum
of the on-shell radiated parton, and $k^2<0$. We define
\begin{equation}
k^\mu=z\,p^\mu-\eta^\mu \,\frac{\abs{\ktvec}^2}{2p\cdot \eta\;
  (1-z)}+\kt^\mu\, ,
\end{equation}
and 
\begin{equation}
\hat{k}_{\sss\rm T}^\mu=\frac{\kt^\mu}{\abs{\ktvec}}\,.
\end{equation}
We have
\begin{eqnarray}
  \mathcal{R}^{(g)} (p) & : & \mathcal{B}_{\mu \nu}^{(g)} (z p) \frac{8 \pi
    \as \CA}{- k^2} \left\{ - 2 \left[ \frac{z}{1 - z} + \frac{1 - z}{z} + z
    (1 - z) \right] g^{\mu \nu} \right.
\nonumber\\
&&\left.
\phantom{ \mathcal{B}_{\mu \nu}^{(g)} (z p) \frac{8 \pi
    \as \CA}{- k^2} \ \,\, }
+ \frac{4 (1 - z)}{z} \left[ \hat{k}_{\sss\rm
      T}^{\mu} \hat{k}_{\sss\rm T}^{\nu} + \frac{g^{\mu \nu}}{2} \right]
  \right\} \nonumber\\
   & = & \mathcal{B}_{\mu \nu}^{(g)} (z p) \frac{8 \pi \as \CA}{- k^2}
  \left\{ - 2 \left[ \frac{z}{1 - z} + z (1 - z) \right] g^{\mu \nu} +
  \frac{4 (1 - z)}{z} \hat{k}_{\sss\rm T}^{\mu} \hat{k}_{\sss\rm T}^{\nu}
  \right\}\,, \\
  \mathcal{R}^{(q)} (p) & : & \mathcal{B}^{(q)} (z p) \frac{8 \pi \as \CF}{-
    k^2} \frac{1 + z^2}{1 - z}\,, \\
  \mathcal{R}^{(q)} (p) & : & \mathcal{B}_{\mu \nu}^{(g)} (z p) \frac{8 \pi
    \as \CF}{- k^2} \left\{ - g^{\mu \nu} \frac{1 + (1 - z)^2}{z} + \frac{4
    (1 - z)}{z} \left[ \hat{k}_{\sss\rm T}^{\mu} \hat{k}_{\sss\rm T}^{\nu} +
    \frac{g^{\mu \nu}}{2} \right] \right\} \nonumber\\ & = & \mathcal{B}_{\mu
    \nu}^{(g)} (z p) \frac{8 \pi \as \CF}{- k^2} \left\{ - g^{\mu \nu} z +
  \frac{4 (1 - z)}{z} \hat{k}_{\sss\rm T}^{\mu} \hat{k}_{\sss\rm T}^{\nu}
  \right\}\,, \\
  \mathcal{R}^{(g)} (p) & : & \mathcal{B}^{(q)} (z p) \frac{8 \pi \as \TF}{-
    k^2} \(z^2 + (1 - z)^2\)\, .
\end{eqnarray}

\subsection{Final-state singularities}
We call $k$ the momentum of the splitting parton, and write
\begin{equation}
k^\mu=\frac{p^\mu}{z}+\eta^\mu \,\frac{z\,\abs{\ktvec}^2}{2p\cdot \eta\;
  (1-z)}+\kt^\mu\,. 
\end{equation}
We have
\begin{eqnarray}
  \mathcal{R}^{(g g)} (p) & : & \mathcal{B}_{\mu \nu}^{(g)}
  \left(\frac{p}{z}\right) \frac{8 \pi 
  \as \CA}{k^2}  \left\{ - 2 \left[ \frac{z}{1 - z} + \frac{1 - z}{z} + z
  (1 - z) \right] g^{\mu \nu} \right.
\nonumber\\
&&
\phantom{ \mathcal{B}_{\mu \nu}^{(g)}
  \left(\frac{p}{z}\right) \frac{8 \pi 
  \as \CA}{k^2} \ \, \,}
\left. + 4 z (1 - z) \left[ \hat{k}_{\sss\rm T}^{\mu} 
  \hat{k}_{\sss\rm T}^{\nu} + \frac{g^{\mu \nu}}{2} \right] \right\}
  \nonumber\\ 
  & = & \mathcal{B}_{\mu \nu}^{(g)} \left(\frac{p}{z}\right) \frac{8 \pi \as
  \CA}{k^2}  
  \left\{ - 2 \left[ \frac{z}{1 - z} + \frac{1 - z}{z} \right] g^{\mu \nu} + 4
  z (1 - z) \hat{k}_{\sss\rm T}^{\mu}  \hat{k}_{\sss\rm T}^{\nu} \right\} \,,\\
\label{eq:qg_split}
  \mathcal{R}^{(q g)} (p) & : & \mathcal{B}^{(q)} \left(\frac{p}{z}\right)
  \frac{8 \pi \as 
  \CF}{k^2}  \frac{1 + (1 - z)^2}{z}\,, \\
  \mathcal{R}^{(q \bar{q})} (p) & : & \mathcal{B}_{\mu \nu}^{(g)}
  \left(\frac{p}{z}\right) 
  \frac{8 \pi \as \TF}{k^2} \left\{ - g^{\mu \nu} \(z^2 + (1 - z)^2\) - 4 z (1
  - z) \left[ \hat{k}_{\sss\rm T}^{\mu} \hat{k}_{\sss\rm T}^{\nu} +
    \frac{g^{\mu \nu}}{2} \right] \right\}\, \nonumber\\
  & = & \mathcal{B}_{\mu \nu}^{(g)} \left(\frac{p}{z}\right) \frac{8 \pi \as
  \TF}{k^2} \left\{ - 
  g^{\mu \nu} - 4 z (1 - z) \hat{k}_{\sss\rm T}^{\mu} \hat{k}_{\sss\rm
    T}^{\nu} \right\}\,.
\end{eqnarray}
In eq.~(\ref{eq:qg_split}), it is assumed that $p$ is the momentum of the
gluon.

\section{Upper bounding functions for FSR}
\label{app:ubfsr}
We use an upper bounding function of the following form (for ease of notation
we write $\as$ instead of $\as^{\rm rad}$)
\begin{equation}
  U(\xi,y) = N \frac{\as (\kt^2)}{\xi (1 - y)}\,,
\end{equation}
with
\begin{equation}
 \label{eq:k2fsr}
  \kt^2 = \frac{s}{2} \,\xi^2 (1 - y)\,,
\end{equation}
$s$ being the partonic CM energy squared, and
\begin{equation}
  \label{eq:alphagenrad}
  \as \(\kt^2\) = \frac{1}{b_0 \log \(\kt^2/\Lambda^2\)}\,.
\end{equation}
The ranges of $\xi$, $y$ and $\phi$ are given by
\begin{equation}
 \label{eq:radrange}
  0 \le \xi \le \xi_{\max}\equiv \frac{s - M^2_{\tmop{rec}}}{s},\qquad - 1 \le
  y \le 1, \qquad 0 \le \phi < 2 \pi\,,
\end{equation}
where $M_{\tmop{rec}}$ is the mass of the system recoiling against the
emitter and emitted partons (see eq.~(5.49) or
ref.~\cite{Frixione:2007vw}). We want to generate $\pt$ uniformly in
\begin{equation}
  \Delta^{(U)} (\pt) = \exp \left[ - \int U(\xi,y)\, \theta\!\(\kt - \pt\)
   \,d \xi \, dy \,d \phi \right].
\end{equation}
Trading $y$ for $\kt^2$ (see eq.~(\ref{eq:k2fsr})), we get
\begin{eqnarray}
 \label{eq:ubsolution1fsr}
  - \log \Delta^{(U)} (\pt) & = & 2 \pi N \int_0^{\xi_{\max}} \frac{d
    \xi}{\xi} \int_{\pt^2}^{\xi^2 s} \frac{d \kt^2}{\kt^2} \,\as (\kt^2)
  \nonumber\\ & = & \frac{\pi N}{b_0} \, \theta \(\xi_{\max}^2 -
  \frac{\pt^2}{s}\) \left[ \log \frac{\xi_{\max}^2 s}{\Lambda^2} \log
    \frac{\log \(\xi_{\max}^2 s/\Lambda^2\)}{\log
      \( \pt^2/\Lambda^2\)} - \log \frac{\xi_{\max}^2 s}{\pt^2} \right].
\phantom{aaaa} 
\end{eqnarray}
The equation $r = \Delta^{(U)} (\pt)$ is translated into $\log r = \log
\Delta^{(U)} (\pt)$, and solved numerically for $\pt$.  Once $\pt$ is
generated, we generate $\xi$ uniformly in $\log \xi$ (see second member of
eq.~(\ref{eq:ubsolution1fsr})) within the limits
\begin{equation}
  \frac{\pt^2}{s} \leqslant \xi \leqslant \xi_{\max}\,,
\end{equation}
and then use
\begin{equation}
  \pt^2 = \frac{s}{2} \, \xi^2 (1 - y)
\end{equation}
to obtain $y$, while $\phi$ is generated uniformly between $0$ and $2 \pi$.

More options are offered in the \POWHEGBOX.  They are both related to the use
of the upper bounding function
\begin{equation}\label{eq:moreoptions}
  U(\xi,y) = N\frac{\as (\kt^2)}{\xi^2 (1 - y) \left( 1 - \frac{\xi}{2} (1 - y)
    \right)^2}\,,
\end{equation}
which yields
 \begin{eqnarray}
\label{eq:ubsolution2fsr}
  - \log \Delta^{(U)} (\pt) \!& = &\! 2 \pi N \int_{\pt^2}^{s \xi_{\max}^2}
  \frac{d \kt^2}{\kt^2} \int_{\sqrt{\kt^2 / s}}^{\xi_{\max}} \frac{d
    \xi}{(\xi - \kt^2 / s)^2} 
\nonumber\\ 
\!& = & \!4 \pi N \left[ \frac{1}{2
      \xi_{\max}} \left( \log \frac{1 - \xi_{\max}}{\xi_{\max}} (1 - 2
    \xi_{\max}) - 2 \right) \right. \nonumber\\ 
\!& - & \!\!\left. \frac{1}{2 p\, \xi_{\max}} \!\left( p \log \(\xi_{\max} -
p^2\) - 2 \!\left(\! p \log \frac{1 - p}{p} + 1 \!\right) \!\xi_{\max} - 2 p
\log p \right) \right]\!, \phantom{aaaaa}
\end{eqnarray}
where $p = \sqrt{\pt^2 / s}$.  The second option implemented in the
\POWHEGBOX{} makes use of the upper bounding function in
eq.~(\ref{eq:moreoptions}) multiplied by a factor of $\xi$. One then uses the
same expression~(\ref{eq:ubsolution2fsr}) for the generation of $\pt$, and
uses a further veto, accepting the event if a given random number is less
than $\xi$, to implement the extra factor of $\xi$.  The third option is
similar to the second, but with an extra factor of $1-\xi(1-y)/2$.

\section{Upper bounding functions for ISR}
\label{app:ubisr}
The upper bounding function is (where $x=1-\xi$, so that the singular limit
is reached when $x\to1$)
\begin{equation}
  U(x,y) = N \frac{\as (\kt^2)}{(1 - x) (1 - y^2)}\,,
\end{equation}
with
\begin{equation}
  \kt^2 = \frac{s_b}{4 x}\, (1 - x)^2 (1 - y^2)\,,
\end{equation}
$s_b$ being the underlying Born CM energy squared. The range of $U(x,y)$
must cover the range of the radiation variables for the given underlying Born
configuration.  A practical restriction for the range of $U(x,y)$ is
\begin{equation}
  \rho \leqslant x \leqslant 1\,,\qquad \kt^2 \leqslant k_{\sss \rm T
    \max}^2\,,
\end{equation}
where
\begin{equation}
  \rho = \frac{s_b}{S}\,, \qquad k_{\sss \rm T \max}^2 = s_b \frac{\(1 -
    \bar{x}_\splus^2\) \(1 - \bar{x}_\sminus^2\)}{( \bar{x}_\splus +
    \bar{x}_\sminus)^2} \,. 
\end{equation}
We want to generate $\pt$ uniformly in
\begin{equation}
  \Delta^{(U)} (\pt) = \exp \left[ - \int U(x,y) \,\theta\!\(\kt - \pt\) d
    x \, d y \, d \phi \right],
\end{equation}
in the given range. We assume $0 \leqslant \phi \leqslant 2 \pi$. Trading $y$
for $\kt^2$ we find
\begin{equation}
  |y| = \sqrt{1 - \frac{4 x}{(1 - x)^2} \frac{\kt^2}{s_b}}\,,
\end{equation}
and
\begin{equation}
  \int d x \int d y \int_0^{2 \pi} d \phi \, U(x,y) \,\theta\!\(\kt - \pt\)
  = 2 \pi N \int_{\rho}^{x_-} d x \int_{\pt^2}^{k_{T \max}^2} \frac{d
  \kt^2}{\kt^2} \frac{\as\!\(\kt^2\)}{\sqrt{(x_+ - x) (x_- - x)}}\,,
\end{equation}
where
\begin{equation}
  x_{\pm} = \left( \sqrt{1 + \frac{\kt^2}{s_b}} \pm \frac{\kt}{s_b} \right)^2.
\end{equation}
The $x$ integration can be performed to yield
\begin{equation}
  \int d x \int d y \int_0^{2 \pi} d \phi \, U(x,y) \,\theta (\kt - \pt) =
  \int_{\pt^2}^{k_{T \max}^2} \frac{d \kt^2}{\kt^2} \, V (\kt^2)\,,
\end{equation}
where
\begin{equation}
  V (\kt^2) = 2 \pi N \as (\kt^2) \log \frac{\sqrt{x_+ - \rho} +
  \sqrt{x_- - \rho}}{\sqrt{x_+ - \rho} - \sqrt{x_- - \rho}}\, .
\end{equation}
We observe that
\begin{equation}
  \log \frac{\sqrt{x_+ - \rho} + \sqrt{x_- - \rho}}{\sqrt{x_+ - \rho} -
  \sqrt{x_- - \rho}} \leqslant \log \frac{\sqrt{x_+} + \sqrt{x_-}}{\sqrt{x_+}
  - \sqrt{x_-}} = \frac{1}{2} \log \frac{\kt^2 + s_b}{\kt^2}\, .
\end{equation}
In ref.~\cite{Nason:2006hf}, it is suggested to use the bound
\begin{equation}
  \frac{1}{2} \log \frac{\kt^2 + s_b}{\kt^2} \le \frac{1}{2} \log
  \frac{q^2}{\kt^2}\,, \qquad \tmop{with}\ q^2 = k_{\sss \rm T \max}^2 +
  s_b\,,
\end{equation}
and to define
\begin{equation}
  \tilde{V} (\kt^2) = 2 \pi N \as (\kt^2) \frac{1}{2} \log
  \frac{q^2}{\kt^2} \geqslant V (\kt^2)\, .
\end{equation}
The $d \kt^2$ integral of $\tilde{V}$ can be performed analytically, yielding
\begin{equation}
  \int_{\pt^2}^{k_{T \max}^2} \frac{d \kt^2}{\kt^2} \, \tilde{V} (\kt^2) =
  \frac{\pi N}{b_0} \left[ \log \frac{q^2}{\Lambda^2} \log \frac{\log
  \(k_{\sss \rm T \max}^2/\Lambda^2\)}{\log \( \pt^2/\Lambda^2\)} - \log
  \frac{k_{\sss \rm T \max}^2}{\pt^2} \right] .
\end{equation}
One generates $\pt$ uniformly in
\begin{equation}
  \tilde{\Delta} (\pt) = \exp \left[ - \int_{\pt^2}^{k_{T \max}^2} \frac{d
  \kt^2}{\kt^2} \, \tilde{V} (\kt^2) \right],
\end{equation}
and then use the veto method to get the $\pt$ distributed according to
$\Delta^{(U)} (\pt)$.

The following variant of this procedure has been introduced in
ref.~\cite{Frixione:2007nw}. We have used the bound
\begin{equation}
  \log \frac{\kt^2 + s_b}{\kt^2} \leqslant \left\{ \begin{array}{ll}
    \log \frac{2 s_b}{\kt^2} & \ \tmop{for}\  \kt^2 < s_b\\
    \log 2 & \ \tmop{for}\  \kt^2 > s_b
  \end{array} \right.
\end{equation}
so
\begin{equation}
  \tilde{V} (\kt^2) = \pi N \as (\kt^2) \left[ \theta (s_b - \kt^2) \log
  \frac{2 s_b}{\kt^2} + \theta (\kt^2 - s_b) \log 2 \right] .
\end{equation}
We then get
\begin{eqnarray}
  \log \tilde{\Delta} (\pt) & = & \theta\!\(s_b - \pt^2\) \frac{\pi N}{b_0}
  \left\{ \theta\(k_{\sss \rm T \max}^2 - s_b\) \left[ \log \frac{2
      s_b}{\Lambda^2} \log \frac{\log \(s_b/\Lambda^2\)}{\log \(
      \pt^2/\Lambda^2\)} - \log \frac{s_b}{\pt^2} \right.\right.\nonumber
\\
 &+& \left.\left.  \log (2) \, \log \frac{ \log
      \( k_{\sss \rm T \max}^2/\Lambda^2\)}{\log \( s_b/\Lambda^2\)} \right]
\right.\nonumber
\\
&+& \left. \theta\! \(s_b - k_{\sss \rm T \max}^2\) \!\!\left[
    \log \frac{2 s_b}{\Lambda^2} \,\log \frac{\log \( k_{\sss \rm T
	  \max}^2 / \Lambda^2\)}{\log \( \pt^2 /\Lambda^2\)} - \log
    \frac{k_{T \max}^2}{\pt^2} \right]\! \right\} \nonumber
\\ & + & \theta \!\(\pt^2 - s_b\) \frac{\pi N}{b_0} \log (2) \,\log \frac{\log
  \( k_{\sss \rm T \max}^2 / \Lambda^2\) }{\log \( \pt^2/\Lambda^2\)}\,.
  \label{eq:logdeltisr}
\end{eqnarray}
To improve the behaviour for small $x$ effects it may be convenient to use
instead
\beq
U(x,y)  = N \frac{\as (\kt^2)}{x (1 - x) (1 - y^2)} 
\eeq
as upper bounding function. As derived in ref.~\cite{Nason:2006hf}
\beq 
\int U(x,y)  \, \theta \!\(\kt  - \pt\) \, d \Phi_r =
\int_{\pt^2}^{k^2_{\sss \rm T \max}} \frac{d \kt^2}{\kt^2}\, V \(\kt^2\)\,, 
\eeq
where
\beq
V \(\kt^2\) = \pi N \as (\kt^2) \log \frac{\sqrt{x_+ - \rho} + \sqrt{x_- -
   \rho}}{\sqrt{x_+ - \rho} - \sqrt{x_- - \rho}} \,. 
\eeq
With the new upper bounding function one gets instead
\beq
 V (\kt^2) = \pi N \as (\kt^2) \left[ \log \frac{2}{\rho} + \log
   \frac{\sqrt{(x_+ - \rho) (x_- - \rho)} + 1 - \frac{\rho}{2} (x_+ +
   x_-)}{x_+ - x_-} \right] . 
\eeq
In this case too $V(\kt^2)$ satisfies a simple upper bound
\beq
V (\kt^2) < \pi N \as (\kt^2) \log \frac{S}{\kt^2}\,, 
\eeq
that can be used for fast generation by vetoing.

\section{Choice of scales}\label{app:scales}
\subsection{Scales and couplings for the inclusive cross section}
In the evaluation of $\tilde{B}$ and of the remnant function, a user
provided, process dependent subroutine
\tmtexttt{set\_fac\_ren\_scales(muf,mur)} sets the factorization and
renormalization scales. It should only depend upon the underlying Born
kinematics (thus, for example, it cannot depend upon the radiation transverse
momentum).  It is called by the \POWHEGBOX{} subroutine
\tmtexttt{setscalesbtilde} (called during the evaluation of the $\tilde{B}$
function and of the remnant cross section) that stores the square of the
factorization scale, the square of the renormalization scale and the strong
coupling constant in three global variables of the \tmtexttt{st\_} common
block: \tmtexttt{st\_muren2}, \tmtexttt{st\_mufact2} and
\tmtexttt{st\_alpha}. By inspecting the subroutine we see that the
factorization and renormalization scales are set equal to the square of the
scales returned by the \tmtexttt{set\_fac\_ren\_scales} subroutine multiplied
by renormalization and factorization scale factors \tmtexttt{st\_facfact} and
\tmtexttt{st\_renfact}. These factors are in turn read from the variables
\tmtexttt{facscfact} and \tmtexttt{renscfact} in the \POWHEG{} data file,
normally during the execution of the user initialization routine
\tmtexttt{init\_phys}. If they are not present in the data file, they are set
by default to~1.  The function \tmtexttt{pwhg\_alphas(mu2,Lambda5,n)} returns
the strong coupling constant with \tmtexttt{n} light flavours, as a function
of the square of the scale and of $\Lambda_{\overline{\rm MS}}^{(5)}$. If
called with \tmtexttt{n} $<0$ it uses, as number of flavours, the number of
quarks with mass less than the input scale.  In the evaluation of the
$\tilde{B}$ function, the number of flavours \tmtexttt{st\_nlight} is
used. This is set by the user in the \tmtexttt{init\_couplings} routine.

\subsection{Scales and couplings for radiation}
The choice of scales and couplings in the generation of radiation requires
particular attention, since the shape of the Sudakov peak, in the radiation
transverse momentum, is deeply affected by it. It is discussed in detail in
ref.~\cite{Frixione:2007vw}. Here we report how this is actually done in the
\POWHEGBOX{}. The relevant code is in the subroutine
\tmtexttt{set\_rad\_scales(ptsq)}. It is typically invoked with the radiation
transverse momentum as argument. With this choice one can achieve, in some
cases, complete next-to-leading logarithmic~(NLL) accuracy in the \POWHEG{}
Sudakov form factor~\cite{Nason:2006hf, Frixione:2007vw}.  By inspecting the
code, we can see that it sets \tmtexttt{st\_mufact2} and
\tmtexttt{st\_muren2} to \tmtexttt{ptsq}. The factorization and
renormalization scale factors, \tmtexttt{st\_facfact} and
\tmtexttt{st\_renfact}, are not used in this context.
The program also takes care that the factorization scale never goes below the
minimum allowed value in the pdf's. The strong coupling is then evaluated,
the number of flavours is taken as the number of quarks with mass below the
\tmtexttt{ptsq} scale. Furthermore, the strong coupling constant is
multiplied by the factor given in formula~(4.32) in
ref.~\cite{Frixione:2007vw}.  This factor improves the NLL accuracy of the
Sudakov form factor.

During the generation of radiation, we need a simplified one loop expression
for the running coupling that is an upper bound of the running coupling that
we use. We choose
\begin{equation}
\as^{\tmop{rad}}=\frac{1}{b_0^{(5)}\log \frac{\mu^2}{\Lambda_{\rm ll}^2}},
\quad\quad b_0^{(5)}=\frac{33-2\times 5}{12\pi}\;,
\end{equation}
and we fix $\Lambda_{\rm ll}$ by requiring
\begin{equation}
\as^{\tmop{rad}}(\mu^0)=\as^{\tmop{PW}}(\mu^0)\;,
\end{equation}
where the scale $\mu_0$ is the minimum allowed value for the renormalization
scale, and is taken equal to $2\Lambda^{(5)}_{\sss \MSB}$. The value of
$\Lambda_{\rm ll}$ is stored in the global variable \tmtexttt{rad\_lamll}.

\section{Miscellaneous features of the code}\label{app:misc}

\subsection{Checking the soft, collinear and soft-collinear limits}
\label{app:check}
In the \POWHEGBOX, the routine \tmtexttt{checklims}, through a call to the
subroutines \tmtexttt{checksoft} and \tmtexttt{checkcoll}, allows the user to
check the limiting behaviours of the real squared amplitudes, against their
soft and collinear approximations.  This routine has been
used as a debug feature in the developing stages of the implementation
of specific processes.
If activated by the flags \tmtexttt{dbg\_softtest} and
\tmtexttt{dbg\_colltest} in the \tmtexttt{init\_phys.f} file, it can be used
now to check if the Born, the colour-correlated and spin-correlated Born
amplitudes (used to build the limiting expressions of the real amplitudes)
are consistent with the real contributions, computed in the kinematic
configurations where a gluon becomes soft or when it becomes collinear to
another parton or when two quarks of opposite flavours become collinear.

The double soft-collinear and collinear-soft limits are also
tested. They do not depend upon the real squared amplitudes, but only
upon the Born amplitudes.  They have been used initially to check the
consistency of the \POWHEGBOX{} code, but they can also be used to
perform some checks of the colour correlated Born amplitudes during
the development of new processes.

\subsection{Names}
In the present work, we have made little references to the names of the
fortran files where the various subroutines are stored. We assume that the
reader can find them out by using standard command-line tools.  However,
while we have not spent much energy in deciding how to organize fortran
files, the include files are rather well organized.  For example, the include
file \tmtexttt{pwhg\_flst.h} declares common-block variables that refer to
flavour structures. All these variables start with a prefix
\tmtexttt{flst\_}.  Other important global variables are the kinematics ones
(\tmtexttt{kn\_}), those involving the scales and the strong coupling
(\tmtexttt{st\_}), those involving the generation of radiation
(\tmtexttt{rad\_}), and so on. In this way, when finding such variables in
the code, by inspecting the include files, the reader can better follow what
is their purpose.

\subsection{Input variables}
The \POWHEGBOX{} gets its input data from the routine \tmtexttt{powheginput}.
Upon its first invocation, this routine looks for a file named
\tmtexttt{powheg.input}. If it does not find it, asks interactively to enter
a prefix, and then looks for the file \tmtexttt{"prefix"-powheg.input}. It
reads all the input file at once.  Then, if invoked in the form
\tmtexttt{powheginput("string")} it returns the (real) value associated to
\tmtexttt{"string"} in the input file.  If no matching string is found, it
prints a message and aborts the program.  If invoked in the form
\tmtexttt{"\#string"}, in case no matching string is found, it returns a very
unlikely value $-10^6$. This last mechanism is used in the \POWHEGBOX{} to
set default values.

\subsection{User files}
The files that contain the user routines are organized in the same
way in all the examples provided with the code of the \POWHEGBOX{}.
They are:
\begin{itemize}
\item \tmtexttt{nlegborn.h} contains the number of legs of the Born process,
  \tmtexttt{nlegborn};
\item 
  \tmtexttt{init\_processes.f} contains the subroutine
  \tmtexttt{init\_processes}, whose major task is to fill the list 
  \tmtexttt{flst\_born} and \tmtexttt{flst\_real};
\item 
  \tmtexttt{init\_couplings.f} contains the subroutine
  \tmtexttt{init\_couplings}, that initializes process-dependent
  couplings;
\item in \tmtexttt{PhysPars.h}, there is a collection of physical variables
  that are in common with many subroutines (masses, electroweak couplings,
  widths\ldots);
\item \tmtexttt{Born.f} contains the \tmtexttt{setborn} subroutine;
\item in \tmtexttt{Born\_phsp.f} the user can found the \tmtexttt{born\_phsp}
subroutine;
\item in \tmtexttt{real.f}, the  \tmtexttt{setreal} routine;
\item and finally, in \tmtexttt{virtual.f}, the \tmtexttt{setvirtual}
routine.
\end{itemize}
A template for the \tmtexttt{analysis} subroutine can be found in
\tmtexttt{pwhg\_analysis.f}.

\bibliography{paper}

\end{document}